\documentclass{article}

\usepackage{algorithm}
\usepackage{algpseudocode}
\usepackage{amssymb}
\usepackage{amsmath}
\usepackage[style=ieee,isbn=false,url=false]{biblatex}
\usepackage{booktabs}
\usepackage[final]{changes}
\usepackage{circuitikz}
\usepackage[margin=1in]{geometry}
\usepackage{glossaries}
\usepackage{graphicx}
\usepackage{hyperref}
\hypersetup{
    colorlinks=true, 
    linkcolor=black,   
    citecolor=black,  
    urlcolor=black, 
}
\usepackage{lineno}
\usepackage{makecell}
\usepackage{orcidlink}
\usepackage{pgfplots}
\usepackage{pgfplotstable}
\pgfplotsset{compat=1.18}
\usepgfplotslibrary{groupplots}
\usepgfplotslibrary{fillbetween}
\usepackage{setspace}
\usepackage{siunitx}
\usepackage{subcaption}
\usepackage{tabularx}
\newcolumntype{C}{>{\centering\arraybackslash}X}
\usepackage{tikz}
\usetikzlibrary[plotmarks]
\usetikzlibrary{arrows.meta,shapes.geometric,patterns,positioning,spy}
\usetikzlibrary{positioning,fit}
\usetikzlibrary{decorations.pathmorphing, shapes.multipart}
\usepackage{todonotes}
\usepackage{url}

\urldef{\urlDWaveAutoScale}\url{https://docs.dwavequantum.com/en/latest/quantum_research/solver_parameters.html#auto-scale}
\urldef{\urlDWaveICEs}\url{https://docs.dwavequantum.com/en/latest/quantum_research/errors.html#ice}

\usepackage{cleveref}

\definecolor{oi-blue}   {RGB}{0,114,178}
\definecolor{oi-orange} {RGB}{230,159,0}
\definecolor{oi-sky}    {RGB}{86,180,233}
\definecolor{oi-vermilion}{RGB}{213,94,0}
\definecolor{oi-green}  {RGB}{0,158,115}
\definecolor{oi-yellow} {RGB}{240,228,66}
\definecolor{oi-purple} {RGB}{204,121,167}
\definecolor{oi-grey}   {RGB}{128,128,128}
\newacronym{ae}{AE}{Fixstars Amplify Annealing Engine}
\newacronym{ba}{BA}{``best approximation''}
\newacronym{cpu}{CPU}{central processing unit}
\newacronym{da}{DA}{digital annealing}
\newacronym{dac}{DAC}{digital-to-analog converter}
\newacronym{fem}{FEM}{finite element method}
\newacronym{gpu}{GPU}{graphics processing unit}
\newacronym{ice}{ICE}{integrated control error}
\newacronym{pubo}{PUBO}{polynomial unconstrained binary optimization}
\newacronym{qa}{QA}{quantum annealing}
\newacronym{qpu}{QPU}{quantum processing unit}
\newacronym{qubo}{QUBO}{quadratic unconstrained binary optimization}
\newacronym{sa}{SA}{simulated annealing}
\newacronym{sdk}{SDK}{software developing kit}
\newacronym{fsi}{FSI}{fluid structure interaction}

\newcommand{\lp}{\left(}
\newcommand{\rp}{\right)}
\newcommand{\lb}{\left[}
\newcommand{\rb}{\right]}
\newcommand{\lcb}{\left\{}
\newcommand{\rcb}{\right\}}
\newcommand{\lab}{\langle}
\newcommand{\rab}{\rangle}

\newcommand{\designVar}{\alpha}
\newcommand{\designVarVec}{\boldsymbol{\designVar}}
\newcommand{\setAdmissibleDesigns}{\mathbb{A}}
\newcommand{\designVarVecBest}{\designVarVec^*}
\newcommand{\nChoices}{n_{c}}

\newcommand{\stateVar}{y}
\newcommand{\stateVarVec}{\boldsymbol{\stateVar}}
\newcommand{\numStateVar}{m}
\newcommand{\stateSpace}{\mathbb{R}^\numStateVar}
\newcommand{\stateVarVecBest}{\stateVarVec^*}

\newcommand{\objective}{\mathcal{F}}
\newcommand{\constraintsEquality}{\mathcal{G}}
\newcommand{\constraintsInequality}{\mathcal{H}}

\newcommand{\hamiltonian}{H}

\newcommand{\hamiltonianIsing}{\hamiltonian_{\text{Ising}}}
\newcommand{\spin}{s}
\newcommand{\spinVec}{\boldsymbol{\spin}}
\newcommand{\isingCoeffQuadratic}{J}
\newcommand{\isingCoeffLinear}{h}

\newcommand{\hamiltonianQUBO}{\hamiltonian_{\text{QUBO}}}
\newcommand{\binary}{x}
\newcommand{\binaryVec}{\boldsymbol{\binary}}
\newcommand{\binaryVecBest}{\binaryVec^*}
\newcommand{\quboMatrixSymbol}{Q}
\newcommand{\quboMatrix}{\boldsymbol{\quboMatrixSymbol}}
\newcommand{\quboMatrixCoeff}{\quboMatrixSymbol}
\newcommand{\numQubits}{\mathcal{N}}
\newcommand{\numQubitsPerNode}{N}

\newcommand{\offsetCoeff}{\tau}
\newcommand{\scaleCoeff}{\lambda}
\newcommand{\repCoeff}{c}

\newcommand{\stateVarMin}[1][i]{\stateVar_{#1,\text{min}}}
\newcommand{\stateVarMax}[1][i]{\stateVar_{#1,\text{max}}}

\newcommand{\relaxationFactor}{\rho}

\newcommand{\structure}{s}

\newcommand{\boundaryDispl}{\boundary^{\displ}}

\newcommand{\displ}{u}

\newcommand{\displPrescribed}{\hat{\displ}}

\newcommand{\setKinematicallyAdmissible}{\mathbb{U}}

\newcommand{\volumeForce}{f}

\newcommand{\strainTensor}{\epsilon}
\newcommand{\elasticityTensor}{C}
\newcommand{\strainEnergy}{U}

\newcommand{\potentialExternalForces}{W}

\newcommand{\potentialEnergy}{\Pi}

\newcommand{\fluid}{f}

\newcommand{\pressure}{p}

\newcommand{\length}{L}
\newcommand{\lengthPiston}{\length_\structure}
\newcommand{\lengthChamber}{\length_\fluid}
\newcommand{\lengthChamberInitial}{\lengthChamber^{(0)}}
\newcommand{\area}{A}
\newcommand{\areaPiston}{\area_\structure}
\newcommand{\areaChamber}{\area_\fluid}

\newcommand{\heatRatio}{\gamma}

\newcommand{\hamiltonianQUBOPotEnergy}[1][]{\hamiltonianQUBO^{\potentialEnergy#1}}

\newcommand{\stressSymbol}{\sigma}
\newcommand{\stressTensor}{\boldsymbol{\stressSymbol}}

\newcommand{\stressTensorComponent}[1]{\stressSymbol_{#1}}

\newcommand{\stress}{\stressSymbol}
\newcommand{\displacementSymbol}{u}
\newcommand{\displacement}{\displacementSymbol}
\newcommand{\displacementPrescribed}{\hat{\displacement}}
\newcommand{\displacementVector}{\boldsymbol{\displacementSymbol}}

\newcommand{\tractionSymbol}{t}
\newcommand{\traction}{\tractionSymbol}

\newcommand{\tractionPrescribed}{\hat{\traction}}

\newcommand{\bodyForceDensity}{f}
\newcommand{\domain}{\Omega}
\newcommand{\boundary}{\Gamma}

\newcommand{\boundaryTraction}{\boundary^{\stressSymbol}}
\newcommand{\normal}{n}

\newcommand{\complianceTensorSymbol}{S}

\newcommand{\complianceTensorComponent}[1]{\complianceTensorSymbol_{#1}}

\newcommand{\superscriptComplementaryEnergy}{*}
\newcommand{\totalComplEnergy}{\Pi^{\superscriptComplementaryEnergy}}

\newcommand{\complStrainEnergy}{U^{\superscriptComplementaryEnergy}}

\newcommand{\complExternal}{W^{\superscriptComplementaryEnergy}}

\newcommand{\setStaticallyAdmissible}{\boldsymbol{\mathbb{S}}}

\newcommand{\designVarSymbol}{\alpha}
\newcommand{\designVarVector}{\boldsymbol{\designVarSymbol}}

\newcommand{\setAdmissibleDesign}{\boldsymbol{\mathbb{A}}}

\newcommand{\rodLength}{L}

\newcommand{\youngModulus}[1][]{E_{#1}}
\newcommand{\crossSectionalArea}[1][]{A_{#1}}
\newcommand{\x}{x}
\newcommand{\elemIndex}{e}
\newcommand{\nElem}{n_e}

\newcommand{\nodeIndex}{i}

\newcommand{\basisFctSymbol}{\phi}
\newcommand{\basisFct}{\basisFctSymbol}
\newcommand{\basisCoeff}{a}
\newcommand{\forceSymbol}{F}
\newcommand{\force}{\forceSymbol}
\newcommand{\forceAnalytic}{\forceSymbol^{*}}

\newcommand{\binariesVecCoeff}{\binaryVec^{\basisCoeff}}

\newcommand{\qubitsVector}{\binaryVec}

\newcommand{\qubitsVectorCoeffs}[1][]{\qubitsVector^{\basisCoeff#1}}

\newcommand{\qubitsVectorDesigns}[1][]{\qubitsVector^{\crossSectionalArea#1}}

\newcommand{\penaltyWeight}{\lambda}
\newcommand{\penaltyTerm}{\pi}

\newcommand{\crossSectionalAreaChoice}[1]{\mathcal{A}_{#1}}

\newcommand{\qubitsVectorAuxiliary}{\boldsymbol{\Hat{\binary}}}

\newcommand{\hamiltonianQUBOComplEnergy}[1][]{\hamiltonianQUBO^{\totalComplEnergy#1}}

\newcommand{\iter}{k}
\newcommand{\maxIter}{k_{\mathrm{max}}}
\newcommand{\relChangeDisplacement}{\Delta_{\displ}^{\!(\iter)}}
\newcommand{\tolRelChangeDisplacement}{\varepsilon_{\mathrm{tol}}}
\newcommand{\penaltyWeightFactor}{\eta}
\newcommand{\objectivePenalized}{\mathcal{J}}
\newcommand{\tolFeasibility}{\varepsilon_{\mathrm{feas}}}
\newcommand{\basisCoeffMin}[1][i]{\basisCoeff_{#1,\text{min}}}
\newcommand{\basisCoeffMax}[1][i]{\basisCoeff_{#1,\text{max}}}

\newcommand{\isingScalingFactor}{\beta}
\newcommand{\isingCoeffLinearScaled}{\hat{\isingCoeffLinear}}
\newcommand{\isingCoeffQuadraticScaled}{\hat{\isingCoeffQuadratic}}

\newcommand{\relErrorHOne}{\epsilon_{H^1}}

\newcommand{\annealingTime}{t_{A}}
\newcommand{\numReads}{n_{\mathrm{reads}}}
\newcommand{\nRuns}{n_{\mathrm{runs}}}

\title{Adaptive Encoding Strategy for Quantum Annealing in Mixed-Variable Engineering Optimization}

\newcommand{\Author}[5]{%
  \begin{minipage}[t]{0.45\textwidth}
    \centering
    #1\if\relax\detokenize{#5}\relax\else\thanks{Corresponding author.}\,\fi\,\orcidlink{#2}\\
    \small #3\\
    \small #4
  \end{minipage}
}

\author{
  \Author{Fabian Key}{0000-0001-6622-4806}{fabian.key@tuwien.ac.at}{TU Wien, Vienna, Austria}{corr}
  \and
  \Author{Lukas Freinberger}{0009-0000-3251-1153}{lukas.freinberger@tuwien.ac.at}{TU Wien, Vienna, Austria}{}
  \and
  \Author{Mayu Muramatsu}{0000-0002-6487-0457}{muramatsu@mech.keio.ac.jp}{Keio University, Yokohama, Kanagawa, Japan}{}
  \and
  \Author{Norbert Hosters}{0000-0003-1174-4446}{hosters@cats.rwth-aachen.de}{RWTH Aachen University, Aachen, Germany}{}
}
\date{}

\begin{document}

%%%%%%%%%%%%%%%%
%% TITLE PAGE %%
%%%%%%%%%%%%%%%%
\maketitle

%% ABSTRACT %%
\begin{abstract}
% Introduction
Mixed discrete–continuous optimization is central to engineering design, exemplified by structural design where discrete choices interact with continuous fields such as displacement and stress. These problems are difficult due to high-dimensional, complex search spaces. To tackle them, \gls{qa}  is promising, yet its native binary nature supports only discrete variables, making accurate and efficient encodings of continuous quantities a central challenge.
% Limitations of existing approaches
Existing approaches either split the coupled problem, mapping discrete decisions to \gls{qa} while solving continuous fields classically, or use fixed-bit-depth encodings. The former compromises \gls{qa}’s global search advantages; the latter can underrepresent dynamic range or inflate the number of binary variables, harming solution quality and scalability. In line with this, we show that simply increasing bit depth can even degrade performance on current \gls{qa} hardware, underscoring the need for alternative encodings.
% Objective 
In response, we introduce an adaptive encoding strategy for continuous variables in \gls{qa} that enables efficient treatment of coupled mixed-variable problems. 
% Methodology
To this end, we propose an update strategy for the representable ranges of the continuous variables. We demonstrate its utility by integrating it into the minimum complementary energy formulation for structural design optimization, which provides a single, coupled constrained problem over stress and design variables. We apply a quadratic penalty method where, at each iteration, we update the representation of the continuous variables while targeting the full original objective, preserving \gls{qa}’s global search capability.
% Results
On a published structural design benchmark, namely the size optimization of a composite rod, our adaptive encoding improves solution quality by orders of magnitude under a fixed binary variable budget, demonstrating a superior precision–resource trade-off. 
% Conclusion 
Since the framework generalizes beyond structural design, it offers practical guidance for encoding continuous variables for \gls{qa} and indicates that adaptive representations can enhance precision on current hardware.
\end{abstract}

%% KEYWORDS %%
\section*{Keywords}
Applied quantum computing; quantum annealing; adaptive encoding strategy; mixed discrete–continuous optimization; structural design optimization; composite rod; computational mechanics
%% DATA AVAILABILITY %%
\section*{Data availability}
The data presented in this study are available at: \url{https://doi.org/10.48436/vmpfx-80w27}.
%% FUNDING %%
\section*{Funding}
This research was funded in whole or in part by the Austrian Science Fund (FWF) 10.55776/ESP2444325. For open access purposes, the author has applied a CC BY public copyright license to any author accepted manuscript version arising from this submission.
%% ACKNOWLEDGMENTS %%
\section*{Acknowledgments}
The authors gratefully acknowledge the Jülich Supercomputing Centre (\url{https://www.fz-juelich.de/ias/jsc}) for funding this project by providing computing time on the D-Wave Advantage™ System JUPSI through the Jülich UNified Infrastructure for Quantum computing (JUNIQ).
\par
The authors acknowledge TU Wien Bibliothek for financial support through its Open Access Funding Program.

\section*{Conflict of interest}
The authors declare no conflicts of interest.

%%%%%%%%%%%%%%%
%% MAIN BODY %%
%%%%%%%%%%%%%%%
% \doublespacing
% \onehalfspacing
% \linenumbers
%% INTRODUCTION %%
\section{Introduction}
\label{sec:introduction}
%% Context and Motivation
Mixed-variable optimization arises widely in engineering design, where combinatorial choices about sizes, shapes, materials, or topology are tightly coupled to continuous fields that define the state of the system (e.g., displacement, stress, pressure, or velocity). Identifying optimal solutions is of great interest since the resulting designs have significant implications for performance, safety, and cost in fields ranging from aerospace and civil infrastructure to energy systems and advanced manufacturing.
%% Problem Framing
However, the resulting search spaces are often high‑dimensional and nonconvex, with strong couplings between discrete decisions and continuous fields, making scalable, high‑precision search difficult for classical methods. 
Local methods, such as gradient-based descent, are prone to trapping in suboptimal solutions while global or multi-start approaches typically suffer unfavorable scaling.
\par
In response, a class of heuristic approaches known as \textit{Ising-machine approaches}~\cite{Mohseni2022} has gained traction. The name stems from the fact that suitable target problems can be cast as Ising or, equivalently, \gls{qubo} models. This class includes approaches like \gls{qa}~\cite{Apolloni1989,Kadowaki1998,Farhi2001}, \gls{da}~\cite{Matsubara2020}, and \gls{sa}~\cite{Kirkpatrick1983,Cerny1985} implemented on classical \glspl{cpu} or \glspl{gpu}~\cite{FixstarsAmplifyAE}. 
Yet, the corresponding model forms natively encode only binary variables. Consequently, any continuous variable must be represented via an explicit encoding, which directly impacts the total number of variables, numerical precision, and ultimately, the quality of the solution~\cite{Endo2024}. 
In this regard, the design of resource‑efficient and accurate encodings for continuous fields becomes a central challenge in mixed‑variable engineering optimization, which serves as the motivation for the present study.
\bigskip\par
%% Prior Work and Limitations
Applications of Ising-machine approaches to mixed‑variable optimization span fluid-device design~\cite{Suzuki2026}, power flow analysis~\cite{Kaseb2024,Kaseb2025}, and structural design~\cite{Ye2023,Honda2024,Key2024, Wang2024, Sukulthanasorn2025}, among others. In the following, we focus on structural design optimization as a canonical case of mixed-variable optimization.
Here, the continuous variables either represent displacements or stresses while the discrete choices determine the structure's size, shape, material or topology.
To handle the mixed nature of the problem, one common approach is partitioning: continuous fields are solved classically, while discrete decisions are mapped to solving a binary problem. In this context, a quantum-classical hybrid methodology has been proposed for solving continuum topology optimization problems, separating field and design variables~\cite{Ye2023}. 
In discrete structural topology optimization, a nested approach is employed that solves a \gls{qubo} problem to update cross-sectional areas and uses classical \gls{fem} to compute continuous (real-valued) stress data~\cite{Wang2024}.
Similarly, a design update strategy for \gls{qa} has been developed for the topology optimization of truss and continuum structures, using the \gls{fem} for structural analysis and \gls{qa} for the design updates~\cite{Sukulthanasorn2025}.
These approaches decouple the problem and make the encoding of continuous variables dispensable, which can yield practical workflows. However, the incremental approach sacrifices joint exploration of the coupled objective. As a result, the global search advantages associated with Ising‑machine solvers are not fully leveraged.
\par
As an alternative, the continuous variables can be encoded using a fixed number of binary variables. 
On the one hand, this enables the conversion of the entire problem into Ising or \gls{qubo} form, facilitating the joint exploration of design–field interactions and utilization of Ising‑machine solvers' strengths. 
On the other hand, two issues arise: First, insufficient bit depth can underrepresent the dynamic range or resolution of the continuous fields, degrading accuracy. 
Second, increasing bit depth for a finer resolution can quickly inflate the binary-variable count. Especially for \gls{qa}, this implies further complications when the problem is mapped to the hardware, including the accumulation of \glspl{ice} and increased embedding complexity as binary counts grow.
For truss structures, continuous nodal displacements have been represented using sums of random coefficients to enable deformation analysis on \gls{qa}~\cite{Honda2024}. Still, the analysis subproblem was treated separately from the discrete design optimization, resulting in the same decoupling limitation as above. 
In contrast, the development of a minimum complementary energy formulation enabled a fully coupled structural design problem to be solved end‑to‑end on \gls{qa}. In a small‑scale size‑optimization study, however, hardware limitations prevented arbitrarily high accuracy~\cite{Key2024}.
\bigskip\par
%% Research Gap
These limitations underscore the need for encodings that (i) provide sufficient dynamic range and resolution under tight binary budgets and 
(ii)
\replaced{retain the fully coupled problem structure, allowing global search over the current encoded domain.}
{preserve the fully coupled problem structure to retain the global search advantages of Ising‑machine solvers.}
Existing approaches either decouple the problem or rely on fixed‑bit encodings that scale poorly and can degrade performance, leaving a gap for adaptive, resource‑aware continuous‑variable encodings that improve solution quality while targeting the full coupled objective on current hardware.
\par
%% Objective
Motivated by these limitations, our objective is twofold: (i) to empirically diagnose the performance limitations of fixed continuous‑variable encodings on \gls{qa} hardware, and%
\replaced{
(ii) to propose and validate an adaptive encoding strategy that retains the coupled objective and improves solution quality under a fixed binary budget on today’s devices.}
{(ii) to propose and validate an adaptive encoding strategy to be integrated into a global search over the fully coupled problem while improving solution quality on today’s devices.}
\added{Particularly, we aim to analyze the strategy in terms of accuracy, convergence behavior, and robustness; we do not claim quantum advantage over classical or hybrid methods, e.g., in computational time, which is beyond the scope of this study.}
\bigskip\par
%% Approach
% To meet these objectives, we first conduct an empirical error analysis on a coupled \gls{fsi} problem, solving the structural subproblem for displacements via total potential energy minimization.
To meet these objectives, we first perform an empirical error analysis of a structural analysis problem formulated as the minimization of total potential energy with respect to the displacements, which serve as the continuous variables. This formulation is then converted via a fixed-range encoding into a pure binary problem suitable for \gls{qa}. We embed this setup in an \gls{fsi} testbed to capture the iterative nature of the coupling.
This stage does not yet involve design optimization but focuses on solution quality for the fixed continuous-variable encoding for \gls{qa}.
% The continuous displacement variables are represented with a fixed binary encoding, and we evaluate the effect of increasing the number of binary variables in the encoding on the solution quality achieved on \gls{qa} hardware.
To this end, we systematically increase the bit depth of the encoding and quantify its effect on the quality of equilibrium solutions returned by the \gls{qa} hardware.
\par
Second, we propose an adaptive encoding strategy that can be embedded in any iterative optimization scheme to refine the representable range of an encoded continuous variable based on its solution history. Finally, we demonstrate its effectiveness in structural design optimization\added{, a representative mixed-variable engineering problem in which accurate continuous fields are essential for reliable design decisions}.
Here, we use the minimum complementary energy formulation from~\cite{Key2024} to pose a single, coupled constrained problem over continuous stress and discrete design variables. 
The continuous stress field is encoded into binary variables, while the discrete design variables are directly represented in binary form, yielding a single optimization problem expressed purely in binary variables.
Then, we apply a quadratic penalty method where in each step we update the representation range for the continuous field variables according to our adaptive strategy. 
\replaced{At each step, the full coupled objective is re-solved over the current encoded domain, enabling the solver to perform a global search within that domain.}
{By re‑solving the full coupled problem at every step, the approach preserves the global search behavior of \gls{qa}.}

\par
%% Contributions
Implementing this approach leads to the following contributions:
\begin{itemize}
    \item an empirical error analysis for a fixed encoding on \gls{qa} hardware when increasing the number of binary variables,
    \item an adaptive strategy for the encoding of continuous variables that refines representable intervals using a constant number of binary variables, and
    \item the demonstration of this strategy's impact on structural design optimization using a benchmark problem from the literature, entirely solved on \gls{qa} hardware.
\end{itemize}
\bigskip\par
%% Results and Impact
Finally, we summarize our main findings and their implications. 
Our empirical error analysis shows that increasing bit depth, while improving representation, can amplify hardware errors degrading \gls{qa} performance, rendering it ineffective for improving continuous‑variable accuracy.
In contrast, the adaptive encoding embedded in a coupling‑preserving scheme substantially improves precision at fixed binary budgets, yielding a superior precision-resource trade-off. 
As we will show, these results provide practical guidance for encoding continuous variables on current \gls{qa} devices and may broaden the applicability of \gls{qa} to mixed‑variable engineering optimization.
% These results provide practical guidance for continuous-variable encodings on today’s \gls{qa} devices and broaden the applicability of \gls{qa} to mixed-variable engineering optimization.

% In our empirical error analysis presented in \Cref{subsec:empiricalErrorAnalysis}, we find that while increasing bit depth improves the resolution, it simultaneously enlarges the overall problem size and coupling complexity, which has the effect of degrading \gls{qa} performance in practice. In contrast, as we will show in \Cref{subsec:structuralDesignOptimizationWithAdaptiveNumberRepresentation} for structural design optimization, the application of the adaptive encoding allows to substantially improve precision at fixed qubit budgets on current \gls{qa} hardware. Together, these results demonstrate a superior precision–resource trade-off for the adaptive encoding and provide practical guidance for continuous‑variable representations on present \gls{qa} devices, thereby widening the application scope of \gls{qa} in mixed‑variable engineering optimization.

%% MATERIALS AND METHODS %%
\section{Materials and Methods}
\label{sec:materialsAndMethods}
This section details the models, encodings, and solver schemes used in our study. 
We begin by characterizing the mixed‑variable optimization problems in which discrete design variables are coupled to continuous field variables. 
We then present the canonical \gls{qa} formulations (Ising and \gls{qubo}) and, because \gls{qa} operates on binary quadratic models, describe how continuous variables can be represented in binary ones, contrasting fixed and adaptive encodings.
Next, we introduce structural formulations grounded in energy principles (minimum potential energy and minimum complementary energy) and express them in a \gls{qa}‑ready form. 
We then lay the groundwork for empirical error analysis by detailing hardware‑related error sources on current devices and define the ``best approximation'', i.e., the optimal solution attainable within a given encoding. Finally, we outline the solution schemes used to analyze the behavior of the different encodings. 
%% MIXED-VARIABLE OPTIMIZATION PROBLEMS %%
\subsection{Mixed-Variable Optimization Problems}
We consider mixed-variable optimization problems in which discrete design variables are coupled to continuous state variables through governing relations and constraints.
Let $\designVarVec\in\setAdmissibleDesigns$ denote discrete design decisions from a set of admissible designs $\setAdmissibleDesigns$ and $\stateVarVec\in\stateSpace$ continuous state variables; for example, in a size-optimization problem for a rod with $\nElem$ elements and $\nChoices$ available cross-section sizes, the discrete design is $\designVarVec=(A_1,\ldots,A_{\nElem})$ with $\setAdmissibleDesign=\{A^{(1)},\ldots,A^{(\nChoices)}\}$, while the continuous state $\stateVarVec$ could collect the nodal displacements.
Given the objective $\objective$ of the optimization problem, a generic form is
\begin{equation}
    \min_{\designVarVec,\stateVarVec} 
    \objective\lp\designVarVec,\stateVarVec\rp\quad\text{s.t.}\quad\constraintsEquality\lp\designVarVec,\stateVarVec\rp=\boldsymbol{0},\,\constraintsInequality\lp\designVarVec,\stateVarVec\rp\le\boldsymbol{0},\,\designVarVec\in\setAdmissibleDesigns,\stateVarVec\in\stateSpace,
    \label{eq:optimizationTask}
\end{equation}
where $\constraintsEquality$ encodes equality constraints, such as governing relations, and $\constraintsInequality$ collects inequality constraints.
\par
We refer to the joint optimization over $(\designVarVec,\stateVarVec)$ as the fully coupled problem.
By contrast, separated (decoupled) schemes alternate between solving for $\stateVarVec$ given $\designVarVec$ (analysis) and updating $\designVarVec$ given $\stateVarVec$ (design update). 
While decoupling can simplify implementation and enable hybrid classical--quantum workflows, it can miss interactions between $\designVarVec$ and $\stateVarVec$ that drive the overall behavior of the system to be optimized, which can, for example, lead to sensitivity to initial designs and degraded performance~\cite{Suzuki2026}.
Thus, preserving the full coupling is of interest to avoid designs with suboptimal performance.
Furthermore, it is advantageous for annealing-based solvers because the joint objective and constraints can be encoded into a single problem instance, exposing cross-coupling between $\designVarVec$ and $\stateVarVec$ to global search.
The subsequent sections instantiate this coupled view in \gls{qa}-ready formulations and compare binary encodings that allow taking these interactions into account.
%% PROBLEM FORMULATIONS FOR QUANTUM ANNEALING %%
\subsection{Problem Formulations for Quantum Annealing}
\label{subsec:problemFormulationsForQA}
In \gls{qa} and related Ising‑machine approaches, the optimization task from \Cref{eq:optimizationTask} is recast as the minimization of another mathematical function, the so-called Hamiltonian function $\hamiltonian$.
One option is to provide $\hamiltonian$ in the classical Ising representation where one employs spin variables $\spin_i\in\{-1,1\}$ and a quadratic function $\hamiltonianIsing$ defining the state of the model:
\begin{equation}
    \hamiltonianIsing\lp\spinVec\rp
    =
    -\sum_{i,j}\isingCoeffQuadratic_{ij}\spin_i\spin_j
    -\sum_i \isingCoeffLinear_i\spin_i,
\end{equation}
with pairwise couplings $\isingCoeffQuadratic_{ij}$ and linear coefficients $\isingCoeffLinear_i$.
Its ground state, i.e., the global minimizer of $\hamiltonianIsing$, encodes the optimal solution to the original optimization problem.
\par
Instead of using spin variables $\spin_i$, it may be more convenient to formulate the optimization problem in binary variables $\binary_i\in\{0,1\}$ and the \gls{qubo} Hamiltonian
\begin{equation}
    \hamiltonianQUBO\lp\binaryVec\rp
    =
    \sum_{i<j}\quboMatrixCoeff_{ij}\binary_i\binary_j
    +\sum_{i}\quboMatrixCoeff_{ii}\binary_i
    =
    \binaryVec^T\quboMatrix\binaryVec,
\end{equation}
with symmetric coefficient matrix $\quboMatrix$, commonly called \gls{qubo} matrix.
Note that both types of formulation are equivalent and can be converted to each other (up to an additive constant that does not affect the minimizer) by the change of variables $\spin_i = 1 -2\binary_i$. 
\par
The remaining question is how, for a specific instance of mixed-variable optimization from~\Cref{eq:optimizationTask}, a corresponding Ising or \gls{qubo} model can be derived. Due to the discrete nature of these model forms, we have to distinguish between discrete and continuous variables in the original optimization problem. While discrete design decisions are natively supported because categorical or binary choices in $\designVarVec$ can be directly mapped to corresponding spin or binary variables, continuous state variables $\stateVarVec \in \stateSpace$, by contrast, must be provided with a binary encoding to admit a combined Ising or \gls{qubo} formulation. 
To this end, we will first consider encoding approaches for continuous variables before we derive specific forms of $\hamiltonianQUBO$ in structural problems.
%% BINARY REPRESENTATIONS OF CONTINUOUS VARIABLES %%
\subsection{Binary Encodings of Continuous Variables}
In the following, we present ways to encode individual continuous state variables $\stateVar_i\in\mathbb{R}$ in corresponding binary vectors $\binaryVec_i\in\{0,1\}^\numQubitsPerNode$ of length 
$\numQubitsPerNode$. We consider two strategies: fixed encodings, which set range and resolution a priori, and adaptive encodings, which can dynamically adjust these during the solution process. 
%% Fixed Encoding
\subsubsection{Fixed Encoding}
For the encoding of continuous variables using a fixed representation, multiple options exist (see, e.g.,~\cite{Endo2024}).
In general, a continuous state variable $\stateVar_i$ can be expressed in terms of binary variables $\binary_{i,l}$ as follows:
\begin{equation}
    \stateVar_i\lp\binary_{i,l}\rp = \offsetCoeff_i + \scaleCoeff_i \sum_{l=0}^{\numQubitsPerNode-1} \repCoeff_{i,l} \binary_{i,l},
    \label{eq:real_number_representation}
\end{equation}
where $\offsetCoeff_i$ is an offset parameter, $\scaleCoeff_i$ scales the range of representable numbers, and $\repCoeff_{i,l}$ specifies the representation method of choice.
For the expression of a variable $\stateVar_i$ in terms of binary variables $\binaryVec_i$, we also write $\stateVar_i\lab\binaryVec_i\rab$.
\par
In the following, we focus on the classical binary representation with $\repCoeff_{i,l} = 2^{l}$.
To represent real numbers in the range $[\stateVarMin, \stateVarMax]$ using $\numQubitsPerNode$ binary variables, we set the offset and scaling parameters $\offsetCoeff_i$ and $\scaleCoeff_i$ as follows:
\begin{equation}
    \offsetCoeff_i = \stateVarMin, 
    \quad 
    \scaleCoeff_i = \frac{\stateVarMax-\stateVarMin}{2^{\numQubitsPerNode}-1}.
\end{equation}
To illustrate this, we consider the following example. 
If we choose $\stateVar_i\in[\stateVarMin,\stateVarMax]=[0,1]$ and $\numQubitsPerNode = 2$,
the expression for the encoding becomes
\begin{align}
    \stateVar_i\lp\binary_{i,l}\rp &= 0 + \frac{1-0}{4-1} \sum_{l=0}^{1} 2^l \binary_{i,l}
        = \frac{1}{3} \binary_{i,0} + \frac{2}{3} \binary_{i,1}.
\end{align}
Consequently, the representable values of $\stateVar_i$ are $\{0,1/3,2/3,1\}$ with corresponding bit strings $[0,0]$, $[1,0]$, $[0,1]$, $[1,1]$ for $\binaryVec_i=[\binary_{i,0},\binary_{i,1}]$, respectively.
\bigskip\par
Once the target range and the number of binary variables are specified, the precision of the encoding is fully determined. To improve precision within the same range, the only way is to increase $\numQubitsPerNode$ for each variable, which enhances the resolution but also raises the total number of binary variables.
For globally coupled Ising-machine approaches, scaling to larger problem sizes tends to degrade performance; for current \gls{qa} hardware, this effect, as we show later, is particularly pronounced.
This motivates an alternative encoding strategy that allows control over the precision of the continuous variables while keeping the number of binary variables constant.
%% Adaptive Encoding
\subsubsection{Adaptive Encoding}
\label{subsubsec:adaptiveEncoding}
In contrast to fixed encoding, where the representable range is chosen a priori and the precision is determined by the number of binary variables, we introduce an adaptive encoding strategy that is able to progressively improve solution accuracy during the solution process while keeping $\numQubitsPerNode$ and, thus, the total binary variable count constant. The key idea is to use a variable representational range for the continuous variables rather than a fixed one. 
At each solution step, the range bounds for the next step are updated based on the solutions from previous steps. 
In particular, for each continuous variable $\stateVar_i$, the bounds of the representable range are adjusted in two stages: (1) a contraction with respect to preceding solution values, and (2) a potential expansion if the last solution saturates one of the bounds.
\par
\begin{figure}
     \centering
     \begin{subfigure}[t]{0.32\textwidth}
        \centering
        \resizebox{\textwidth}{!}{%
        \begin{tikzpicture}
\tikzstyle{every node}=[font=\small]
% \node [font=\normalsize] at (11.5,13.75) {};

\node [font=\normalsize] at (4.25,18.5) {};
\node [font=\normalsize] at (4.25,18.5) {};
\draw [->, >=latex] (2.75,16.25) -- (2.75,21.25)node[pos=0.5,left]{$\stateVar_i$};
\draw [->, >=latex] (2.75,16.25) -- (7.75,16.25);

% k-2 at x = 3.75
\node[rotate around={90:(0,0)}] at (3.75,16.25) {\scalebox{1.5}{\pgfuseplotmark{-}}};
\node [font=\normalsize] at (3.75,15.75) {$k-2$};

% k-1 at x = 5
\node[rotate around={90:(0,0)}] at (5,16.25) {\scalebox{1.5}{\pgfuseplotmark{-}}};
\node [font=\normalsize] at (5,15.75) {$k-1$};
\node [font=\small, left] at (5,16.75) {$\stateVarMin^{(k-1)}$};
\node [font=\small, left] at (5,21) {$\stateVarMax^{(k-1)}$};

\draw [dashed, thick] (5,21) -- (5,16.75);
\node[rotate around={0:(0,0)}, thick] at (5,16.75) {\scalebox{1.5}{\pgfuseplotmark{-}}};
\node[rotate around={0:(0,0)}, thick] at (5,21) {\scalebox{1.5}{\pgfuseplotmark{-}}};

% k at x = 6.25
\node[rotate around={90:(0,0)}] at (6.25,16.25) {\scalebox{1.5}{\pgfuseplotmark{-}}};
\node [font=\normalsize] at (6.25,15.75) {$k$};

\draw [dashed, color=oi-orange, thick] (6.25,20) -- (6.25,16.75);
\node[rotate around={0:(0,0)}, color=oi-orange, thick] at (6.25,16.75) {\scalebox{1.5}{\pgfuseplotmark{-}}};
\node[font=\small, right, color=oi-orange] at (6.35,16.75) {$\stateVarMin^{(k)}$};
\node[rotate around={0:(0,0)}, color=oi-orange, thick] at (6.25,20) {\scalebox{1.5}{\pgfuseplotmark{-}}};
\node[font=\small, below right, color=oi-orange] at (6.35,20) {$\stateVarMax^{(k)}$};

\draw[dotted, color=oi-green, thick] (3.75,20) -- (6.25,20);
\draw[dotted, color=oi-green, thick] (5,21) -- (6.25,21);
\draw [<->, >=latex, color=oi-green, thick] (6.25,20) -- (6.25,21)node[pos=0.5, right]{$\delta_{i,\text{max}}^{(k-1)}$};

\draw [ ->, >=latex, color=oi-orange, thick] (5.25,20.85) -- (6,20.15);
\draw [->, >=latex, color=oi-orange, thick] (5.25,16.75) -- (6,16.75);

\node at (3.75,20) [circ, color=oi-blue] {};
\node at (5,18.75) [circ, color=oi-blue] {};
\draw [very thick, color=oi-blue] (3.75,20) -- (5,18.75);
\node [font=\small, color=oi-blue] at (3.8,20) [left] {$\stateVar_i^{(k-2)}$};
\node [font=\small, color=oi-blue] at (4.95,18.75) [left] {$\stateVar_i^{(k-1)}$};
\end{tikzpicture}
        }%
        \caption{Case 1: $\stateVar_i^{(\iter-1)} < \stateVar_i^{(\iter-2)}$. The upper bound is decreased; the lower bound remains unchanged.}
         \label{fig:rangeUpdate1}
     \end{subfigure}
     \hfill
     \begin{subfigure}[t]{0.32\textwidth}
        \centering
        \resizebox{\textwidth}{!}{%
        \begin{tikzpicture}
\tikzstyle{every node}=[font=\small]
% \node [font=\normalsize] at (11.5,13.75) {};

\node [font=\normalsize] at (4.25,18.5) {};
\node [font=\normalsize] at (4.25,18.5) {};
\draw [->, >=latex] (2.75,16.25) -- (2.75,21.25)node[pos=0.5,left]{$\stateVar_i$};
\draw [->, >=latex] (2.75,16.25) -- (7.75,16.25);

% k-2 at x = 3.75
\node[rotate around={90:(0,0)}] at (3.75,16.25) {\scalebox{1.5}{\pgfuseplotmark{-}}};
\node [font=\normalsize] at (3.75,15.75) {$k-2$};

% k-1 at x = 5
\node[rotate around={90:(0,0)}] at (5,16.25) {\scalebox{1.5}{\pgfuseplotmark{-}}};
\node [font=\normalsize] at (5,15.75) {$k-1$};
\node [font=\small, left] at (5,16.75) {$\stateVarMin^{(k-1)}$};
\node [font=\small, left] at (5,21) {$\stateVarMax^{(k-1)}$};

\draw [dashed, thick] (5,21) -- (5,16.75);
\node[rotate around={0:(0,0)}, thick] at (5,16.75)  {\scalebox{1.5}{\pgfuseplotmark{-}}};
\node[rotate around={0:(0,0)}, thick] at (5,21) {\scalebox{1.5}{\pgfuseplotmark{-}}};

% k at x = 6.25
\node[rotate around={90:(0,0)}] at (6.25,16.25) {\scalebox{1.5}{\pgfuseplotmark{-}}};
\node [font=\normalsize] at (6.25,15.75) {$k$};

\draw [dashed, color=oi-orange, thick] (6.25,21) -- (6.25,17.5);
\node[rotate around={0:(0,0)}, color=oi-orange, thick] at (6.25,17.5)  {\scalebox{1.5}{\pgfuseplotmark{-}}};
\node[font=\small, above right, color=oi-orange] at (6.35,17.5) {$\stateVarMin^{(k)}$};
\node[rotate around={0:(0,0)}, color=oi-orange, thick] at (6.25,21) {\scalebox{1.5}{\pgfuseplotmark{-}}};
\node[font=\small, right, color=oi-orange] at (6.35,21) {$\stateVarMax^{(k)}$};

\draw [ ->, >=latex, color=oi-orange, thick] (5.25,21) -- (6,21);
\draw [ ->, >=latex, color=oi-orange, thick] (5.25,16.85) -- (6,17.4);

\draw[dotted, color=oi-green, thick] (5,16.75) -- (6.25,16.75);
\draw[dotted, color=oi-green, thick] (3.75,17.5) -- (6.25,17.5);
\draw [<->, >=latex, color=oi-green, thick] (6.25,16.75) -- (6.25,17.5)node[pos=0.5,right]{$\delta_{i,\text{min}}^{(k-1)}$};

\node at (3.75,17.5) [circ, color=oi-blue] {};
\node at (5,18.75) [circ, color=oi-blue] {};
\draw [very thick, color=oi-blue] (3.75,17.5) -- (5,18.75);
\node [font=\small, color=oi-blue] at (3.8,17.5) [left] {$\stateVar_i^{(k-2)}$};
\node [font=\small, color=oi-blue] at (4.95,18.75) [left] {$\stateVar_i^{(k-1)}$};
% \draw [dashed] (3.75,16.75) -- (6.25,16.75);
\end{tikzpicture}
        }%
        \caption{Case 2: $\stateVar_i^{(\iter-1)} > \stateVar_i^{(\iter-2)}$. The lower bound is increased; the upper bound remains unchanged.}
         \label{fig:rangeUpdate2}
     \end{subfigure}
     \hfill
     \begin{subfigure}[t]{0.32\textwidth}
        \centering
        \resizebox{\textwidth}{!}{%
        \begin{tikzpicture}
\tikzstyle{every node}=[font=\small]
% \node [font=\normalsize] at (11.5,13.75) {};

\node [font=\normalsize] at (4.25,18.5) {};
\node [font=\normalsize] at (4.25,18.5) {};
\draw [->, >=latex] (2.75,16.25) -- (2.75,21.25)node[pos=0.5,left]{$\stateVar_i$};
\draw [->, >=latex] (2.75,16.25) -- (7.75,16.25);

% k-2 at x = 3.75
\node[rotate around={90:(0,0)}] at (3.75,16.25) {\scalebox{1.5}{\pgfuseplotmark{-}}};
\node [font=\normalsize] at (3.75,15.75) {$k-2$};
% k-1 at x = 5
\node[rotate around={90:(0,0)}] at (5,16.25) {\scalebox{1.5}{\pgfuseplotmark{-}}};
\node [font=\normalsize] at (5,15.75) {$k-1$};

\draw [dashed, thick] (5,21) -- (5,16.75);
\node[rotate around={0:(0,0)}, thick] at (5,16.75)  {\scalebox{1.5}{\pgfuseplotmark{-}}};
\node[rotate around={0:(0,0)}, thick] at (5,21) {\scalebox{1.5}{\pgfuseplotmark{-}}};
\node [font=\small, left] at (5,16.75) {$\stateVarMin^{(k-1)}$};
\node [font=\small, left] at (5,21) {$\stateVarMax^{(k-1)}$};

% k at x = 6.25
\node[rotate around={90:(0,0)}] at (6.25,16.25) {\scalebox{1.5}{\pgfuseplotmark{-}}};
\node [font=\normalsize] at (6.25,15.75) {$k$};

\draw [dashed, color=oi-orange, thick] (6.25,20) -- (6.25,17.75);
\node[rotate around={0:(0,0)}, color=oi-orange, thick] at (6.25,17.75) {\scalebox{1.5}{\pgfuseplotmark{-}}};
\node[font=\small, below right, color=oi-orange] at (6.35,17.75) {$\stateVarMin^{(k)}$};
\node[rotate around={0:(0,0)}, color=oi-orange, thick] at (6.25,20) {\scalebox{1.5}{\pgfuseplotmark{-}}};
\node[font=\small, above right, color=oi-orange] at (6.35,20) {$\stateVarMax^{(k)}$};

\draw [color=oi-orange, thick, ->, >=latex] (5.15,20.9) -- (6.1,20.1);
\draw [ color=oi-orange, thick, ->, >=latex] (5.15,16.8) -- (6.1,17.65);

%\draw [<->, >=latex] (4.5,21) -- (4.5,18.75)node[pos=0.5,left]{$l_1$};
%\draw [<->, >=latex] (4.5,18.75) -- (4.5,16.75)node[pos=0.5,left]{$l_2$};
\draw[dotted, color=oi-blue, thick] (5,18.75) -- (6.45,18.75);
\draw [<->, >=latex, color=oi-green,thick] (6.45,18.75) -- (6.45,20)node[pos=0.5,right]{$\tfrac{1}{4}\Delta_i^{(k-1)}$};
\draw [<->, >=latex, color=oi-green,thick] (6.45,17.75) -- (6.45,18.75)node[pos=0.5,right]{$\tfrac{1}{4}\Delta_i^{(k-1)}$};

\draw[dotted, color=oi-green, thick] (3.15,16.75) -- (5,16.75);
\draw[dotted, color=oi-green, thick] (3.15,21) -- (5,21);
\draw [<->, >=latex, color=oi-green,thick] (3.15,16.75) -- (3.15,21)node[pos=0.75,right]{$\Delta_i^{(k-1)}$};

\node at (3.75,18.75) [circ, color=oi-blue] {};
\node at (5,18.75) [circ, color=oi-blue] {};
\draw [very thick,color=oi-blue] (3.75,18.75) -- (5,18.75);
\node [font=\small, color=oi-blue] at (3.8,18.75) [below] {$\stateVar_i^{(k-2)}$};
\node [font=\small, color=oi-blue] at (4.95,18.75) [below] {$\stateVar_i^{(k-1)}$};
\end{tikzpicture}
        }%
        \caption{Case 3:  $\stateVar_i^{(\iter-1)} = \stateVar_i^{(\iter-2)}$.  The interval is contracted symmetrically around the current solution $\stateVar_i^{(\iter-1)}$.}
         \label{fig:rangeUpdate3}
     \end{subfigure}
     \hfill  
        \caption{Illustration of the three contraction cases for the representable interval of $\stateVar_i$ (relaxation factor $\relaxationFactor=1$).}
        \label{fig:rangeUpdateContraction}
\end{figure}
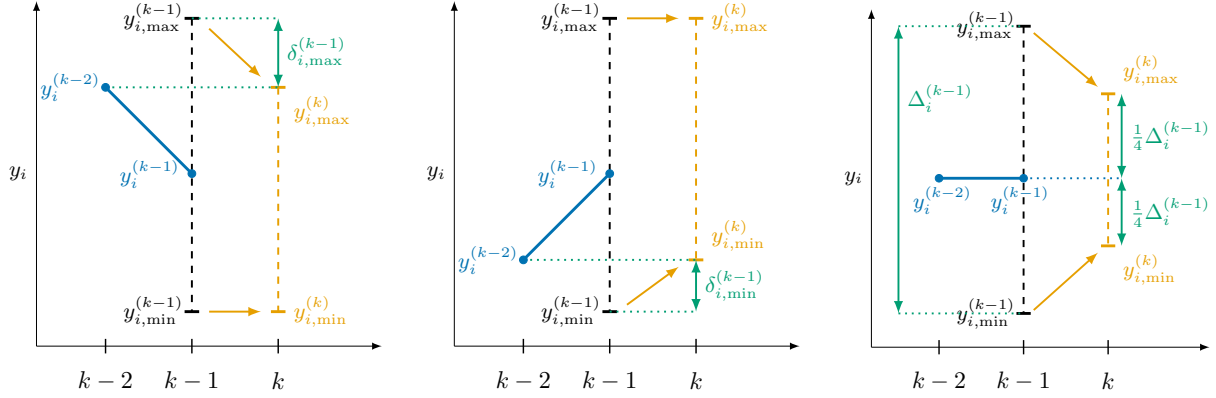

We first formalize three contraction cases and two expansion cases, and then illustrate their effect on the representable range.
Let $\stateVar_i^{(\iter)}$ denote the state variable in iteration $\iter$ with corresponding representable interval $[\stateVarMin^{(\iter)},\stateVarMax^{(\iter)}]$ of width $\Delta_i^{(\iter)} = \stateVarMax^{(\iter)} - \stateVarMin^{(\iter)}$. 
Additionally, we define $\delta_{i,\text{min}}^{(\iter)} = \stateVar_i^{(\iter-1)} - \stateVarMin^{(\iter)} \ge 0$ and
$\delta_{i,\text{max}}^{(\iter)} = \stateVarMax^{(\iter)} -\stateVar_i^{(\iter-1)} \ge 0$ as the distances between the previous solution and current interval bounds.
First, we apply the contraction, based on the evolution of $\stateVar_i$ in the two previous steps, i.e., $\stateVar_i^{(\iter-1)}$ and $\stateVar_i^{(\iter-2)}$. 
When $\stateVar_i$ decreases (Case 1), we tighten the interval by lowering the upper bound; when it increases (Case 2), we tighten by raising the lower bound. If no change is observed over the last two steps (Case 3), we perform a symmetric contraction around the current value.
For all cases, the local contraction spreads the same number of binary variables over a smaller interval, so each encoding represents a smaller step size and the precision increases near the expected solution.
The update rules for Cases~1-3 are given as follows:
\begin{align}
\lb\stateVarMin^{(\iter)},\stateVarMax^{(\iter)}\rb
&=
\begin{cases}
% \lb\stateVarMin^{(\iter-1)},\, (1-c)\,\stateVarMax^{(\iter-1)} + c\,\stateVar_i^{(\iter-1)}\rb,
% \lb\stateVarMin^{(\iter-1)},\, \stateVarMax^{(\iter-1)} - c\lp \stateVarMax^{(\iter-1)}-\stateVar_i^{(\iter-1)}\rp\rb,
\lb\stateVarMin^{(\iter-1)},\, \stateVarMax^{(\iter-1)} - \relaxationFactor\,\delta_{i,\text{max}}^{(\iter-1)}\rb,
& \text{if } \stateVar_i^{(\iter-1)} < \stateVar_i^{(\iter-2)} \quad \text{(Case 1)},\\[4pt]
% \lb(1-c)\,\stateVarMin^{(\iter-1)} + c\,\stateVar_i^{(\iter-1)},\, \stateVarMax^{(\iter-1)}\rb,
\lb\stateVarMin^{(\iter-1)} + \relaxationFactor\,\delta_{i,\text{min}}^{(\iter-1)},\, \stateVarMax^{(\iter-1)}\rb,
& \text{if } \stateVar_i^{(\iter-1)} > \stateVar_i^{(\iter-2)} \quad \text{(Case 2)},\\[4pt]
\lb\stateVar_i^{(\iter-1)} - \tfrac{1+\lp1-\relaxationFactor\rp}{4}\,\Delta_i^{(\iter-1)},\,
     \stateVar_i^{(\iter-1)} + \tfrac{1+\lp1-\relaxationFactor\rp}{4}\,\Delta_i^{(\iter-1)}\rb,
& \text{if } \stateVar_i^{(\iter-1)} = \stateVar_i^{(\iter-2)} \quad \text{(Case 3)}.
\end{cases}
\end{align}
Here, $\relaxationFactor\in(0,1]$ is a relaxation factor that controls the strength of the contraction; in Case 3 the contraction is scaled by a fixed factor of $\tfrac{1}{4}$, which could be chosen differently, but we keep this constant to limit notation complexity.
The three contraction cases are also illustrated in \Cref{fig:rangeUpdateContraction} for the special case $\relaxationFactor=1$.
\par
Second, we expand the interval if the current binary sample saturates at either extreme, all zeros (Case 4) or all ones (Case 5), which indicates that the optimizer seeks values beyond the current bounds.
The corresponding update rules for Cases~4-5 are as follows:
\begin{align}
    \lb\stateVarMin^{(\iter)},\stateVarMax^{(\iter)}\rb
    &=
    \begin{cases}
    \lb\stateVarMin^{(\iter-1)} - \tfrac{1}{4}\,\Delta_i^{(\iter-1)},\, \stateVarMax^{(\iter-1)}\rb,
    & \text{if } \binary_{i,l}^{(\iter-1)} = 0 \ \forall\, l=1,\dots,\numQubitsPerNode \quad \text{(Case 4)},\\[4pt]
    \lb\stateVarMin^{(\iter-1)},\, \stateVarMax^{(\iter-1)} + \tfrac{1}{4}\,\Delta_i^{(\iter-1)}\rb,
    & \text{if } \binary_{i,l}^{(\iter-1)} = 1 \ \forall\, l=1,\dots,\numQubitsPerNode \quad \text{(Case 5)},\\[4pt]
    \end{cases}
\end{align}
where the expansion is scaled by the fixed factor $\tfrac{1}{4}$, as above.
% \begin{align}
% \text{If } \binary_{i,l}^{(\iter-1)} &= 0 \ \forall\, l=1,\dots,\numQubitsPerNode, \text{ then }
% \stateVarMin^{(\iter)} \leftarrow \stateVarMin^{(\iter-1)} - \tfrac{1}{4}\,\Delta_i^{(\iter-1)}
% \quad \text{(Case 4)},
% \\
% \text{If } \binary_{i,l}^{(\iter-1)} &= 1 \ \forall\, l=1,\dots,\numQubitsPerNode, \text{ then }
% \stateVarMax^{(\iter)} \leftarrow \stateVarMax^{(\iter-1)} + \tfrac{1}{4}\,\Delta_i^{(\iter-1)}
% \quad \text{(Case 5)}.
% \end{align}

\begin{algorithm}

\algrenewcommand\algorithmicrequire{\textbf{Input:}}
\algrenewcommand\algorithmicensure{\textbf{Output:}}
\begin{algorithmic}[1]
\Require iteration index $\iter\ge2$, relaxation factor $\relaxationFactor\in(0,1]$
\Require for each continuous variable $\stateVar_i$: previous state values $\stateVar_i^{(\iter-1)}$ and $\stateVar_i^{(\iter-2)}$, previous bounds $\bigl[\stateVarMin^{(\iter-1)},\stateVarMax^{(\iter-1)}\bigr]$, binary sample of previous state value $\{\binary_{i,l}^{(\iter-1)}\}_{l=1}^{\numQubitsPerNode}$
\Ensure updated bounds $\bigl[\stateVarMin^{(\iter)},\stateVarMax^{(\iter)}\bigr]$

\For{$\nodeIndex = 1,2,\dots,m$}
    \State $\Delta_i^{(\iter-1)} \gets \stateVarMax^{(\iter-1)} - \stateVarMin^{(\iter-1)}$
    \State $\delta_{i,\text{min}}^{(\iter)} \gets \stateVar_i^{(\iter-1)} - \stateVarMin^{(\iter-1)}$
    \State $\delta_{i,\text{max}}^{(\iter)} \gets \stateVarMax^{(\iter-1)} - \stateVar_i^{(\iter-1)}$

    \Comment{Stage 1: contraction (Cases 1–3)}
    \If{$\stateVar_i^{(\iter-1)} < \stateVar_i^{(\iter-2)}$} \Comment{Case 1: decreasing}
      \State $\stateVarMin^{(\iter)} \gets \stateVarMin^{(\iter-1)}$
      \State $\stateVarMax^{(\iter)} \gets \stateVarMax^{(\iter-1)} - \relaxationFactor\,\delta_{i,\text{max}}^{(\iter)}$
    \ElsIf{$\stateVar_i^{(\iter-1)} > \stateVar_i^{(\iter-2)}$} \Comment{Case 2: increasing}
      \State $\stateVarMin^{(\iter)} \gets \stateVarMin^{(\iter-1)} + \relaxationFactor\,\delta_{i,\text{min}}^{(\iter)}$
      \State $\stateVarMax^{(\iter)} \gets \stateVarMax^{(\iter-1)}$
    \Else \Comment{Case 3: unchanged}
      \State $\stateVarMin^{(\iter)} \gets \stateVar_i^{(\iter-1)} - \tfrac{1+(1-\relaxationFactor)}{4}\,\Delta_i^{(\iter-1)}$
      \State $\stateVarMax^{(\iter)} \gets \stateVar_i^{(\iter-1)} + \tfrac{1+(1-\relaxationFactor)}{4}\,\Delta_i^{(\iter-1)}$
    \EndIf

    \Comment{Stage 2: potential expansion (Cases 4–5)}
    \If{$\binary_{i,l}^{(\iter-1)} = 0\ \forall\, l=1,\dots,\numQubitsPerNode$} \Comment{Case 4: all zeros}
    \State $\stateVarMin^{(\iter)} \gets \stateVarMin^{(\iter)} - \tfrac{1}{4}\,\Delta_i^{(\iter-1)}$
    \ElsIf{$\binary_{i,l}^{(\iter-1)} = 1\ \forall\, l=1,\dots,\numQubitsPerNode$} \Comment{Case 5: all ones}
    \State $\stateVarMax^{(\iter)} \gets \stateVarMax^{(\iter)} + \tfrac{1}{4}\,\Delta_i^{(\iter-1)}$
    \EndIf

\EndFor
\end{algorithmic}
\caption{Adaptive encoding strategy for continuous variables in an iterative scheme}
\label{alg:adaptiveEncoding}
\end{algorithm}

\par
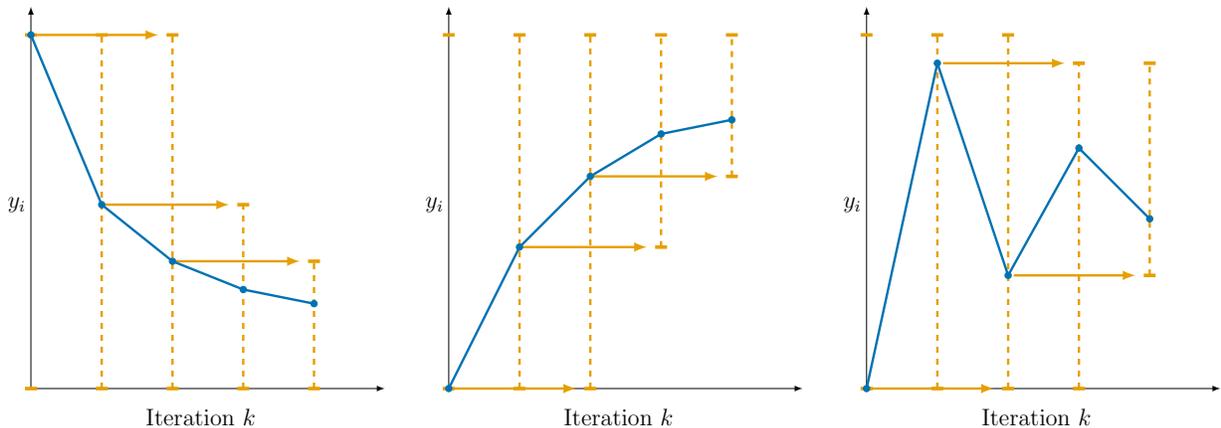
\begin{figure}
     \centering
     \begin{subfigure}[t]{0.32\textwidth}
        \centering
        \raisebox{0pt}{\resizebox{1.\textwidth}{!}{%
        \begin{tikzpicture}
\tikzstyle{every node}=[font=\LARGE]
\draw [->, >=latex] (2.5,13.75) -- (2.5,20.5);
\draw [->, >=latex] (2.5,13.75) -- (8.75,13.75);
\draw [dashed, color=oi-orange, very thick] (3.75,13.75) -- (3.75,20);
\draw [dashed, color=oi-orange, very thick] (5,13.75) -- (5,20);
\draw [dashed, color=oi-orange, very thick] (6.25,13.75) -- (6.25,17);
\draw [dashed, color=oi-orange, very thick] (7.5,13.75) -- (7.5,16);
\node [font=\large] at (5.5,13.25) {Iteration $\iter$};
\node [font=\large] at (2.25,17) {$\stateVar_i$};
% \draw [ color={rgb,255:red,255; orange,38; blue,0}, line width=0.5pt, dashed] (2.5,14.75) -- (8.75,14.75);

\node[color=oi-orange, very thick] at (2.5,13.75){\scalebox{1.5}{\pgfuseplotmark{-}}};
\node[color=oi-orange, very thick] at (3.75,13.75){\scalebox{1.5}{\pgfuseplotmark{-}}};
\node[color=oi-orange, very thick] at (5,13.75){\scalebox{1.5}{\pgfuseplotmark{-}}};
\node[color=oi-orange, very thick] at (6.25,13.75){\scalebox{1.5}{\pgfuseplotmark{-}}};
\node[color=oi-orange, very thick] at (7.5,13.75){\scalebox{1.5}{\pgfuseplotmark{-}}};

\node[color=oi-orange, very thick] at (2.5,20){\scalebox{1.5}{\pgfuseplotmark{-}}};
\node[color=oi-orange, very thick] at (3.75,20){\scalebox{1.5}{\pgfuseplotmark{-}}};
\node[color=oi-orange, very thick] at (5,20){\scalebox{1.5}{\pgfuseplotmark{-}}};
\node[color=oi-orange, very thick] at (6.25,17){\scalebox{1.5}{\pgfuseplotmark{-}}};
\node[color=oi-orange, very thick] at (7.5,16){\scalebox{1.5}{\pgfuseplotmark{-}}};

% \node [font=\LARGE, color={rgb,255:red,45; orange,160; blue,44}, rotate around={-45:(0,0)}] at (3.75,13.75) {\textbf{+}};
% \node [font=\LARGE, color={rgb,255:red,45; orange,160; blue,44}, rotate around={-45:(0,0)}] at (5,20) {\textbf{+}};
% \node [font=\LARGE, color={rgb,255:red,45; orange,160; blue,44}, rotate around={-45:(0,0)}] at (2.5,13.75) {\textbf{+}};
% \node [font=\LARGE, color={rgb,255:red,45; orange,160; blue,44}, rotate around={-45:(0,0)}] at (7.5,13.75) {\textbf{+}};
% \node [font=\LARGE, color={rgb,255:red,45; orange,160; blue,44}, rotate around={-45:(0,0)}] at (2.5,20) {\textbf{+}};
% \node [font=\LARGE, color={rgb,255:red,45; orange,160; blue,44}, rotate around={-45:(0,0)}] at (6.25,13.75) {\textbf{+}};
% \node [font=\LARGE, color={rgb,255:red,45; orange,160; blue,44}, rotate around={-45:(0,0)}] at (6.25,17) {\textbf{+}};
% \node [font=\LARGE, color={rgb,255:red,45; orange,160; blue,44}, rotate around={-45:(0,0)}] at (5,13.75) {\textbf{+}};
% \node [font=\LARGE, color={rgb,255:red,45; orange,160; blue,44}, rotate around={-45:(0,0)}] at (7.5,16) {\textbf{+}};
% \node [font=\LARGE, color={rgb,255:red,45; orange,160; blue,44}, rotate around={-45:(0,0)}] at (3.75,20) {\textbf{+}};

% \draw [ color=oi-orange, very thick, ->, >=latex] (6.25,13.75) -- (7.25,13.75);
% \draw [ color=oi-orange, very thick, ->, >=latex] (3.75,13.75) -- (4.75,13.75);
% \draw [ color=oi-orange, very thick, ->, >=latex] (3.75,20) -- (4.75,20);
\draw [ color=oi-orange, very thick, ->, >=latex] (2.5,20) -- (4.75,20);
% \draw [ color=oi-orange, very thick, ->, >=latex] (5,13.75) -- (6,13.75);
% \draw [ color=oi-orange, very thick, ->, >=latex] (2.5,13.75) -- (3.5,13.75);
\draw [ color=oi-orange, very thick, ->, >=latex] (3.75,17) -- (6,17);
\draw [ color=oi-orange, very thick, ->, >=latex] (5,16) -- (7.25,16);
\draw [ color=oi-blue, very thick] (2.5,20) -- (3.75,17);
\draw [ color=oi-blue, very thick] (3.75,17) -- (5,16);
\draw [ color=oi-blue, very thick] (5,16) -- (6.25,15.5);
\draw [ color=oi-blue, very thick] (6.25,15.5) -- (7.5,15.25);
\node at (2.5,20) [circ, color=oi-blue] {};
\node at (3.75,17) [circ, color=oi-blue] {};
\node at (5,16) [circ, color=oi-blue] {};
\node at (6.25,15.5) [circ, color=oi-blue] {};
\node at (7.5,15.25) [circ, color=oi-blue] {};

\end{tikzpicture}
        }}
         \caption{Monotonically decreasing solution: repeated Case 1 lowers the upper bound; Case 4 expands the lower bound if saturation occurs.}
         \label{fig:adaptiveScenario1}
     \end{subfigure}
     \hfill
     \begin{subfigure}[t]{0.32\textwidth} 
        \centering
        \raisebox{0pt}{\resizebox{1.\textwidth}{!}{%
        \begin{tikzpicture}
\tikzstyle{every node}=[font=\normalsize]
\draw [->, >=latex] (2.5,13.75) -- (2.5,20.5);
\draw [->, >=latex] (2.5,13.75) -- (8.75,13.75);
\draw [dashed, color=oi-orange, very thick] (3.75,13.75) -- (3.75,20);
\draw [dashed, color=oi-orange, very thick] (5,13.75) -- (5,20);
\draw [dashed, color=oi-orange, very thick] (6.25,16.25) -- (6.25,20);
\draw [dashed, color=oi-orange, very thick] (7.5,17.5) -- (7.5,20);
\node [font=\large] at (5.5,13.25) {Iteration $\iter$};
\node [font=\large] at (2.25,17) {$\stateVar_i$};

\node[color=oi-orange, very thick] at (2.5,13.75){\scalebox{1.5}{\pgfuseplotmark{-}}};
\node[color=oi-orange, very thick] at (3.75,13.75){\scalebox{1.5}{\pgfuseplotmark{-}}};
\node[color=oi-orange, very thick] at (5,13.75){\scalebox{1.5}{\pgfuseplotmark{-}}};
\node[color=oi-orange, very thick] at (6.25,16.25){\scalebox{1.5}{\pgfuseplotmark{-}}};
\node[color=oi-orange, very thick] at (7.5,17.5){\scalebox{1.5}{\pgfuseplotmark{-}}};

\node[color=oi-orange, very thick] at (2.5,20){\scalebox{1.5}{\pgfuseplotmark{-}}};
\node[color=oi-orange, very thick] at (3.75,20){\scalebox{1.5}{\pgfuseplotmark{-}}};
\node[color=oi-orange, very thick] at (5,20){\scalebox{1.5}{\pgfuseplotmark{-}}};
\node[color=oi-orange, very thick] at (6.25,20){\scalebox{1.5}{\pgfuseplotmark{-}}};
\node[color=oi-orange, very thick] at (7.5,20){\scalebox{1.5}{\pgfuseplotmark{-}}};

% \node [font=\LARGE, color={rgb,255:red,45; orange,160; blue,44}, rotate around={-45:(0,0)}] at (5,13.75) {\textbf{+}};
% \node [font=\LARGE, color={rgb,255:red,45; orange,160; blue,44}, rotate around={-45:(0,0)}] at (3.75,20) {\textbf{+}};
% \node [font=\LARGE, color={rgb,255:red,45; orange,160; blue,44}, rotate around={-45:(0,0)}] at (2.5,13.75) {\textbf{+}};
% \node [font=\LARGE, color={rgb,255:red,45; orange,160; blue,44}, rotate around={-45:(0,0)}] at (6.25,20) {\textbf{+}};
% \node [font=\LARGE, color={rgb,255:red,45; orange,160; blue,44}, rotate around={-45:(0,0)}] at (2.5,20) {\textbf{+}};
% \node [font=\LARGE, color={rgb,255:red,45; orange,160; blue,44}, rotate around={-45:(0,0)}] at (3.75,13.75) {\textbf{+}};
% \node [font=\LARGE, color={rgb,255:red,45; orange,160; blue,44}, rotate around={-45:(0,0)}] at (7.5,20) {\textbf{+}};
% \node [font=\LARGE, color={rgb,255:red,45; orange,160; blue,44}, rotate around={-45:(0,0)}] at (7.5,17.5) {\textbf{+}};
% \node [font=\LARGE, color={rgb,255:red,45; orange,160; blue,44}, rotate around={-45:(0,0)}] at (6.25,16.25) {\textbf{+}};
% \node [font=\LARGE, color={rgb,255:red,45; orange,160; blue,44}, rotate around={-45:(0,0)}] at (5,20) {\textbf{+}};
% \draw [ color=oi-orange, very thick, ->, >=latex] (2.5,13.75) -- (3.5,13.75);
% \draw [ color=oi-orange, very thick, ->, >=latex] (5,20) -- (6,20);
% \draw [ color=oi-orange, very thick, ->, >=latex] (6.25,20) -- (7.25,20);
% \draw [ color=oi-orange, very thick, ->, >=latex] (3.75,20) -- (4.75,20);
\draw [ color=oi-orange, very thick, ->, >=latex] (2.5,13.75) -- (4.75,13.75);
\draw [ color=oi-orange, very thick, ->, >=latex] (5,17.5) -- (7.25,17.5);
% \draw [ color=oi-orange, very thick, ->, >=latex] (2.5,20) -- (3.5,20);
\draw [ color=oi-orange, very thick, ->, >=latex] (3.75,16.25) -- (6,16.25);
\draw [ color=oi-blue, very thick] (2.5,13.75) -- (3.75,16.25);
\draw [ color=oi-blue, very thick] (3.75,16.25) -- (5,17.5);
\draw [ color=oi-blue, very thick] (5,17.5) -- (6.25,18.25);
\draw [ color=oi-blue, very thick] (6.25,18.25) -- (7.5,18.5);
\node at (7.5,18.5) [circ, color=oi-blue] {};

% \draw [ color={rgb,255:red,255; orange,38; blue,0}, line width=0.5pt, dashed] (2.5,19) -- (8.75,19);
\node at (2.5,13.75) [circ, color=oi-blue] {};
\node at (3.75,16.25) [circ, color=oi-blue] {};
\node at (5,17.5) [circ, color=oi-blue] {};
\node at (6.25,18.25) [circ, color=oi-blue] {};
\end{tikzpicture}
        }}
         \caption{Monotonically increasing solution: repeated Case 2 raises the lower bound; Case 5 expands the upper bound if saturation occurs.}
         \label{fig:adaptiveScenario2}
     \end{subfigure}
     \hfill
     \begin{subfigure}[t]{0.32\textwidth}
        \centering
        \raisebox{0pt}{\resizebox{1.\textwidth}{!}{%
        \begin{tikzpicture}
\tikzstyle{every node}=[font=\LARGE]
\draw [->, >=latex] (2.5,13.75) -- (2.5,20.5);
\draw [->, >=latex] (2.5,13.75) -- (8.75,13.75);

\node [font=\large] at (5.5,13.25) {Iteration $\iter$};
\node [font=\large] at (2.25,17) {$\stateVar_i$};

\draw [dashed, color=oi-orange, very thick] (3.75,13.75) -- (3.75,20);
\draw [dashed, color=oi-orange, very thick] (5,13.75) -- (5,20);
\draw [dashed, color=oi-orange, very thick] (6.25,13.75) -- (6.25,19.5);
\draw [dashed, color=oi-orange, very thick] (7.5,15.75) -- (7.5,19.5);

\node[color=oi-orange, very thick] at (2.5,13.75){\scalebox{1.5}{\pgfuseplotmark{-}}};
\node[color=oi-orange, very thick] at (3.75,13.75){\scalebox{1.5}{\pgfuseplotmark{-}}};
\node[color=oi-orange, very thick] at (5,13.75){\scalebox{1.5}{\pgfuseplotmark{-}}};
\node[color=oi-orange, very thick] at (6.25,13.75){\scalebox{1.5}{\pgfuseplotmark{-}}};
\node[color=oi-orange, very thick] at (7.5,15.75){\scalebox{1.5}{\pgfuseplotmark{-}}};

\node[color=oi-orange, very thick] at (2.5,20){\scalebox{1.5}{\pgfuseplotmark{-}}};
\node[color=oi-orange, very thick] at (3.75,20){\scalebox{1.5}{\pgfuseplotmark{-}}};
\node[color=oi-orange, very thick] at (5,20){\scalebox{1.5}{\pgfuseplotmark{-}}};
\node[color=oi-orange, very thick] at (6.25,19.5){\scalebox{1.5}{\pgfuseplotmark{-}}};
\node[color=oi-orange, very thick] at (7.5,19.5){\scalebox{1.5}{\pgfuseplotmark{-}}};

% \node [font=\LARGE, color={rgb,255:red,45; orange,160; blue,44}, rotate around={-45:(0,0)}] at (5,13.75) {\textbf{+}};
% \node [font=\LARGE, color={rgb,255:red,45; orange,160; blue,44}, rotate around={-45:(0,0)}] at (3.75,20) {\textbf{+}};
% \node [font=\LARGE, color={rgb,255:red,45; orange,160; blue,44}, rotate around={-45:(0,0)}] at (2.5,13.75) {\textbf{+}};
% \node [font=\LARGE, color={rgb,255:red,45; orange,160; blue,44}, rotate around={-45:(0,0)}] at (5,20) {\textbf{+}};
% \node [font=\LARGE, color={rgb,255:red,45; orange,160; blue,44}, rotate around={-45:(0,0)}] at (2.5,20) {\textbf{+}};
% \node [font=\LARGE, color={rgb,255:red,45; orange,160; blue,44}, rotate around={-45:(0,0)}] at (6.25,19.5) {\textbf{+}};
% \node [font=\LARGE, color={rgb,255:red,45; orange,160; blue,44}, rotate around={-45:(0,0)}] at (3.75,13.75) {\textbf{+}};
% \node [font=\LARGE, color={rgb,255:red,45; orange,160; blue,44}, rotate around={-45:(0,0)}] at (7.5,15.75) {\textbf{+}};
% \node [font=\LARGE, color={rgb,255:red,45; orange,160; blue,44}, rotate around={-45:(0,0)}] at (7.5,19.5) {\textbf{+}};
% \node [font=\LARGE, color={rgb,255:red,45; orange,160; blue,44}, rotate around={-45:(0,0)}] at (6.25,13.75) {\textbf{+}};

% \draw [ color=oi-orange, thick, ->, >=latex] (2.6,20) -- (3.5,20);
% \draw [ color=oi-orange, thick, ->, >=latex] (2.6,13.75) -- (3.5,13.75);
\draw [ color=oi-orange, very thick, ->, >=latex] (2.6,13.75) -- (4.75,13.75);
% \draw [ color=oi-orange, thick, ->, >=latex] (3.85,20) -- (4.75,20);
% \draw [ color=oi-orange, thick, ->, >=latex] (5.1,13.75) -- (6,13.75);
\draw [ color=oi-orange, very thick, ->, >=latex] (3.85,19.5) -- (6,19.5);
% \draw [ color=oi-orange, thick, ->, >=latex] (6.35,19.5) -- (7.25,19.5);
\draw [ color=oi-orange, very thick, ->, >=latex] (5.1,15.75) -- (7.25,15.75);

% \draw [ color={rgb,255:red,255; orange,38; blue,0}, line width=0.5pt, dashed] (2.5,17) -- (8.75,17);
\node at (2.5,13.75) [circ, color=oi-blue] {};
\node at (3.75,19.5) [circ, color=oi-blue] {};
\node at (5,15.75) [circ, color=oi-blue] {};
\node at (6.25,18) [circ, color=oi-blue] {};
\node at (7.5,16.75) [circ, color=oi-blue] {};
\draw [ color=oi-blue, very thick] (2.5,13.75) -- (3.75,19.5);
\draw [ color=oi-blue, very thick] (3.75,19.5) -- (5,15.75);
\draw [ color=oi-blue, very thick] (5,15.75) -- (6.25,18);
\draw [ color=oi-blue, very thick] (6.25,18) -- (7.5,16.75);
\end{tikzpicture}
        }}
         \caption{Oscillatory solution: Cases 1–2 alternate to tighten the interval; Case 3 applies when two consecutive values coincide.}
         \label{fig:adaptiveScenario3}
     \end{subfigure}
     \hfill
     \caption{Adaptive encoding across three representative iterative behaviors (relaxation factor $\relaxationFactor=1$).}
     \label{fig:adaptiveScenarios}
\end{figure}

This adaptive encoding strategy integrates seamlessly into iterative schemes, as outlined in \Cref{alg:adaptiveEncoding}. 
For monotone solution trends, repeated Case 1 (decreasing, cf.~\Cref{fig:adaptiveScenario1}) or Case 2 (increasing, cf.~\Cref{fig:adaptiveScenario2}) tightens the interval from the appropriate side.
For oscillatory but contracting behavior (cf.~\Cref{fig:adaptiveScenario3}), alternating applications of Cases 1 and 2 progressively narrow the interval and, when consecutive values coincide, Case 3 yields a symmetric shrink around the current value. If the optimization repeatedly saturates at an extreme (all zeros or all ones), the expansion rules in Cases 4–5 enlarge the corresponding bound to restore coverage; this also handles short bursts of diverging oscillations.
\par
Taken together with the step‑wise scenarios, the adaptive encoding serves as a modular component for Ising/QUBO‑based solvers in mixed‑variable optimization: it updates each continuous variable independently, concentrates resolution via contraction, and restores coverage via saturation‑triggered expansion, thereby improving local precision at a constant number of binary variables.
With the adaptive encoding in place, we now turn to energy formulations for structural analysis and design optimization, where the encoded variables will be embedded in \gls{qubo} objectives.

%% ENERGY FORMULATIONS IN STRUCTURAL ANALYSIS AND DESIGN OPTIMIZATION %%
\subsection{Energy Formulations in Structural Analysis and Design Optimization}
\label{subsec:energyFormulationsInStructuralAnalysisAndDesignOptimization}
This section presents the energy-based formulations used for our two primary tasks: structural analysis and design optimization. For each task, we specify the objective function, identify the target continuous variables, and describe the process of encoding them into a pure binary \gls{qubo} problem. This lays the groundwork for applying our adaptive encoding strategy.
\par
We employ two established variational principles from structural mechanics that are naturally suited for the minimization-based \gls{qubo} framework:
\begin{enumerate}
    \item For structural analysis, we use the principle of minimum potential energy to solve for the continuous displacement field. This field defines the equilibrium response of the structure to applied loads for a given, fixed design.
    \item For design optimization, we employ the principle of minimum complementary energy. This allows a simultaneous search for the optimal discrete design variables and the associated continuous stress field that together minimize the structural compliance. This stress field represents the internal distribution of forces within the material that must balance the applied loads for the optimal design.
\end{enumerate}
The choice of complementary energy for the design stage is particularly advantageous. Under linear elasticity, it yields a pure minimization (min-min) problem over both design and stress variables, avoiding a more complex saddle-point formulation. This structure aligns directly with the minimization-based nature of the Ising/\gls{qubo} framework.
\par
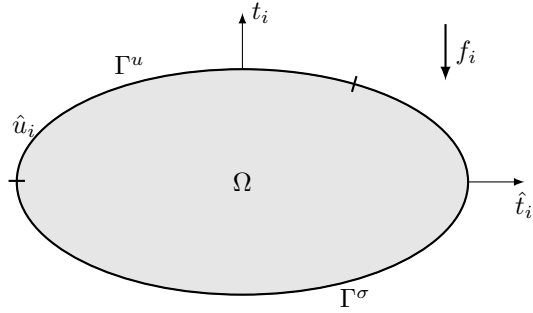
\begin{figure}
    \centering
    \begin{tikzpicture}[scale=1.5]
    \draw [fill =oi-grey!20] (0,0) ellipse (2 and 1);
    \node at (0,0) {$\domain$};
    % Displacement boundary
    \draw[thick, |-|, black] (-2,0) node [shift=({0.1,0.75})]{$\displPrescribed_i$} arc (-180:-300:2 and 1) node [midway, above] {$\boundaryDispl$};
    \draw [-latex] (0,1) --++ (0, 0.5) node [right] {$\traction_i$};
    % Traction boundary
    \draw[thick, black] (-2,0) arc (180:420:2 and 1) node [midway, below] {$\boundaryTraction$};
    \draw[ -latex, black](2,0) --++ (0.5,0) node [below] {$\tractionPrescribed_i$};
    % Volume force
    \draw[thick, -latex] (1.8,1.4) --++ (0,-0.5) node [midway, right] {$\volumeForce_i$};

\end{tikzpicture}
    \caption{Elastic body $\domain$ with boundary partition $\boundary=\boundaryDispl\cup\boundaryTraction$: displacements $\displPrescribed_i$ are prescribed on $\boundaryDispl$, and surface tractions $\tractionPrescribed_i$ on $\boundaryTraction$. A volume force density $\volumeForce_i$ acts in $\domain$.}
    \label{fig:elastic_body}
\end{figure}

To formalize these principles, we first establish the standard mechanical setup.
We consider an elastic body $\domain \subset \mathbb{R}^d$ (see \Cref{fig:elastic_body}) with displacement field $\displ_i$ and surface traction denoted by $\traction_i$. A volume force density $\volumeForce_i$ may act in $\domain$. The boundary $\boundary = \partial\domain$ is partitioned into the disjoint subsets $\boundaryDispl$ and $\boundaryTraction$, on which displacements $\displPrescribed_i$ and surface traction $\tractionPrescribed_i$ are prescribed, respectively.
%% Principle of Minimum Potential Energy for Structural Analysis

\begin{figure}
    \centering
     % \begin{subfigure}[t]{0.49\textwidth} 
     %    \centering
     %    % \raisebox{0pt}{\resizebox{1.\textwidth}{!}{%
     %    \input{fig/materials_and_methods/rod_composed_piston}
     %    % }}
     %     \caption{}
     %     \label{fig:rodComposedPiston}
     % \end{subfigure}
     % \hfill
     % \begin{subfigure}[t]{0.49\textwidth}
     %    \centering
     %    % \raisebox{0pt}{\resizebox{1.\textwidth}{!}{%
     %    \input{fig/materials_and_methods/rod_composed}
     %    % }}
     %     \caption{}
     %     \label{fig:rodComposed}
     % \end{subfigure}
         \begin{tikzpicture}
        % \fill[pattern=north east lines] (0,0) rectangle (3,0.5);
        % \draw[thick] (0,0) -- (3,0);
        \draw[] (0,0) -- (3,0);
        \draw[] (0,-5) -- (3,-5);
        \draw[-latex] (0.25,0) -- (0.25, -0.75) node [right] {$\x$};
        \draw[latex-latex] (0.75,0) -- node[left]{\rodLength}(0.75,-5);
        
        % \draw (1,0) -- (1,-5) -- (2,-5) -- (2,0);
        % \draw[-latex] (1.5, -0.1) -- (1.5, -0.9);
        \newcommand{\xa}{3.4}
        \newcommand{\xb}{3.5}
        \newcommand{\xm}{3.45}
        % e = 1 
        \draw (1,0) rectangle (2,-1);
        \node at (1.5,-0.5) {$1$};
        % \node at (2.1, -0.5) [right] {$\elemIndex = 1$};
        \node at (\xm,0) [circ] {};
        \node at (\xm,-1) [circ] {};
        \draw (\xa,-0) -- (\xb,-0) node [right] {$\x_1=0$};
        \draw (\xa,-1) -- (\xb,-1) node [right] {$\x_2$};
        \draw (1.25,-1) -- (1.25,-1.25);
        \draw (1.75,-1) -- (1.75,-1.25);
        \draw [dashed] (1.25,-1.25) -- (1.25, -1.5);
        \draw [dashed] (1.75,-1.25) -- (1.75, -1.5);

        \draw[dashed] (1.25,-1.5) -- (1.25,-1.75);
        \draw[dashed] (1.75,-1.5) -- (1.75,-1.75);
        \draw (1.25,-1.75) -- (1.25,-2);
        \draw (1.75,-1.75) -- (1.75,-2);
        % e
        \node at (\xm,-2) [circ] {};
        \draw (\xa,-2) -- (\xb,-2) node [right] {$\x_{\nodeIndex}$};
        \draw (1.1,-2) rectangle (1.9,-3);
        \node at (1.5,-2.5) {$\elemIndex$};
        % \node at (2.1, -2.5) [right] {$\elemIndex$};
        \node at (\xm,-3) [circ] {};
        \draw (\xa,-3) -- (\xb,-3) node [right] {$\x_{\nodeIndex+1}$};
        \draw (1,-3) -- (2,-3);
        \draw (1,-3) -- (1,-3.25);
        \draw (2,-3) -- (2,-3.25);
        \draw[dashed] (1,-3.25) -- (1,-3.5);
        \draw[dashed] (2,-3.25) -- (2,-3.5);
        \draw[dashed] (1,-3.5) -- (1,-3.75);
        \draw[dashed] (2,-3.5) -- (2,-3.75);
        \draw (1,-3.75) -- (1,-4);
        \draw (2,-3.75) -- (2,-4);
        % e = n_e
        \draw (1,-4) -- (2,-4);
        \node at (\xm,-4) [circ] {};
        \draw (\xa,-4) -- (\xb,-4) node [right] {$\x_{\nElem}$};
        \draw (1.25,-4) rectangle (1.75,-5);
        \node at (1.5,-4.5) {$\nElem$};

        % \node at (2.1, -4.5) [right] {$\elemIndex = \nElem$};
        \node at (\xm,-5) [circ] {};
        \draw (\xa,-5) -- (\xb,-5) node [right] {$\x_{\nElem+1}=\rodLength$};
        \draw (\xm,0) -- (\xm,-1); 
        \draw[dashed] (\xm,-1) -- (\xm,-2);
        \draw (\xm,-2) -- (\xm,-3);
        \draw[dashed] (\xm,-3) -- (\xm,-4);
        \draw (\xm,-4) -- (\xm,-5);
        \foreach \y in {-1,-2,-3,-4,-5} {
            % \draw[-latex] (1.5, \y+0.8) -- (1.5, \y+0.2);
        }
        \draw[-latex] (2.5, -1.25) -- node [right] {$\bodyForceDensity$}(2.5, -2.15) ;

        % \draw  (1.75,-5.5) node [below right] {$\crossSectionalArea$} -- (1.5,-5.1);
  
    \end{tikzpicture}
    \caption{One-dimensional rod composed of $\nElem$ elements; element $\elemIndex$ connects nodes at $\x_{\nodeIndex}$ and $\x_{\nodeIndex+1}$.}
    \label{fig:rodComposed}
\end{figure}
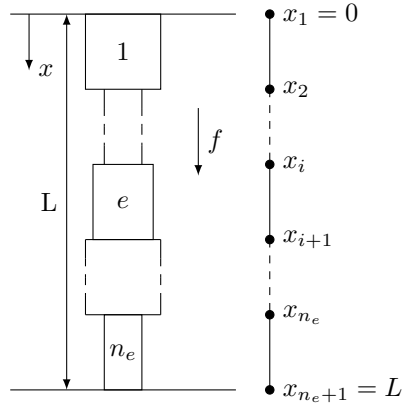
\subsubsection{Principle of Minimum Potential Energy for Structural Analysis}
\label{subsubsec:principleOfMinimumPotentialEnergyForStructuralAnalysis}
The principle of minimum potential energy  \cite[Ch.~4.3]{Reddy2017}  can be used to determine the static equilibrium configuration of a structure under external loads. For stating the principle, it is necessary to restrict our consideration to only kinematically admissible displacement fields $\displ_i$. A displacement field $\displ_i$ is kinematically admissible, if it is continuously differentiable and it satisfies the displacement boundary conditions:
\begin{equation}
    \displ_i = \displPrescribed_i \quad \text{on }\boundaryDispl.
    \label{eq:displacementBoundaryCondition}
\end{equation}
For an elastic problem, the equilibrium configuration is obtained by minimizing the total potential energy $\potentialEnergy \lb \displacementVector \rb$ over all kinematically admissible displacement fields,
\begin{equation}
    \min_{\displacementVector \in \setKinematicallyAdmissible}\left\{
    \potentialEnergy \lb \displacementVector \rb
    \right\},
    \label{eq:structuralAnalysisProblemPotE}
\end{equation}
where $\setKinematicallyAdmissible$ denotes the space of kinematically admissible displacement fields.
The correspondence between the variables, objective, and constraints in this formulation and those of the general mixed‑variable optimization template in \Cref{eq:optimizationTask} is summarized in \Cref{tab:mixedVarFormulations}.
\par
% The total potential energy $\potentialEnergy$ is defined as the sum of the total strain energy $\strainEnergy$ and the total potential energy of external forces $\potentialExternalForces$
% \begin{equation}
%     \potentialEnergy = \strainEnergy + \potentialExternalForces.
%     \label{eq:total_potential_energy}
% \end{equation}
% For a linear elastic body, the strain energy $\strainEnergy$ can be expressed as
% \begin{equation}
%     \strainEnergy
%     = 
%     \frac{1}{2} \int_{\domain} \elasticityTensor_{ijkl} \strainTensor_{ij} \strainTensor_{kl} \, \mathrm{d}{\domain},
%     \label{eq:strain_energy}
% \end{equation}
% where $\elasticityTensor_{ijkl}$ is the elasticity tensor and $\strainTensor_{ij}$ is the strain tensor. 
% The total potential of external forces $\potentialExternalForces$ consists of the contributions from the volume forces and the surface tractions:
% \begin{equation}
%     \potentialExternalForces
%     =
%     \potentialExternalVolumeForces 
%     + 
%     \potentialExternalSurfaceForces.
%     \label{eq:total_external_potential}
% \end{equation}
% If the external forces are independent of the displacement field, their potential is given by
% \begin{equation}
%     \potentialExternalVolumeForcesDensity 
%     = 
%     \int_{\domain} 
%     - \volumeForce_i \displ_i \, \mathrm{d}{\domain}, 
%     \quad 
%     \int_{\domain} 
%     \potentialExternalSurfaceForcesDensity
%     = 
%     - \tractionPrescribed_i \displ_i \, \mathrm{d}{\domain}.
%     \label{eq:external_potentials}
% \end{equation}
The total potential energy splits into internal and external parts, denoted by $\strainEnergy[\displacementVector]$ and $\potentialExternalForces[\displacementVector]$, respectively. For linear elasticity, the internal contribution is quadratic in the strains through the elasticity tensor, and the external contribution accounts for body forces and prescribed traction:
\begin{equation}
\potentialEnergy[\displacementVector]
=
\underbrace{
    \frac{1}{2}\int_{\domain} \elasticityTensor_{ijkl}\,\strainTensor_{ij}(\displacementVector)\,\strainTensor_{kl}(\displacementVector)\,\mathrm{d}\domain
}_{\strainEnergy[\displacementVector]}
\underbrace{
    -
    \int_{\domain} \volumeForce_i\,\displacement_i\,\mathrm{d}\domain
    -
    \int_{\boundaryTraction} \tractionPrescribed_i\,\displacement_i\,\mathrm{d}\boundary
}_{\potentialExternalForces[\displacementVector]},
\label{eq:totalPotentialEnergyLinear}
\end{equation}
% where $\elasticityTensor_{ijkl}$ is the elasticity tensor, $\strainTensor_{ij}(\displacementVector)=\tfrac{1}{2}(\displacement_{i,j} + \displacement_{j,i})$ is the strain tensor ( $(\cdot)_{,j}$ denotes partial differentiation with respect to the $j$-th coordinate), and $\boundaryTraction$ denotes the traction boundary with prescribed tractions $\tractionPrescribed_i$.
where $\elasticityTensor_{ijkl}$ is the elasticity tensor, $\strainTensor_{ij}(\displacementVector)=\tfrac{1}{2}(\displacement_{i,j} + \displacement_{j,i})$ is the (small) strain tensor, with $(\cdot)_{,j} := \partial(\cdot)/\partial \x_j$, and $\boundaryTraction$ denotes the traction boundary with prescribed tractions $\tractionPrescribed_i$. Prescribed displacements are enforced through the admissible space for $\displacementVector$ and do not enter the functional explicitly.
\par
% For the following explanations, we will consider the specific case of a one-dimensional linear elastic rod that is composed into elements $\elemIndex$. In analogy to \Cref{eq:structuralAnalysisProblemPotE}, the structural analysis problem for the element-wise scalar displacements  $\displ_{\elemIndex}\lp\x\rp$ can be written as
% \begin{equation}
%     \min_{\displ_{\elemIndex} \in \setKinematicallyAdmissible_e}\left\{
%     \potentialEnergy \lb \displ_{\elemIndex} \rb
%     \right\}.
%     \label{eq:structuralAnalysisProblemPotEElem}
% \end{equation}
For the following explanations, we consider a one-dimensional, linearly elastic rod partitioned into elements $\elemIndex$ (see \Cref{fig:rodComposed}). Let $\displ(x)$ denote the scalar displacement field and $\displ_{\elemIndex}(x)$ its restriction to element $\elemIndex$. The structural analysis problem is given by \Cref{eq:structuralAnalysisProblemPotE}  
% \begin{equation}
%     \min_{\displ \in \setKinematicallyAdmissible} \lcb\potentialEnergy\lb\displ\rb\rcb,
%     \label{eq:structuralAnalysisProblemPotEElem}
% \end{equation}
with the additive decomposition
$
    \potentialEnergy[\displ] = \sum_{\elemIndex}\potentialEnergy_{\elemIndex}[\displ_{\elemIndex}]
$
and the admissible space
$
    \setKinematicallyAdmissible = \{ \displ \big| \displ_{\elemIndex} \in \setKinematicallyAdmissible_{\elemIndex}\ \forall \elemIndex\}.
$
% \begin{equation}
%     \potentialEnergy\lb\displ\rb = \sum_{\elemIndex} \potentialEnergy_{\elemIndex}\lb\displ_{\elemIndex}\rb,
%     \qquad
%     \setKinematicallyAdmissible = \lcb \displ \big| \displ|_{\elemIndex} \in \setKinematicallyAdmissible_{\elemIndex}\ \forall \elemIndex\rcb.
% \end{equation}
% \begin{equation}
%     \min_{\displ_{\elemIndex} \in \setKinematicallyAdmissible_{\elemIndex}}\lcb\potentialEnergy\lb\displ_{\elemIndex}\rb\rcb,
%     \label{eq:structuralAnalysisProblemPotEElem}
% \end{equation}
% with $\potentialEnergy[\displ]=\sum_{\elemIndex}\potentialEnergy_{\elemIndex}[\displ_{\elemIndex}]$.
To obtain a finite-dimensional representation, we approximate $\displ_{\elemIndex}\lp\x\rp$ by linear interpolation functions $\basisFct^{\mathrm{I/II}}_\elemIndex(\x)$ associated with the element nodes $\mathrm{I}$ and $\mathrm{II}$, and real nodal coefficients $\basisCoeff^{\mathrm{I/II}}_{\elemIndex}\in\mathbb{R}$:
\begin{equation}
    \displ_{\elemIndex}\lp\x\rp
    \approx
    \basisCoeff^{\mathrm{I}}_\elemIndex \basisFct^{\mathrm{I}}_\elemIndex(\x)
    +
    \basisCoeff^{\mathrm{II}}_{\elemIndex} \basisFct^{\mathrm{II}}_{\elemIndex}(\x).
    \label{eq:ansatzDisplacement}
\end{equation}
We couple elements by enforcing displacement continuity across their interfaces, which yields global $C^0$ continuity and, equivalently, the kinematic admissibility conditions on each element: the displacement at an element’s endpoints is prescribed as Dirichlet data (see \Cref{eq:displacementBoundaryCondition}) to the shared nodal values provided by its neighbors. With the linear nodal interpolation functions in use, this identifies element-local degrees of freedom with common nodal coefficients,
\begin{equation}
   \basisCoeff^{\mathrm{II}}_{\elemIndex-1}
   =
   \basisCoeff^{\mathrm{I}}_{\elemIndex}
   =
   \basisCoeff_{\nodeIndex},
   \label{eq:nodalBasisCoeff}
\end{equation}
and physical boundary conditions are imposed by fixing the corresponding nodal coefficient(s).
% To ensure kinematic admissibility on each element, the continuity of the displacement field across elements is achieved by design by sharing the nodes of neighboring elements. Hence, at the interface between elements $\elemIndex - 1$ and $\elemIndex$, the coefficients satisfy
% \begin{equation}
%    \basisCoeff^{\mathrm{II}}_{\elemIndex-1}
%    = 
%    \basisCoeff^{\mathrm{I}}_{\elemIndex}
%    = 
%    \basisCoeff_{\nodeIndex}
%    \label{eq:nodalBasisCoeff}
% \end{equation}
\par
The global displacement field is therefore fully characterized by the set of continuous nodal coefficients $ \basisCoeff_{\nodeIndex}$.
As explained above, we encode each nodal coefficient using binary variables: for node $\nodeIndex$, the continuous variable $\basisCoeff_{\nodeIndex}$ is mapped to a binary vector $\qubitsVectorCoeffs_{\nodeIndex}$ with $\binary^{\basisCoeff}_{i,l}\in\{0,1\}$, $l=1,\dots,\numQubitsPerNode$, according to \Cref{eq:real_number_representation}. We then collect all nodal binaries $\qubitsVectorCoeffs_{\nodeIndex}$ into a single vector $\qubitsVectorCoeffs$.
\par
Since, for linear elasticity, the total potential energy is quadratic in the nodal coefficients and linear in the applied loads (see \Cref{eq:totalPotentialEnergyLinear}), replacing each nodal coefficient by its binary encoding transforms the energy into a quadratic polynomial in the binary variables.
By the minimum potential energy principle, the structural analysis problem thus becomes a \gls{qubo} problem:
\begin{equation}
        \min_{\binariesVecCoeff} \hamiltonianQUBOPotEnergy\lp\binariesVecCoeff\rp,
\end{equation}
where $\hamiltonianQUBOPotEnergy(\binariesVecCoeff)$ is the quadratic polynomial obtained by substituting the binary encoding of the nodal coefficients into the total potential energy.
\Cref{fig:transformationContinuousQUBO} illustrates the entire transformation process from the continuous structural analysis problem to the final \gls{qubo} formulation.
\begin{table}
    \centering
    \begin{tabularx}{\textwidth}{lCCCCC}
    % \begin{tabular}{lccccc}
        \toprule
         & $\designVarVec$ & $\stateVarVec$ & $\objective$ & $\constraintsEquality$ & $\constraintsInequality$ \\
        \midrule
        Min. Pot. Energy & -- & $\displacementVector$ & $\potentialEnergy$ & kinematic admissibility & -- \\
        \makecell[c]{Min. Compl. Energy\\\& Min. Compliance}
        & \makecell[c]{$\designVarVec$}
        & \makecell[c]{$\stressTensor$}
        & \makecell[c]{$\totalComplEnergy$}
        % & \makecell[c]{static admissibility}
        & static admissibility
        & \makecell[c]{--} \\
        \bottomrule
        \label{tab:mixedVarFormulations}
    \end{tabularx}
    \caption{Mixed-variable optimization formulations cast in the general form given by \Cref{eq:optimizationTask}. The table specifies the design variables $\designVarVec$, state variables $\stateVarVec$, objective $\objective$, and equality and inequality constraints $(\constraintsEquality,\constraintsInequality)$ for the minimum potential energy principle (\Cref{subsubsec:principleOfMinimumPotentialEnergyForStructuralAnalysis}) and for the minimum complementary energy principle combined with a minimum‑compliance objective (\Cref{subsubsec:principleOfMinimumComplementaryEnergyForStructuralDesignOptimization}).}
    \label{tab:mixedVarFormulations}
\end{table}
\begin{figure}
\centering
\begin{tikzpicture}[
    node distance=1.5cm and 2cm,
    arrow/.style={-Latex, thick},
    box/.style={draw, rectangle, rounded corners, thick, text centered, 
    text height = 1.75 cm,
    fill=oi-grey!10},
    desc/.style={font=\small, text centered, text width=5cm}
]

% --- 1. Continuous Field ---
\node (cont_field) [box] {
    \begin{tikzpicture}[scale=0.5] 
        \draw[thick, rounded corners=0pt, fill=white] (0,0) rectangle (5.,0.5); % The rod
        \draw[oi-blue, very thick, domain=0.:5., samples=25] plot (\x, {0.25 + 0.5*sin(360*\x/5 )});
        \node[above] at (2.5, 1.2) {$\displ(x)$};
    \end{tikzpicture}
};
\node(cont_desc)[desc, below=0.1cm of cont_field] {\textbf{(a) Continuous Problem} \\ Continuous displacement field $\displ(x)$ over the rod.};

% --- 2. Discretization ---
\node (discrete_field) [box, right=of cont_field] {
    \begin{tikzpicture}[scale=0.5]
        \draw[thick, rounded corners=0pt, fill=white] (0,0) rectangle (5,0.5); % The rod
        % Piecewise linear displacement
        \draw[oi-orange, very thick] (0,0.25) -- (1.25,0.75) -- (2.5,0.25) -- (3.75,-0.25) -- (5,0.25);
        % Nodes
        \foreach \i/\y in {0/0.25, 1.25/0.75, 2.5/0.25, 3.75/-0.25, 5/0.25} {
            \filldraw (\i, \y) circle (2pt);
        }
        \node[above] at (1.25, 1.2) {$\basisCoeff_{\nodeIndex}$};
        % \node[above] at (3.75, 1.2) {$\basisCoeff_{\nodeIndex+2}$};
    \end{tikzpicture}
};
\node(discrete_desc)[desc, below=0.1cm of discrete_field] {\textbf{(b) Discretization} \\ Field is approximated by continuous nodal coefficients $\basisCoeff_{\nodeIndex} \in \mathbb{R}$.};

% --- 3. Binary Encoding ---
\node (encoding) [box, right=of discrete_field] {
    \begin{tikzpicture}[scale=0.5, font=\small]
        % \node (ui) at (0, 0.8) {Nodal Value $\basisCoeff_{\nodeIndex}\langle\qubitsVectorCoeffs_{\nodeIndex}\rangle$};
        % \node[draw, rectangle, minimum width=2.5cm, minimum height=0.6cm] (bin) at (0, -0.8) {$\{\binary^{\basisCoeff}_{i,1}, \binary^{\basisCoeff}_{i,2}, \dots, \binary^{\basisCoeff}_{i,\numQubitsPerNode}\}$};
        % \draw[thick, -Stealth] (ui) -- (bin) node[midway, right, align=left] {Binary\\Encoding};
        % \node (ai) {$\basisCoeff_{\nodeIndex}\langle\qubitsVectorCoeffs_{\nodeIndex}\rangle\rightarrow\{\binary^{\basisCoeff}_{i,1}, \binary^{\basisCoeff}_{i,2}, \dots, \binary^{\basisCoeff}_{i,\numQubitsPerNode}\}$};
        \node (ai) {$\basisCoeff_{\nodeIndex}\langle\qubitsVectorCoeffs_{\nodeIndex}\rangle$};
        \node (arrow) [below= of ai, yshift=+1.5cm]{$\downarrow$};
        \node [below= of arrow, yshift=+1.5cm]{$\{\binary^{\basisCoeff}_{i,1}, \binary^{\basisCoeff}_{i,2}, \dots, \binary^{\basisCoeff}_{i,\numQubitsPerNode}\}$};
    \end{tikzpicture}
};
\node(encoding_desc)[desc, below=0.1cm of encoding] {\textbf{(c) Binary Encoding} \\ Each $\basisCoeff_{\nodeIndex}$ is mapped to a vector of binary variables $\qubitsVectorCoeffs_{\nodeIndex}\in\{0,1\}^\numQubitsPerNode$.};

% Arrows connecting the stages
\draw [arrow,  thick] (cont_field) -- (discrete_field) node[midway, above] {Discretize};
\draw [arrow,  thick] (discrete_field) -- (encoding) node[midway, above] {Encode};

\node (min_pot)[below = 0.5 cm of cont_desc]{
    $\displaystyle
    \min_{\displ \in \setKinematicallyAdmissible}\left\{
    \potentialEnergy \lb \displ \rb
    \right\}$
};

\node (min_QUBO_pot) [below = 0.5 cm of encoding_desc]
{
    $\displaystyle
    \min_{\qubitsVectorCoeffs} \hamiltonianQUBOPotEnergy\lp\qubitsVectorCoeffs\rp
    $
};
\draw [arrow,  thick] (min_pot) -- (min_QUBO_pot) node[midway, above] {Discretize + Encode};
\node[desc, 
% below=0.1cm of qubo_form, 
below= 0.5 cm of $(min_pot)!0.5!(min_QUBO_pot)$,
text width= 15 cm] {\textbf{(d) Final Formulation} \\ The potential energy minimization over the continuous field becomes a QUBO problem over binary variables.};

\end{tikzpicture}
\caption{Illustration of the transformation from a continuous structural analysis problem to a QUBO formulation for the one-dimensional rod. (a) The problem is defined over a continuous displacement field $\displ(x)$. (b) The field is discretized into a finite set of continuous nodal coefficients $\basisCoeff_i$. (c) Each continuous coefficient is then mapped to a vector of binary variables $\qubitsVectorCoeffs_{\nodeIndex}$. (d) This process transforms the total potential energy into a quadratic polynomial of binary variables, resulting in a \gls{qubo} problem.}
\label{fig:transformationContinuousQUBO}
\end{figure}
%% Principle of Minimum Complementary Energy for Structural Design Optimization
\subsubsection{Principle of Minimum Complementary Energy for Structural Design Optimization}
\label{subsubsec:principleOfMinimumComplementaryEnergyForStructuralDesignOptimization}
Having treated the analysis problem for a fixed design, we now address compliance-driven structural design optimization in which the optimal design and the physical state are determined simultaneously. To this end, we adopt a \gls{qubo} formulation based on the principle of minimum complementary energy~\cite{Key2024}, which we outline below.
\par
The principle of minimum complementary energy~\cite[Ch.~4.5]{Reddy2017} yields a minimization problem posed in the stress field $\stressTensorComponent{ij}$, constrained to the set of statically admissible stresses $\setStaticallyAdmissible$. Statically admissible stresses are in equilibrium with the volume force density $\volumeForce_i$ and satisfy the prescribed traction boundary conditions:
\begin{align}
        \stressTensorComponent{ij,j}
        + \bodyForceDensity_{i}
        &=
        0
        \quad\text{in } \domain,
        \label{eq:equilibrium}
        \\
        \stressTensorComponent{ij}\normal_{j}
        &=
        \tractionPrescribed_i
        \quad\text{on } \boundaryTraction,
        \label{eq:tractionBC}
\end{align}
where $\normal_j$ is the outward unit normal on $\boundary$.
We write the total complementary energy $\totalComplEnergy[\stressTensor]$ as a functional of the stresses, and the elastic problem is to minimize $\totalComplEnergy$ over statically admissible fields:
\begin{equation}
    \min_{\stressTensor\in\setStaticallyAdmissible}
    \lcb
        \totalComplEnergy\lb\stressTensor\rb
    \rcb.
    \label{eq:structuralAnalysisProblemStress}
\end{equation}
The total complementary energy splits into internal and external parts, denoted by $\complStrainEnergy[\stressTensor]$ and $\complExternal[\stressTensor]$, respectively. For linear elasticity, the internal contribution is quadratic in the stresses through the compliance tensor, and the external contribution accounts for body forces and prescribed displacements:
\begin{equation}
\totalComplEnergy[\stressTensor]
=
\underbrace{
    \frac{1}{2}\int_{\domain} \complianceTensorComponent{ijkl}\,\stressTensorComponent{ij}\,\stressTensorComponent{kl}\,\mathrm{d}\domain
}_{\complStrainEnergy[\stressTensor]}
\underbrace{
    -
    \int_{\domain} 
        \displacement_i\,\volumeForce_i\,\mathrm{d}\domain
    -
    \int_{\boundaryDispl} 
        \displacementPrescribed_i\,(\stressTensorComponent{ij}\normal_j)\,\mathrm{d}\boundary
%         \displacementPrescribed_i\,\traction_i\,\mathrm{d}\boundary
}_{\complExternal[\stressTensor]}\!,
\label{eq:totalComplEnergyLinear}
\end{equation}
where $\complianceTensorComponent{ijkl}$ is the compliance tensor and $(\stressTensorComponent{ij}\normal_j)=\traction_i$ is the surface traction.
To pose minimum-compliance design, we minimize the complementary energy jointly over the design and the statically admissible stress field (cf.~\cite[Eq.~1.7]{Bendsoe2004}). So, the coupled optimization problem, whose relation to the general mixed-variable optimization template is summarized in \Cref{tab:mixedVarFormulations},  reads
\begin{equation}
    \min_{\designVarVector\in\setAdmissibleDesign}\min_{\stressTensor\in\setStaticallyAdmissible}
    \left\{
        \totalComplEnergy\left[\stressTensor;\designVarVector\right]
    \right\}.
    \label{eq:designOptimizationProblem}
\end{equation}
\par
As above, we use the one-dimensional composed rod to derive a concrete \gls{qubo} formulation (in 1D, the Cauchy stress tensor reduces to the scalar field 
$\stress(\x)$). For each element $\elemIndex$, the combined optimization in the scalar stress field $\stress_{\elemIndex}(\x)$ and the design $\designVarVector_{\elemIndex}$ reads
\begin{equation}
    \min_{\designVarVector_{\elemIndex}\in\setAdmissibleDesign_{\elemIndex}}
    \min_{\stress_{\elemIndex}\in\setStaticallyAdmissible_{\elemIndex}}\totalComplEnergy\lb\stress_{\elemIndex};\designVarVector_{\elemIndex}\rb.
    \label{eq:designOptimizationProblemElem}
\end{equation}
Following \cite{Key2024}, we work with the axial element force
\begin{equation}
    \force_{\elemIndex}(\x) 
    =
    \lvert\stress_{\elemIndex}(\x) \cdot \normal \rvert
    \crossSectionalArea[\elemIndex],
    \label{eq:force}
\end{equation}
and approximate it with linear interpolation functions, ensuring continuity by common nodal coefficients:
\begin{equation}
    \force_{\elemIndex}\lp\x\rp
    \approx
    \basisCoeff^{\mathrm{I}}_\elemIndex \basisFct^{\mathrm{I}}_\elemIndex(\x)
    +
    \basisCoeff^{\mathrm{II}}_{\elemIndex} \basisFct^{\mathrm{II}}_{\elemIndex}(\x),
    \qquad
    \basisCoeff^{\mathrm{II}}_{\elemIndex-1}
    = 
    \basisCoeff^{\mathrm{I}}_{\elemIndex}
    = 
    \basisCoeff_{\nodeIndex}.
    \label{eq:ansatzForce}
\end{equation}
The nodal coefficients $\basisCoeff_{\nodeIndex}$ are, again, the continuous variables. As before, we encode them using \Cref{eq:real_number_representation} with binary variables $\binaryVec^{a}_{i}$, collected in $\binaryVec^{a}$ over all nodes. 
For the discrete design, each cross-sectional area takes one of two values, $\setAdmissibleDesign_{\elemIndex}=\{\crossSectionalAreaChoice{1},\,\crossSectionalAreaChoice{2}\}$, represented by a single binary variable $\binary^{\crossSectionalArea}_{\elemIndex}$ per element, collected in $\binaryVec^{\crossSectionalArea}$ over all elements.
% \begin{equation}
%     \frac{1}{\crossSectionalArea[\elemIndex]}%\lb\qubitDesign_e\rb 
%     = 
%     \frac{1}{\crossSectionalAreaChoice{1}} 
%     +
%     \lp
%         \frac{1}{\crossSectionalAreaChoice{2}}-\frac{1}{\crossSectionalAreaChoice{1}}
%     \rp\binary^{\crossSectionalArea}_{\elemIndex},
%     \label{eq:representationDesign}
% \end{equation}
\par
Note that the traction boundary condition for static admissibility can be enforced by prescribing the corresponding nodal force coefficient (i.e., setting the appropriate $\basisCoeff_{\nodeIndex}$). Additionally, the equilibrium constraint from \Cref{eq:equilibrium} is incorporated via a penalty term $\penaltyTerm$ that aggregates the squared residuals over all elements with penalty weight $\penaltyWeight$. The resulting penalized objective is
\begin{equation}
    \objectivePenalized\lp\qubitsVectorCoeffs,\qubitsVectorDesigns\rp 
    = 
    \totalComplEnergy \lp\qubitsVectorCoeffs,\qubitsVectorDesigns\rp 
    + 
    \penaltyWeight \penaltyTerm\lp\qubitsVectorCoeffs,\qubitsVectorDesigns\rp, 
    \qquad \penaltyWeight>0.
    \label{eq:objectivePenalized}
\end{equation}
\par
Substituting the discretized force and design parametrization into \Cref{eq:objectivePenalized} yields a cubic polynomial in the binary variables. We therefore apply a degree-reduction (quadratization) technique~\cite{Dattani2019}, introducing auxiliary variables $\qubitsVectorAuxiliary$, to provide $\objectivePenalized$ as \gls{qubo} objective $\hamiltonianQUBOComplEnergy$. 
Collecting all binaries into the single vector $\binaryVec = [\,\qubitsVectorCoeffs;\,\qubitsVectorDesigns;\,\qubitsVectorAuxiliary\,]$, the problem takes the final \gls{qubo} form
\begin{equation}
    \min_{\binaryVec}\hamiltonianQUBOComplEnergy(\binaryVec).
    % \min_{\qubitsVectorCoeffs,\,\qubitsVectorDesigns,\,\qubitsVectorAuxiliary}\hamiltonianQUBOComplEnergy\lp\qubitsVectorCoeffs,\qubitsVectorDesigns,\qubitsVectorAuxiliary\rp .
\end{equation}
% So, the final \gls{qubo} problem reads
% \begin{equation}
%     \min_{\qubitsVectorCoeffs,\,\qubitsVectorDesigns,\,\qubitsVectorAuxiliary} \hamiltonianQUBO\lp\qubitsVectorCoeffs,\qubitsVectorDesigns,\qubitsVectorAuxiliary\rp,
% \end{equation}
% where $\qubitsVectorCoeffs$ encodes the continuous nodal coefficients, $\qubitsVectorDesigns$ encodes the element-wise design choices, and $\qubitsVectorAuxiliary$ collects the auxiliary binaries introduced by the degree reduction. 
This \gls{qubo} model is implemented in the open-source software \textit{EngiOptiQA}~\cite{EngiOptiQA}, which leverages the \textit{Fixstars Amplify} \gls{sdk}~\cite{FixstarsAmplifySDK} for problem formulation and supports solution, e.g., via the \textit{\gls{ae}}~\cite{FixstarsAmplifyAE} or \textit{D\mbox{-}Wave’s Ocean} \gls{sdk}~\cite{DWaveOceanSDK}.
%% SOLUTION SCHEMES %%
\subsection{Solution Schemes}
\label{subsec:solutionSchemes}
In the following, we present two specific solution strategies that integrate with both the adaptive encoding strategy and the energy formulations:
(i) as an illustrative testbed for subsequent empirical error analysis, we couple the structural‑only potential energy formulation to a fluid subproblem under a fixed encoding, yielding an \gls{fsi} problem solved iteratively; and (ii) to assess the adaptive encoding in a representative mixed‑variable optimization setting, we employ a quadratic‑penalty method to embed a \gls{qubo} derived from the complementary energy principle for structural design optimization.
% First, we couple the  structural-only potential energy formulation to a fluid subproblem, yielding a \gls{fsi} problem, under a fixed encoding; this serves as an illustrative example and enables the empirical error analysis reported later. 
% To provide an illustrative testbed for the empirical error analysis reported later, we first couple the structural-only potential energy formulation to a fluid subproblem under a fixed encoding, yielding an \gls{fsi} problem.
% Second, we employ a quadratic penalty method that embeds the \gls{qubo} formulation based on the complementary energy principle for structural design optimization. This serves as the basis for assessing the impact of the adaptive encoding in a representative mixed‑variable engineering optimization setting.
% To provide a representative mixed‑variable optimization setting for assessing the adaptive encoding, we then employ a quadratic penalty method to embed a \gls{qubo} formulation derived from the complementary energy principle for structural design optimization.
%% Coupling Scheme for Fluid-Structure-Interaction
\subsubsection{Coupling Scheme for Fluid-Structure-Interaction}
\label{subsubsec:couplingSchemeForFSI}
We now detail the partitioned \gls{fsi} coupling that induces an iterative workflow compatible with our encoding and the \gls{qubo} formulation of the minimum potential energy principle. The fluid and structural subproblems are treated separately and exchange interface quantities in a Gauss-Seidel fixed‑point fashion with iteration index $\iter = 1,2,\dots$.
\par
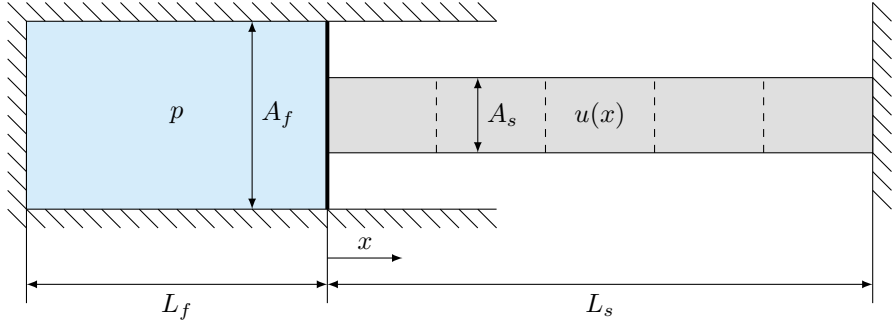
\begin{figure}
    \centering
    \begin{circuitikz}
\begin{scope}[xscale=-1]
\tikzset{normaltext/.style={xscale=1}}
\tikzstyle{every node}=[font=\normalsize]
\draw [short] (3.75,21) -- (3.75,17);
\draw [short] (3.75,20.75) -- (3.5,20.5);
\node [font=\normalsize] at (3.5,21.25) {};
\draw [short] (3.75,21) -- (3.5,20.75);
%\node (tikzmaker) [shift={(1, -0)}] at (3.5,20.5) {\includegraphics[width=2cm]{}};
\draw [short] (3.75,20.5) -- (3.5,20.25);
\draw [short] (3.75,20) -- (3.5,19.75);
\draw [short] (3.75,20.25) -- (3.5,20);
\draw [short] (3.75,19.75) -- (3.5,19.5);
\node [font=\normalsize] at (3.75,18) {};
\draw [short] (3.75,19.25) -- (3.5,19);
\node [font=\normalsize] at (3.75,18.5) {};
\node [font=\normalsize] at (3.75,18.5) {};
\node [font=\normalsize] at (3.75,18.75) {};
\node [font=\normalsize] at (3.75,19) {};
\draw [short] (3.75,19.5) -- (3.5,19.25);
\draw [short] (3.75,19) -- (3.5,18.75);
\node [font=\normalsize] at (4.75,18) {};
\draw [short] (3.75,18.75) -- (3.5,18.5);
\node [font=\normalsize] at (5.25,17.75) {};
\draw [short] (3.75,18.5) -- (3.5,18.25);
\draw [short] (8.75,20.75) -- (11,20.75);
\draw [short] (8.75,18.25) -- (11,18.25);
\draw [short] (15,19) -- (15,17);
\draw [<-, >=latex] (10,17.6) -- (11,17.6)node[normaltext,pos=0.5,above]{$\x$};
% \draw [->, >=latex] (7.75,20.5) -- (6.75,20.5)node[normaltext,pos=0.5,above]{$\displ(x)$};
% \draw [ fill=oi-sky!25] (11,20.75) rectangle  node[normaltext] {\normalsize $\domainFluid$} (15,18.25);
\draw [ fill=oi-sky!25] (11,20.75) rectangle  node[normaltext] {\normalsize $\pressure$} (15,18.25);
\draw [<->, >=latex] (3.75,17.25) -- (11,17.25)node[normaltext,pos=0.5,below]{$\lengthPiston$};
\draw [short] (11,18.25) -- (11,17);
\draw [<->, >=latex] (11,17.25) -- (15,17.25)node[normaltext,pos=0.5,below]{$\lengthChamber$};
% \draw [ fill=oi-grey!25] (3.75,20) rectangle  node[normaltext] {\normalsize $\domainStructure$} (11,19);
\draw [ fill=oi-grey!25] (3.75,20) rectangle node[normaltext] {\normalsize $\displ(x)$} (11,19);
\draw [line width=1.5pt, short] (11,20.75) -- (11,18.25);
\draw [short] (15.25,21) -- (15,20.75);
\node [font=\normalsize] at (4,20.75) {};
\draw [short] (9,21) -- (8.75,20.75);
%\node (tikzmaker) [shift={(1, -0)}] at (3.75,20) {\includegraphics[width=2cm]{}};
\draw [short] (15.25,20.75) -- (15,20.5);
\draw [short] (15.25,20.25) -- (15,20);
\draw [short] (15.25,20.5) -- (15,20.25);
\draw [short] (15.25,20) -- (15,19.75);
\draw [short] (15.25,19.5) -- (15,19.25);
\draw [short] (15.25,19.75) -- (15,19.5);
\draw [short] (15.25,19.25) -- (15,19);
\draw [short] (15.25,19) -- (15,18.75);
\draw [short] (15.25,18.75) -- (15,18.5);
\draw [short] (15.25,18.5) -- (15,18.25);
\node [font=\normalsize] at (9.75,20.75) {};
\draw [short] (9.25,21) -- (9,20.75);
\draw [short] (9.5,21) -- (9.25,20.75);
\draw [short] (9.75,21) -- (9.5,20.75);
\draw [short] (10,21) -- (9.75,20.75);
\draw [short] (10.25,21) -- (10,20.75);
\draw [short] (10.5,21) -- (10.25,20.75);
\draw [short] (10.75,21) -- (10.5,20.75);
\draw [short] (11,21) -- (10.75,20.75);
\draw [short] (11.25,21) -- (11,20.75);
\draw [short] (11.5,21) -- (11.25,20.75);
\draw [short] (11.75,21) -- (11.5,20.75);
\draw [short] (12,21) -- (11.75,20.75);
\draw [short] (12.25,21) -- (12,20.75);
\draw [short] (12.5,21) -- (12.25,20.75);
\draw [short] (12.75,21) -- (12.5,20.75);
\draw [short] (13,21) -- (12.75,20.75);
\draw [short] (13.25,21) -- (13,20.75);
\draw [short] (13.5,21) -- (13.25,20.75);
\draw [short] (13.75,21) -- (13.5,20.75);
\draw [short] (14,21) -- (13.75,20.75);
\draw [short] (14.25,21) -- (14,20.75);
\draw [short] (14.5,21) -- (14.25,20.75);
\draw [short] (14.75,21) -- (14.5,20.75);
\node [font=\normalsize] at (15.25,20.25) {};
\draw [short] (15,21) -- (14.75,20.75);
\draw [short] (15.25,18.25) -- (15,18);
\draw [short] (9,18.25) -- (8.75,18);
\draw [short] (9.25,18.25) -- (9,18);
\draw [short] (9.5,18.25) -- (9.25,18);
\draw [short] (9.75,18.25) -- (9.5,18);
\draw [short] (10,18.25) -- (9.75,18);
\draw [short] (10.25,18.25) -- (10,18);
\draw [short] (10.5,18.25) -- (10.25,18);
\draw [short] (10.75,18.25) -- (10.5,18);
\draw [short] (11,18.25) -- (10.75,18);
\draw [short] (11.25,18.25) -- (11,18);
\draw [short] (11.5,18.25) -- (11.25,18);
\draw [short] (11.75,18.25) -- (11.5,18);
\draw [short] (12,18.25) -- (11.75,18);
\draw [short] (12.25,18.25) -- (12,18);
\draw [short] (12.5,18.25) -- (12.25,18);
\draw [short] (12.75,18.25) -- (12.5,18);
\draw [short] (13,18.25) -- (12.75,18);
\draw [short] (13.25,18.25) -- (13,18);
\draw [short] (13.5,18.25) -- (13.25,18);
\draw [short] (13.75,18.25) -- (13.5,18);
\draw [short] (14,18.25) -- (13.75,18);
\draw [short] (14.25,18.25) -- (14,18);
\draw [short] (14.5,18.25) -- (14.25,18);
\draw [short] (14.75,18.25) -- (14.5,18);
\draw [short] (15,18.25) -- (14.75,18);

\draw [<->, >=latex] (9,19) -- (9,20)node[normaltext,pos=0.5,right]{$\areaPiston$};
\draw [<->, >=latex] (12,18.25) -- (12,20.75)node[normaltext,pos=0.5,right]{$\areaChamber$};

\draw[dashed] (5.2,19) -- (5.2,20);
\draw[dashed] (6.65,19) -- (6.65,20);
\draw[dashed] (8.1,19) -- (8.1,20);
\draw[dashed] (9.55,19) -- (9.55,20);

\end{scope}
\end{circuitikz}
    \caption{Schematic of the 1D static piston problem: a linear elastic rod of length $\lengthPiston$ and cross‑sectional area $\areaPiston$, here composed of $\nElem=5$ elements, seals a gas‑filled chamber of length $\lengthChamber$ and cross‑sectional area $\areaChamber$. The gas pressure $\pressure$ acts on the piston interface at $\x=0$, inducing the rod displacement field $\displ(\x)$.}
    \label{fig:staticPistonProblem}
\end{figure}
\begin{figure}
    \centering
     \begin{subfigure}[b]{0.49\textwidth}
        \centering
        % \raisebox{0pt}{\resizebox{1.\textwidth}{!}{%
        % \begin{tikzpicture}
% \tikzstyle{every node}=[font=\footnotesize]
% \draw  (6,21.75) ellipse (1cm and 0.5cm) node {\small Start} ;
% \draw [ fill=oi-sky!50 ] (4.5,20.5) rectangle  node {\small Solve Fluid} (7.5,19.5);
% \draw [ fill=oi-grey!30 ] (4.5,18.75) rectangle  node {\small Solve Structure} (7.5,17.75);
% \draw  (4.5,16.25) -- (6,15.5) -- (7.5,16.25) -- (6,17) -- cycle;
% \node [font=\small] at (6,16.25) {Converged?};
% \draw  (6,14.25) ellipse (1cm and 0.5cm) node {\small Stop} ;
% \draw [->, >=latex] (6,19.5) -- (6,18.75);
% \draw [->, >=latex] (6,21.25) -- (6,20.5);
% \draw [->, >=latex] (6,17.75) -- (6,17);
% \draw [->, >=latex] (6,15.5) -- (6,14.75);
% \draw [short] (7.5,16.25) -- (8.5,16.25);
% \draw [short] (8.5,16.25) -- (8.5,20)node[pos=0.3,right, rotate around={90:(0,0)}, fill=white]{$\iter=\iter+1$};
% \draw [->, >=latex] (8.5,20) -- (7.5,20);
% \node [font=\small] at (6.25,19.25) {$\forcePrescribed$};
% \node [font=\small] at (6.5,17.5) {$\displ(x)$};
% \node [font=\small, rotate around={90:(0,0)}] at (3.25,18.25) {Fixed-Point Iteration};
% \draw [ dashed] (3.75,21) rectangle  (9,15);
% \node [font=\small] at (6.5,15.25) {Yes};
% \node [font=\small] at (8,16.5) {No};
% \end{tikzpicture}

\begin{tikzpicture}[
  node distance=10mm and 10mm,
  >=latex,
  every node/.style={font=\small},
  block/.style={draw, align=center, minimum width=30mm, , minimum height=10mm, inner sep=10pt},
  decision/.style={draw, diamond, aspect=2, align=center, inner sep=1.5pt, minimum width=30mm},
  line/.style={->, thick}
]
% Nodes
\node[block, ellipse] (start) {Start};
\node[block, below=of start,fill=oi-sky!25] (fluid) {Solve Fluid};
\node[block, below=of fluid, fill=oi-green!25] (encoding) {Encoding};
\node[block, below=of encoding, fill=oi-orange!25] (struct) {$\min_{\binariesVecCoeff} \hamiltonianQUBOPotEnergy[\!,(\iter)](\binariesVecCoeff)$};
\node[block, below=of struct, fill=oi-green!25] (decoding) {Decoding};
\node[decision, below=of decoding, fill=oi-vermilion!25] (conv) {$\relChangeDisplacement<\tolRelChangeDisplacement$\\or\\$\iter=\maxIter$?};
\node[block, ellipse, below=of conv] (stop) {Stop};
\node[block, right=10mm of conv, fill=oi-grey!15] (update) {
Update ranges \\$\lb\basisCoeffMin^{(k+1)},\basisCoeffMin^{(k+1)}\rb$
};
% Dashed box around the fixed-point iteration
% \node[draw, dashed, fit=(fluid)(struct)(conv), inner sep=6mm] (fpibox) {};
% \node[left=7mm of fpibox, rotate=90] {Fixed-Point Iteration};
% Edges
\draw[line] (start) -- (fluid);
\draw[line] (fluid) -- node[fill=white, right] {$\pressure$}(encoding);
\draw[line] (encoding) -- node[fill=white, right] {$\displ\lab\binariesVecCoeff\rab$} (struct);
\draw[line] (struct) -- node[fill=white, right] {$\qubitsVectorCoeffs[,(k)]$}(decoding);
\draw[line] (decoding) -- node[fill=white, right] {$\basisCoeff_{\nodeIndex}^{(\iter)}$} (conv);
\draw[line] (conv) -- node[right, xshift=2mm]{Yes} (stop);
% \draw[line] (conv.east) -- node[above]{No}++(10mm,0)   |- node[pos=0.3, rotate around={90:(0,0)}, fill=white] {$\iter=\iter+1$} (fluid.east);
\draw[line] (conv.east) -- node[above]{No} (update.west);
\draw[line] (update.north) |- node [pos=0.25,rotate around={90:(0,0)}, fill=white] {$\iter=\iter+1$} (fluid.east);
\end{tikzpicture}
        % }}
         \caption{Fixed-point iteration for \gls{fsi} where the \gls{qubo} formulation of the principle of minimum potential energy is used for structural analysis.}
         \label{fig:partitioned_approach}
     \end{subfigure}
     \hfill
     \begin{subfigure}[b]{0.49\textwidth} 
        \centering
        % \raisebox{0pt}{\resizebox{1.\textwidth}{!}{%
        \begin{tikzpicture}[
  node distance=10mm and 10mm,
  >=latex,
  every node/.style={font=\small},
  block/.style={draw, align=center, minimum width=30mm, minimum height=10mm, inner sep=10pt},
  decision/.style={draw, diamond, aspect=2, align=center, inner sep=1.5pt, minimum width=30mm},
  line/.style={->, thick}
]
\node[block,ellipse] (init) {Start%\\
%$k=0$, $[\stateVarMin^{(0)},\stateVarMax^{(0)}]$, $\penaltyWeight^{(0)}>0$, $\gamma>1$
};
\node[block, below= of init, fill=oi-green!25] (encoding) {Encoding};
% \node[block, below=of encoding] (penobj) {%Penalized \acrshort{qubo} objective\\
% $
% \objectivePenalized^{(\iter)}
% = \totalComplEnergy
% + \penaltyWeight^{(\iter)}\,\penaltyTerm
% $
% };
\node[block, below=of encoding, fill=oi-orange!25] (subprob) {%Solve \acrshort{qubo} problem\\
$
\min_{\binaryVec} \;\hamiltonianQUBOComplEnergy[\!,(\iter)](\binaryVec)%\Rightarrow(\designVarVec^{(k)},\stateVarVec^{(k)}) 
$
};
\node[block, below=of subprob, fill=oi-green!25] (decoding) {Decoding};
\node[decision, below=of decoding, fill=oi-vermilion!25] (feas) {
$\penaltyTerm \le \tolFeasibility$
\\
or 
\\
$k = k_{\text{max}}$?
};
\node[block, right=10mm of feas, fill=oi-grey!15] (update) {Increase penalty\\
$\penaltyWeight^{(k+1)} = \penaltyWeightFactor\,\penaltyWeight^{(k)}$\\
Update ranges \\$\lb\basisCoeffMin^{(k+1)},\basisCoeffMax^{(k+1)}\rb$
};
\node[block, ellipse, below=of feas] (stop) {Stop};
% Edges
\draw[line] (init) --  (encoding);
\draw[line] (encoding) -- node [fill=white, right]{$\crossSectionalArea[e]\lab\qubitsVectorDesigns\rab,\,\basisCoeff_{\nodeIndex}\lab\qubitsVectorCoeffs\rab$} (subprob);
% \draw[line] (penobj) -- (subprob);
\draw[line] (subprob) -- node [fill=white, right] {$\qubitsVectorDesigns[,(k)],\,\qubitsVectorCoeffs[,(k)]$} (decoding);
\draw[line] (decoding) -- node [fill=white, right] {$\crossSectionalArea[e]^{(\iter)},\,\basisCoeff_{\nodeIndex}^{(\iter)}$}(feas);
\draw[line] (feas) -- node[right, xshift=2mm]{Yes} (stop);
\draw[line] (feas.east) -- node[above]{No} (update.west);
\draw[line] (update.north) |- node [pos=0.25,rotate around={90:(0,0)}, fill=white] {$\iter=\iter+1$} (encoding.east);
\end{tikzpicture}
        % }}
         % \caption{Quadratic penalty method with the adaptive encoding strategy (\Cref{subsubsec:adaptiveEncoding}) for the \gls{qubo}-based structural design optimization problem (\Cref{subsubsec:principleOfMinimumComplementaryEnergyForStructuralDesignOptimization}).}
         \caption{Quadratic penalty method with the adaptive encoding strategy for the \gls{qubo}-based structural design optimization problem.}
         \label{fig:quadraticPenaltyMethodFlowchart}
     \end{subfigure}
    \caption{Examples for iterative solution schemes that integrate with both the structural energy formulations and the adaptive encoding strategy.}
    \label{fig:examplesIterativeSolutionSchemes}
\end{figure}
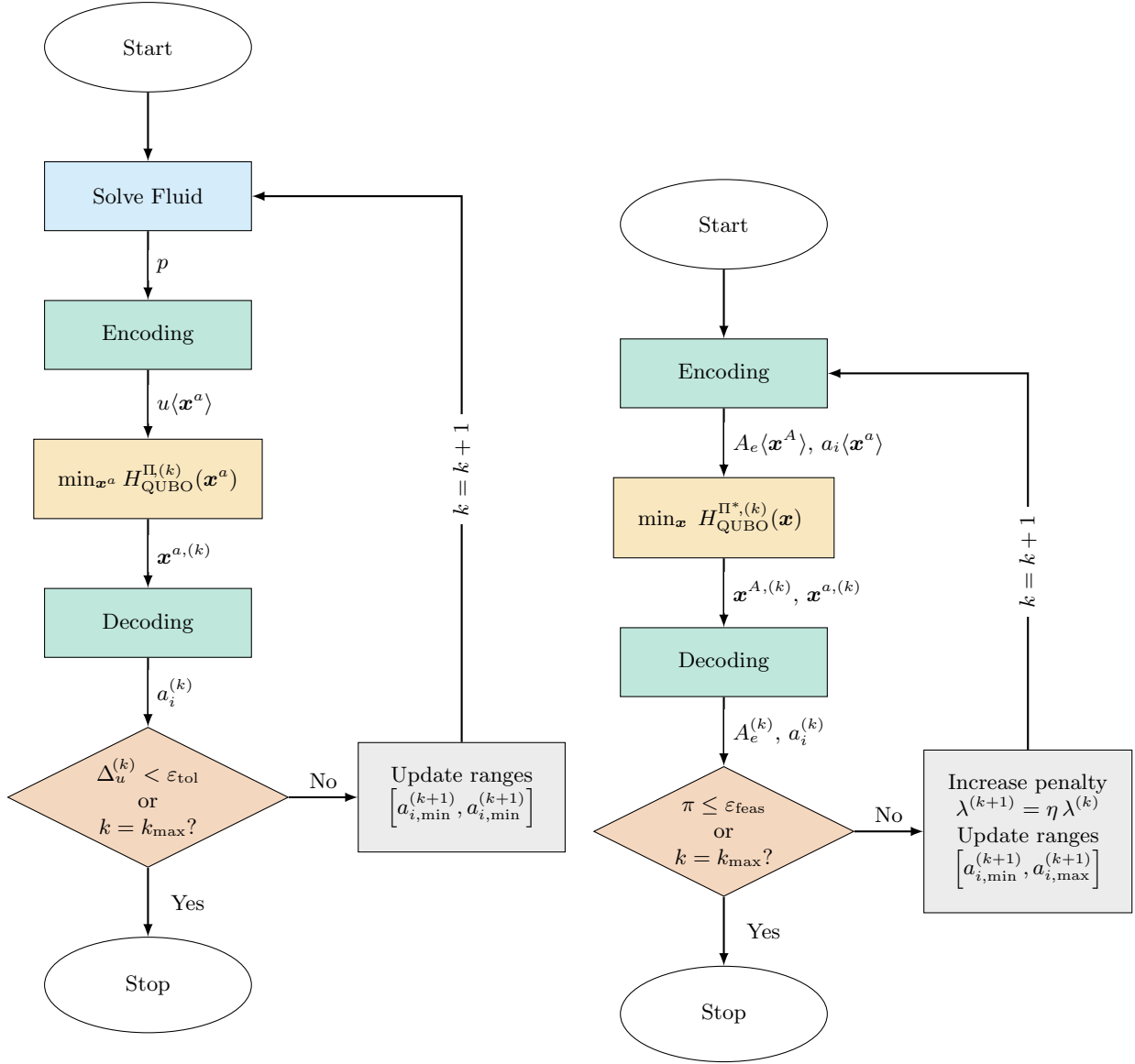

Concretely, we consider a static piston problem in which a one‑dimensional linear elastic rod of length $\lengthPiston$, cross‑sectional area $\areaPiston$, and Young’s modulus $\youngModulus$, composed of $\nElem$ elements, seals a gas‑filled chamber of length $\lengthChamber$ and cross‑sectional area $\areaChamber$ (see~\Cref{fig:staticPistonProblem}). The structural subproblem in each iteration follows the minimum potential energy principle and is solved via its \gls{qubo} formulation from \Cref{subsubsec:principleOfMinimumPotentialEnergyForStructuralAnalysis} for the scalar displacement $\displ^{(\iter)}(\x)$. 
The fluid obeys the ideal‑gas law with specific heat ratio $\heatRatio$, yielding the pressure update $\pressure^{(\iter+1)}$ as a function of the initial chamber length $\lengthChamber^{(0)}$ and the current piston displacement at the interface ($\x=0$):
% The fluid is modeled via the ideal gas law with specific heat ratio $\heatRatio$, which provides the pressure $\pressure^{(\iter)}$ as a function of the chamber volume determined by the current piston displacement at the interface ($\x=0$):
\begin{equation}
    \pressure^{(\iter+1)} = \pressure^{(\iter)} \left(\frac{\lengthChamber^{(0)}}{\lengthChamber^{(0)} + \displ^{(k)}(0)}\right)^{\!\heatRatio}.
    \label{eq:p_2_finished}
\end{equation}
This pressure, in turn, acts as the load on the piston rod, accounted for by the prescribed surface traction in \Cref{eq:totalPotentialEnergyLinear}. A detailed description is provided in~\cite[Ch.~3]{Freinberger2025}.
The coupling proceeds as a fixed‑point iteration (see \Cref{fig:partitioned_approach}) and terminates when the relative change in the displacement field,
\begin{equation}
\relChangeDisplacement = \frac{\lVert \displ^{(\iter)}(\x) - \displ^{(\iter-1)}(\x) \rVert_{H^1}}{\lVert \displ^{(\iter-1)}(\x) \rVert_{H^1}},
\end{equation}
falls below the tolerance $\tolRelChangeDisplacement$ or when the maximum number of iterations $\maxIter$ is reached.
\par
This lightweight \gls{fsi} setting serves two purposes: (i) it provides an illustrative testbed to study the interaction between encoding and solver accuracy, and (ii) it yields an iterative process in which adaptive encoding can be exercised and evaluated.

%% Quadratic Penalty Method for Structural Design Optimization
\subsubsection{Quadratic Penalty Method for Structural Design Optimization}
\label{subsubsec:quadraticPenaltyMethodForStructuralDesignOptimization}
% \begin{figure}
%     \centering
%     \input{fig/materials_and_methods/quadratic_penalty_method}
%     \caption{Quadratic penalty method with the adaptive encoding strategy (\Cref{subsubsec:adaptiveEncoding}) for the \gls{qubo}-based structural design optimization problem (\Cref{subsubsec:principleOfMinimumComplementaryEnergyForStructuralDesignOptimization}).}
%     \label{fig:quadraticPenaltyMethodFlowchart}
% \end{figure}
To handle the constrained structural design optimization problem in \Cref{eq:designOptimizationProblem}, we now turn to penalty methods.
Penalty methods replace a constrained optimization problem as given in \Cref{eq:optimizationTask}
by a sequence of unconstrained minimization problems, with outer iterations indexed by $\iter\ge0$. 
Their corresponding objectives include additional nonnegative constraint-violation terms that vanish for feasible solutions and add a positive penalty otherwise (see, e.g., \cite[Ch.~17.1]{Nocedal2006}).
A common quadratic choice is
\begin{equation}
\objectivePenalized^{(k)}\lp\designVarVec,\stateVarVec\rp
= \objective\lp\designVarVec,\stateVarVec\rp
+ \frac{\penaltyWeight^{(k)}}{2}\,\big\|\constraintsEquality\lp\designVarVec,\stateVarVec\rp\big\|_2^2
+ \frac{\penaltyWeight^{(k)}}{2}\,\big\|\,[\,\constraintsInequality\lp\designVarVec,\stateVarVec\rp\,]_+\,\big\|_2^2,  
\end{equation}
where $[z]_+ := \max(z,0)$ is applied componentwise and the penalty weight is progressively increased by a growth factor $\penaltyWeightFactor>1$ via $\penaltyWeight^{(k+1)}=\penaltyWeightFactor\,\penaltyWeight^{(k)}$. This discourages infeasibility and drives the minimizers of $\objectivePenalized^{(k)}$ toward the feasible region.
%As an additional measure, we keep the penalty weight constant ($\penaltyWeightFactor=1$) whenever the constraint violation worsens, which safeguards the adaptive number representation against becoming trapped in infeasible ranges.
\par
The stepwise nature of the method enables integrating the adaptive encoding strategy and its application in our \gls{qubo}-based structural design setting. 
% In each penalty iteration $k$, we apply the encoding of $(\designVarVec^{(k)},\stateVarVec^{(k)})$ with the current ranges $[\stateVarMin^{(k)},\stateVarMax^{(k)}]$ to solve the \gls{qubo} problem for the binary variables $\binaryVec$ with the penalized objective
In each penalty iteration $k$, we apply the encoding of the cross-sectional areas and nodal force coefficients $(\crossSectionalArea[e]\lab\qubitsVectorDesigns\rab,\basisCoeff_{\nodeIndex}\lab\qubitsVectorCoeffs\rab)$ with the current ranges $[\basisCoeffMin^{(k)},\basisCoeffMax^{(k)}]$ for $\basisCoeff_{\nodeIndex}$ to construct the penalized objective
\begin{equation}
\objectivePenalized^{(\iter)}\lp\qubitsVectorDesigns,\qubitsVectorDesigns\rp
= \totalComplEnergy\lp\qubitsVectorCoeffs,\qubitsVectorDesigns\rp 
+ \penaltyWeight^{(k)}\,\penaltyTerm\lp\qubitsVectorCoeffs,\qubitsVectorDesigns\rp .    
\end{equation}
After applying the degree reduction, we solve the \gls{qubo} problem,
\begin{equation}
    \min_{\binaryVec}\hamiltonianQUBOComplEnergy[\!,(k)](\binaryVec),
\end{equation}
for the binary variables $\binaryVec = [\,\qubitsVectorCoeffs;\,\qubitsVectorDesigns;\,\qubitsVectorAuxiliary\,]$.
Given the (approximate) minimizer $\binaryVec^{(k)}$, we update the design and state variables accordingly:
\begin{equation}
% (\designVarVec^{(k)},\stateVarVec^{(k)})
\lp\crossSectionalArea[e]^{(k)},\basisCoeff_{\nodeIndex}^{(k)}\rp
% := \lp\designVarVec\lp\binaryVec^{(k)}\rp,\stateVarVec\lp\binaryVec^{(k)}\rp\rp.    
= \lp\crossSectionalArea[e]\lp\qubitsVectorDesigns[,(k)]\rp,\basisCoeff_{\nodeIndex}\lp\qubitsVectorCoeffs[,(k)]\rp\rp.    
\end{equation}
We declare feasibility if $\penaltyTerm(\qubitsVectorDesigns[,(k)],\qubitsVectorCoeffs[,(k)]) \le \varepsilon_{\mathrm{feas}}$; otherwise, we increase the penalty parameter, update the encoding ranges $[\basisCoeffMin^{(k+1)},\basisCoeffMax^{(k+1)}]$ based on $(\basisCoeff_{\nodeIndex}^{(k-1)},\basisCoeff_{\nodeIndex}^{(k)})$, and repeat until the feasibility tolerance is met or a maximum number of iterations $\maxIter$ is reached.
The overall procedure is summarized in the flow chart of \Cref{fig:quadraticPenaltyMethodFlowchart}, which illustrates the adaptive encoding, \gls{qubo} solve, feasibility check, and penalty update within each iteration.
\par
The quadratic penalty method and the coupling scheme for \gls{fsi} (\Cref{subsubsec:couplingSchemeForFSI}) jointly illustrate how the adaptive encoding can be embedded in iterative algorithms using \gls{qubo}-based formulations. We now shift from algorithmic design to implementation, focusing on hardware-induced effects and error measures when solving the \gls{qubo} problems on \gls{qa} devices.

%% HARDWARE EFFECTS AND ERROR MEASURE
\subsection{Hardware Effects and Error Measure}
\label{subsec:hardwareEffectsAndErrorMeasure}
This subsection examines hardware-specific effects on \gls{qa} solvers and introduces the error metrics used in our evaluation. In particular, we address the following practical aspects of executing the \gls{qubo} problems on \gls{qa} hardware. 
First, we discuss scaling for \gls{qa} hardware, whereby the coefficients are normalized and rescaled to respect device-specific range limits. 
Second, we examine how the programmed Hamiltonian differs from the logical one, i.e., the scaled version of $\hamiltonianIsing$ resulting from $\hamiltonianQUBO$, because of hardware-related fidelity limitations.
Third, to disentangle representational inaccuracies due to the encoding from shortcomings in the solver performance, we introduce the ``best approximation'': the best solution representable under a given encoding, i.e., the minimizer of $\hamiltonianQUBO$ over the encoded domain. This serves as a reference for reporting solution quality, so that deviations from this reference can be attributed solely to solver performance.
Finally, we introduce the $H^1$ error to evaluate the accuracy of the solution for the state variables.
\bigskip\par
Starting from the \gls{qubo} Hamiltonian functions $\hamiltonianQUBO$ derived in \Cref{subsec:energyFormulationsInStructuralAnalysisAndDesignOptimization}, two subsequent transformations are applied to obtain a device-ready problem formulation. 
First, we perform a change of variables, $\binaryVec\rightarrow\spinVec$, to convert the \gls{qubo} formulation to its Ising form, as explained in \Cref{subsec:problemFormulationsForQA}.
Second, since each \gls{qpu} supports only a specific range for the linear and quadratic Ising coefficients $\isingCoeffLinear_{i}$ and $\isingCoeffQuadratic_{ij}$, these coefficients are uniformly scaled by a common factor to fit the available range:
\begin{equation}
    \isingCoeffLinearScaled_i = \frac{1}{\isingScalingFactor}\, \isingCoeffLinear_i, \qquad
    \isingCoeffQuadraticScaled_{ij} = \frac{1}{\isingScalingFactor}\, \isingCoeffQuadratic_{ij}. 
\end{equation}
The computation of the scaling factor $\isingScalingFactor$ is described in the D-Wave documentation.\footnote{\urlDWaveAutoScale}
\par
After scaling, the logical Ising problem must be embedded into the hardware connectivity graph. Because the logical problem structure may not match the hardware graph, a minor‑embedding is performed in which each logical spin variable is mapped to a connected subgraph, i.e., ``chain'', of physical qubits. Qubits within a chain are coupled with strength $J_{\mathrm{chain}}<0$ to encourage consistency, so that ideally all qubits in the chain align and represent a single logical value. A chain break occurs when not all qubits in a chain agree, and the chain‑break fraction is the proportion of broken chains per sample. Choosing $J_{\mathrm{chain}}$ involves a trade‑off: stronger chains reduce breakage but must fit within device coefficient bounds alongside the problem couplings, which can force additional global rescaling; longer and more numerous chains also increase resource usage and susceptibility to errors. 
\bigskip\par
Beyond this deterministic scaling and embedding step (with a possible additional rescaling to accommodate chain couplers), the coefficients realized on the \gls{qpu} may deviate from their scaled targets, resulting in a slightly altered Hamiltonian:
\begin{equation}
\hamiltonianIsing^\delta 
= 
\sum_i \lp\isingCoeffLinearScaled_i + \delta \isingCoeffLinear_i\rp s_i 
+ 
\sum_{i < j} \lp\isingCoeffQuadraticScaled_{ij} + \delta \isingCoeffQuadratic_{ij}\rp \spin_i \spin_j,
\label{eq:h_ising_ice}
\end{equation}
where $\delta \isingCoeffLinear_i$ and $\delta \isingCoeffQuadratic_{ij}$ denote deviations from the intended $\isingCoeffLinearScaled_i$ and $\isingCoeffQuadraticScaled_{ij}$ values, respectively.
These implementation infidelities are commonly referred to as \glspl{ice}.\footnote{\urlDWaveICEs}
Dominant contributors include background susceptibility, flux noise in qubits, I/O system effects, and \gls{dac} quantization and control errors. 
For example, \gls{dac} quantization errors arise when the user-defined (scaled) coefficients $\isingCoeffLinearScaled$ and $\isingCoeffQuadraticScaled$ are converted to analog signals to program the \gls{qpu}. 
Because the perturbations $\delta \isingCoeffLinear_i$ and $\delta \isingCoeffQuadratic_{ij}$ are summed over the number of qubits $\numQubits$, their aggregate impact typically grows with problem size. 
\bigskip\par
To isolate hardware-related effects from errors due to the finite resolution of the encoding of continuous variables, we define the best approximation for a given encoding as \added{the solution obtained by solving the induced binary optimization problem exactly:}
\begin{equation}
\lp\designVarVecBest,\stateVarVecBest\rp 
= 
\lp\designVarVec\lp\binaryVecBest\rp,\stateVarVec\lp\binaryVecBest\rp\rp, 
\qquad
\binaryVecBest = \arg\min_{\binaryVec} \;\hamiltonianQUBO(\binaryVec).    
\end{equation}
\added{Here, the decoding maps the binary vector $\binaryVecBest$ to the corresponding discrete and continuous variables $\designVarVecBest$ and $\stateVarVecBest$.} 
\replaced{Consequently, this}{This} reference represents the best solution representable under the chosen encoding (i.e., absent hardware and solver imperfections); deviations from it are attributed to solver/\gls{qpu} effects.
\par 
Finally, to quantify the solution quality for the continuous state variables (e.g., displacement or force), we use the relative error in the $H^1$ norm,
\begin{equation}
    % \epsilon_{\mathrm{H^1}} 
    \relErrorHOne[f]
    =
    \frac{\lVert f(x)-f^*(x) \rVert_{H^1}}{\lVert f^*(x) \rVert_{H^1}},
    \label{eq:relErrorH1}
\end{equation}
where $f$ denotes the computed field and $f^*$ the reference solution.

%% RESULTS %%
\section{Results and Discussion}
\label{sec:results}
With the methodology in place, we now turn to empirical evaluation. The results address two complementary questions: how well fixed encodings for continuous variables perform on QA hardware, and how beneficial and robust the adaptive encoding strategy is. 
First, we conduct an empirical error analysis to quantify how the binary budget per continuous variable $\numQubitsPerNode$ in a fixed encoding affects solution accuracy on hardware‑embedded \gls{qa}. While increasing $\numQubitsPerNode$ improves the encoding‑optimal solution, the \gls{qa} accuracy improves only up to moderate bit depths and then plateaus. 
Second, motivated by these observations, we assess how the adaptive encoding, under fixed binary budgets, affects solution quality in structural design optimization and examine its sensitivity to key hyperparameters (the relaxation factor $\relaxationFactor$, the initial range scale, and the number of \gls{qa} reads) and conduct a scalability analysis with a quantum-inspired solver to probe its performance on larger problem instances. 
Together, the results show why simply adding bits may not translate to better hardware performance and how the adaptive encoding can robustly deliver improved accuracy with the same number of binary variables.
\subsection{Empirical Error Analysis}
\label{subsec:empiricalErrorAnalysis}
In a fixed, uniform encoding with $\numQubitsPerNode$ binary variables per continuous variable, increasing $\numQubitsPerNode$ improves local resolution and tightens the best attainable (encoding‑optimal) approximation. On hardware‑embedded \gls{qa}, however, larger $\numQubitsPerNode$ also increases the total problem size $\numQubits$, which can amplify \glspl{ice} and thereby the aggregate deviation of the implemented Hamiltonian from its target. Moreover, larger problems typically require embeddings with more and/or longer chains. We therefore assess how solution accuracy evolves with $\numQubitsPerNode$ on the illustrative \gls{fsi} problem described in \Cref{subsubsec:couplingSchemeForFSI}. The parameters for the test case are collectively presented in \Cref{tab:piston_problem_physical_parameters}.
We first examine the solution quality in a single coupling step, measured by the relative $H^1$ error in the displacement $\relErrorHOne[\displ]$ computed with respect to the analytical solution, as $\numQubitsPerNode$ increases. We then show how the adaptive encoding further improves accuracy over the full coupling process.
\begin{table}
    \begin{tabularx}{\textwidth}{CCCCCCCC}
    \toprule
    $\nElem$ &
    $\lengthPiston$	& 
    $\areaPiston$	& 
    $\youngModulus$ & 
    $\lengthChamberInitial$ & 
    $\areaChamber$ & 
    $\heatRatio$ &  
    $\pressure^{(0)}$\\
    \midrule
    $2$ &
    $1.0$ & 
    $\lcb 1.0,1.0 \rcb$ & 
    $\lcb 1.0,1.0 \rcb$ 
    & $1.0$
    & $2.0$ 
    & $1.4$ 
    & $0.25$ \\
    \bottomrule
    \end{tabularx}
    \caption{Static piston problem: parameters for the static piston problem.}
    \label{tab:piston_problem_physical_parameters}
\end{table}
\bigskip\par
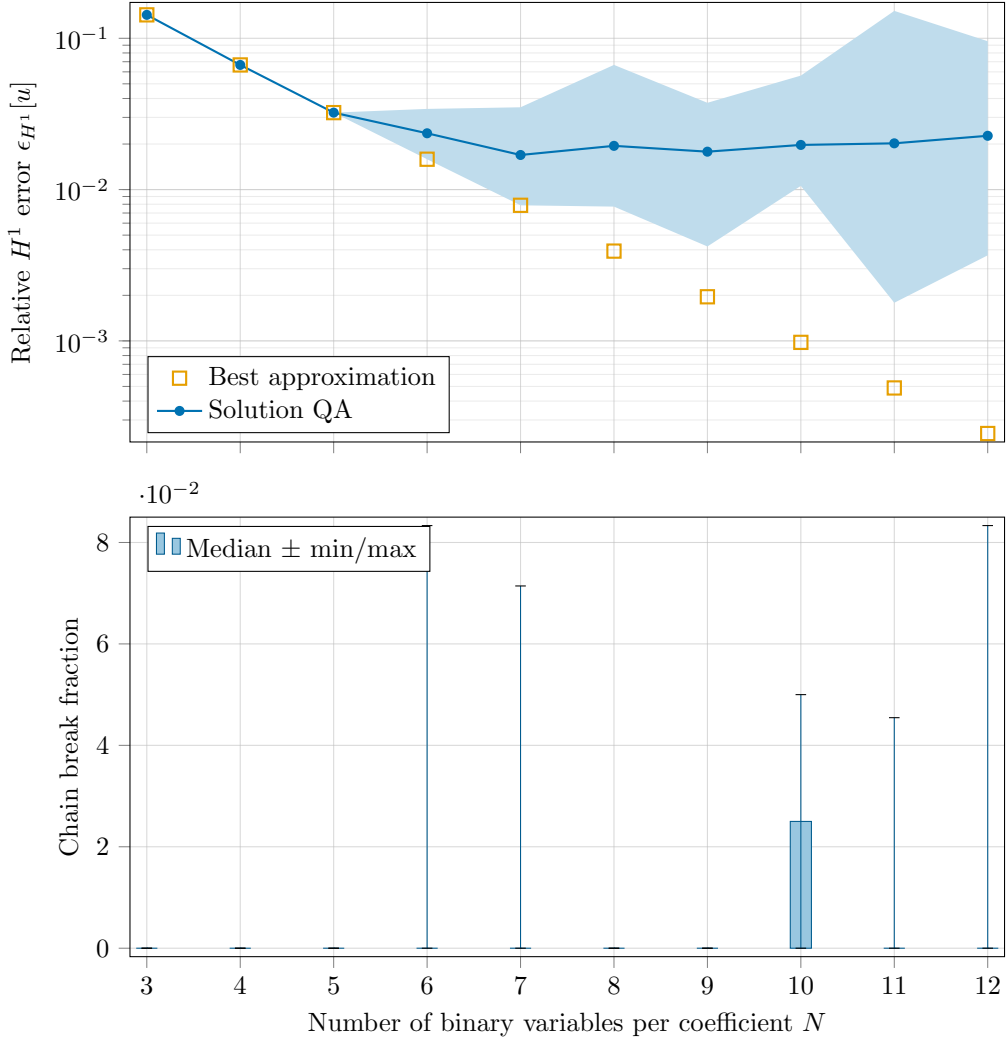
\begin{figure}
    \centering
    \begin{tikzpicture}
\begin{groupplot}[group style={group size=1 by 3}]
\nextgroupplot[
    width=0.8\linewidth,
    height=0.45\linewidth,
    ylabel={Relative $H^1$ error $\relErrorHOne[\displ]$},
    ymode=log,
    log basis y=10,
    grid=both,
    grid style={opacity=0.6},
    minor grid style={opacity=0.3},
    legend style={at={(0.02,0.02)}, anchor=south west, cells={anchor=west}},
    legend columns=1,
    tick align=outside,
    tick pos=left,
    enlargelimits=0.02,
    xticklabel=\empty
    ]
    % Best approximation error
    \addplot[only marks, mark=square, mark size=2.5pt, thick, color=oi-orange]
      table[x=N, y=error_ba_median, col sep=comma] {data/empirical_error_analysis/error_over_num_binary_variables_aggregated.csv};
    \addlegendentry{Best approximation}

    % shaded min-max band
    \addplot[name path=low, draw=none, forget plot]
      table[x=N, y=error_min, col sep=comma]
      {data/empirical_error_analysis/error_over_num_binary_variables_aggregated.csv};
    \addplot[name path=high, draw=none, forget plot]
      table[x=N, y=error_max, col sep=comma]
      {data/empirical_error_analysis/error_over_num_binary_variables_aggregated.csv};
    \addplot[oi-blue!25, forget plot] fill between[of=low and high];  
    % Median error
    \addplot[mark=*, mark size=1.5pt, thick, color=oi-blue]
      table[x=N, y=error_median, col sep=comma] {data/empirical_error_analysis/error_over_num_binary_variables_aggregated.csv};
    \addlegendentry{Solution \acrshort{qa}}

% \nextgroupplot[
%  width=0.8\linewidth,
%     height=0.45\linewidth,
%     xlabel={Number of binary variables per coefficient $\numQubitsPerNode$},
%     ylabel={Chain break fraction},
%     grid=both,
%     grid style={opacity=0.6},
%     minor grid style={opacity=0.3},
%     legend style={at={(0.02,0.02)}, anchor=south west, cells={anchor=west}},
%     legend columns=1,
%     tick align=outside,
%     tick pos=left,
%     enlargelimits=0.02,
%     ]
%     % shaded min-max band
%     \addplot[name path=low, draw=none, forget plot]
%       table[x=N, y=chain_break_min, col sep=comma]
%       {data/empirical_error_analysis/error_over_num_binary_variables_aggregated.csv};
%     \addplot[name path=high, draw=none, forget plot]
%       table[x=N, y=chain_break_max, col sep=comma]
%       {data/empirical_error_analysis/error_over_num_binary_variables_aggregated.csv};
%     \addplot[oi-blue!25, forget plot] fill between[of=low and high]; 
%     % Chain break fraction
%     \addplot[mark=*, mark size=1.5pt, thick, color=oi-blue]
%       table[x=N, y=chain_break_median, col sep=comma] {data/empirical_error_analysis/error_over_num_binary_variables_aggregated.csv};

\nextgroupplot[
  width=0.8\linewidth,
  height=0.45\linewidth,
  xlabel={Number of binary variables per coefficient $\numQubitsPerNode$},
  ylabel={Chain break fraction},
  grid=both,
  grid style={opacity=0.6},
  minor grid style={opacity=0.3},
  legend style={at={(0.02,0.98)}, anchor=north west},
  legend columns=1,
  tick align=outside,
  tick pos=left,
  enlargelimits=0.02,
  ybar,
  bar width=8pt,
  %xtick=data,
  %enlarge x limits=0.2
]
\addplot[
  draw=oi-blue!80!black,
  fill=oi-blue!40,
  error bars/.cd,
    y dir=both,
    y explicit,
]
table[
  x=N,
  y=chain_break_median,
  y error plus expr=\thisrow{chain_break_max}-\thisrow{chain_break_median},
  y error minus expr=\thisrow{chain_break_median}-\thisrow{chain_break_min},
  col sep=comma
]{data/empirical_error_analysis/error_over_num_binary_variables_aggregated.csv};
\addlegendentry{Median ± min/max}

% \nextgroupplot[
%   width=0.8\linewidth,
%   height=0.45\linewidth,
%   xlabel={Number of binary variables per coefficient $\numQubitsPerNode$},
%   ylabel={Maximum chain length },
%   grid=both,
%   grid style={opacity=0.6},
%   minor grid style={opacity=0.3},
%   legend style={at={(0.02,0.98)}, anchor=north west},
%   legend columns=1,
%   tick align=outside,
%   tick pos=left,
%   enlargelimits=0.02,
%   ybar,
%   bar width=8pt,
%   %xtick=data,
%   %enlarge x limits=0.2
% ]
% \addplot[
%   draw=oi-blue!80!black,
%   fill=oi-blue!40,
%   error bars/.cd,
%     y dir=both,
%     y explicit,
% ]
% table[
%   x=N,
%   y=max_chain_length_median,
%   y error plus expr=\thisrow{max_chain_length_max}-\thisrow{max_chain_length_median},
%   y error minus expr=\thisrow{max_chain_length_median}-\thisrow{max_chain_length_min},
%   col sep=comma
% ]{data/empirical_error_analysis/error_over_num_binary_variables_aggregated.csv};
% \addlegendentry{Median ± min/max}
\end{groupplot}

\end{tikzpicture}
    \caption{Static piston problem: empirical error analysis versus bit‑depth $\numQubitsPerNode$ across $\nRuns=10$ runs for one iteration in the coupling scheme. Top: median best approximation error and median \gls{qa} solution error; shaded bands indicate the min–max range. Bottom: median chain‑break fraction with min/max bars.}
    \label{fig:empiricalErrorAnalysis}
\end{figure}

We vary $\numQubitsPerNode \in \{3,\dots,12\}$ and use the range $[0,1]$ for all nodal displacement coefficients $\basisCoeff_{\nodeIndex}$ defining $\displ(\x)$.
All \gls{qa} runs are executed on D\mbox{-}Wave’s \texttt{Advantage\_system4.1} with an annealing time of $\annealingTime=\SI{10}{\micro\second}$ and $\numReads=500$ reads, i.e.,
500 independent annealing samples of the same problem Hamiltonian $\hamiltonianQUBOPotEnergy[\!,(\iter)]$ per coupling step.
\added{The best approximation is computed by exhaustively enumerating all binary configurations and selecting the configuration with minimum $\hamiltonianQUBOPotEnergy[\!,(\iter)]$. This yields the exact optimum of the corresponding encoded binary problem.}
We report aggregated results over $\nRuns=10$ runs for (i) the best approximation error, (ii) the error of the \gls{qa} solutions, and (iii) the corresponding chain‑break fractions; see \Cref{fig:empiricalErrorAnalysis}. As expected, the best approximation error decreases monotonically with $\numQubitsPerNode$. The \gls{qa} solution tracks this trend only up to $\numQubitsPerNode \le 5$, where it frequently attains the encoding‑optimal approximation. For $\numQubitsPerNode \ge 6$, the \gls{qa} error stops improving and plateaus. The chain‑break fraction fluctuates but its median remains low; notably, for $\numQubitsPerNode \in \{8,9\}$ we observe only runs with zero chain‑break fraction, but the best approximation still cannot be reached.
\par
This divergence beyond $\numQubitsPerNode=5$ indicates that hardware effects dominate the gains from finer quantization. Because the shortfall occurs even without chain breaks, it cannot be attributed to chaining alone; rather, it is consistent with \glspl{ice} that grow with problem size, making larger instances more vulnerable. 
In summary, beyond a moderate bit‑depth, simply increasing the binary count per continuous variable fails to improve accuracy on current hardware.
This observation motivates the application of the proposed adaptive encoding strategy, which concentrates resolution where needed under a fixed binary budget.
\bigskip\par
\begin{figure}
    \centering
    \begin{tikzpicture}

\begin{axis}[
    width=0.8\linewidth,
    height=0.45\linewidth,
    ylabel={Relative $H^1$ error $\relErrorHOne[\displ]$},
    ymode=log,
    log basis y=10,
    grid=both,
    grid style={opacity=0.6},
    minor grid style={opacity=0.3},
    legend style={at={(0.02,0.02)}, anchor=south west, cells={anchor=west}},
    legend columns=1,
    tick align=outside,
    tick pos=left,
    enlargelimits=0.02,
    xlabel={Iteration $\iter$},
    ]
    % Fixed range
    % Shaded IQR band
    \addplot[name path=low, draw=none, forget plot]
      table[x=iteration, y=error_fixed_q25, col sep=comma]
      {data/empirical_error_analysis/error_over_iterations_aggregated.csv};
    \addplot[name path=high, draw=none, forget plot]
      table[x=iteration, y=error_fixed_q75, col sep=comma]
      {data/empirical_error_analysis/error_over_iterations_aggregated.csv};
    \addplot[oi-blue!25, forget plot] fill between[of=low and high]; 
    % Median
    \addplot[ mark=*, mark size=1.5pt, thick, color=oi-blue]
      table[x=iteration, y=error_fixed_median, col sep=comma] {data/empirical_error_analysis/error_over_iterations_aggregated.csv};
    \addlegendentry{Fixed encoding}
    % Adaptive range
    % Shaded IQR band
    \addplot[name path=low, draw=none, forget plot]
      table[x=iteration, y=error_adaptive_q25, col sep=comma]
      {data/empirical_error_analysis/error_over_iterations_aggregated.csv};
    \addplot[name path=high, draw=none, forget plot]
      table[x=iteration, y=error_adaptive_q75, col sep=comma]
      {data/empirical_error_analysis/error_over_iterations_aggregated.csv};
    \addplot[oi-orange!25, forget plot] fill between[of=low and high]; 
    % Median    
    \addplot[mark=square, mark size=1.5pt, thick, color=oi-orange]
      table[x=iteration, y=error_adaptive_median, col sep=comma] {data/empirical_error_analysis/error_over_iterations_aggregated.csv};
    \addlegendentry{Adaptive encoding}

\end{axis}

\end{tikzpicture}
    \caption{Static piston problem: median relative $H^1$ error $\relErrorHOne$ versus coupling iteration $\iter$ for fixed and adaptive encodings across $\nRuns=10$ independent runs. Shaded bands denote the interquartile range.}
    \label{fig:empiricalErrorAnalysisHistory}
\end{figure}
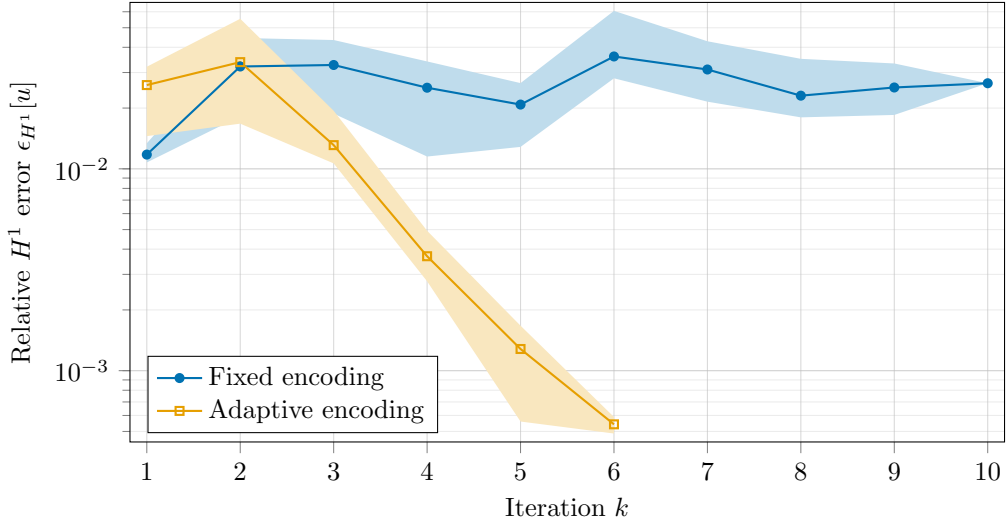

To quantify this effect, we compare the relative $H^1$ error over coupling iterations for fixed versus adaptive encodings across $\nRuns = 10$ independent runs under identical settings: $\numQubitsPerNode = 8$, initial coefficient ranges $[0,1]$ for all nodes, and relaxation factor $\relaxationFactor = 1$. The coupling terminates when the relative change in displacement falls below $\tolRelChangeDisplacement = 2\times 10^{-2}$ or when $\maxIter = 15$ iterations are reached. The results, summarized in \Cref{fig:empiricalErrorAnalysisHistory}, show that with a fixed encoding the error remains approximately constant at a few percent throughout the coupling process. In contrast, the adaptive encoding exhibits a monotonic decrease in error over iterations, achieving a final relative $H^1$ error of about $5\times 10^{-4}$ under the same binary budget and, typically, requiring fewer coupling steps to meet the convergence criterion. These observations demonstrate that adaptive encoding strategy can significantly improve accuracy without increasing the number of binary variables.

\subsection{Structural Design Optimization with Adaptive Encoding Strategy}
\label{subsec:structuralDesignOptimizationWithAdaptiveNumberRepresentation}
To assess the impact of the adaptive encoding strategy on solution accuracy under a fixed budget of binary variables in structural design optimization, we revisit the compound‑rod size optimization from the literature~\cite[Sec.~3.2]{Key2024}, which also uses \gls{qa} for the optimization. In this example, a one‑dimensional rod with $\nElem=2$ elements and under self‑weight loading is studied. In particular, the optimal choice of element cross‑sectional areas for minimum compliance is sought. 
We solve this optimization using the quadratic‑penalty method described in \Cref{subsubsec:quadraticPenaltyMethodForStructuralDesignOptimization}.
We use the same geometry, material properties, boundary conditions, and load case.
The design variables remain the element‑wise cross‑sectional areas $\crossSectionalArea[\elemIndex]\in\setAdmissibleDesign_{\elemIndex}=\{\crossSectionalAreaChoice{1},\,\crossSectionalAreaChoice{2}\}=\{0.25,0.5\}$. 
In the reference, one binary variable per $\crossSectionalArea[\elemIndex]$ was used and a uniform fixed‑precision encoding was employed with $\numQubitsPerNode=3$ for the real‑valued coefficients $\basisCoeff_{\nodeIndex}\in[0,1]$, yielding $\numQubits=26$ binary variables (including those introduced by degree reduction). 
\bigskip\par
In the present study, we retain the same binary budget ($\numQubitsPerNode=3\rightarrow\numQubits=26$) but replace the uniform encoding with the proposed adaptive scheme that assigns variable‑specific ranges within the quadratic‑penalty formulation. 
As in the reference, we assess solution quality by comparing the internal force distribution $\force(\x)$.
To ensure comparability, all experiments are conducted on the same \gls{qa} hardware used in~\cite{Key2024}, namely D\mbox{-}Wave’s \texttt{Advantage\_system4.1}. 
\added{The best approximation is again determined by exhaustive enumeration of all binary configurations, yielding the exact optimum of the corresponding encoded binary problem.}
For the adaptive encoding, we initialize the ranges to $[0,1]$, as used for the fixed encoding in the reference, and use a relaxation factor $\relaxationFactor=0.5$ for updates.
In the quadratic penalty method, the initial penalty weight is set to $\penaltyWeight^{(0)}=5$ (matching the reference) with update factor $\penaltyWeightFactor=1.5$; the feasibility tolerance is $\tolFeasibility=\num{1e-9}$, and we allow up to $\maxIter=50$ iterations.
Within a penalty iteration, each \gls{qa} run  uses an annealing time $\annealingTime=\SI{10}{\micro\second}$ and $\numReads=800$.
\bigskip\par
% --- Read CSV and capture final error as a plain string ---
\pgfplotstableread[col sep=comma]{data/structural_design_optimization/error_over_iterations.csv}\errtable
\pgfplotstablegetrowsof{\errtable}
\pgfmathtruncatemacro{\nrows}{\pgfplotsretval}
\pgfmathtruncatemacro{\lastrow}{\nrows-1}

% Immediately freeze the CSV cell into a macro (plain string)
\pgfplotstablegetelem{\lastrow}{error}\of{\errtable}
\edef\EFINALSTR{\pgfplotsretval}

% Define the reference error as a plain string (literal)
\edef\EREFSTR{0.015873015873015817}

% --- Compute derived quantities using FPU on the strings ---
\pgfkeys{/pgf/fpu=true}

% Parse both strings into pgf float tokens
\pgfmathfloatparsenumber{\EFINALSTR}\let\efloat\pgfmathresult
\pgfmathfloatparsenumber{\EREFSTR}\let\erefFloat\pgfmathresult

% Improvements, ratio, orders (still floats)
\pgfmathparse{(\erefFloat - \efloat)/\erefFloat}   \let\relimprFloat\pgfmathresult
\pgfmathparse{100*\relimprFloat}                   \let\relimprPctFloat\pgfmathresult
\pgfmathparse{\erefFloat/\efloat}                  \let\ratioFloat\pgfmathresult
\pgfmathparse{log10(\erefFloat) - log10(\efloat)}  \let\ordersFloat\pgfmathresult

% Convert floats to printable strings
\pgfmathprintnumberto{\efloat}{\EFINAL}              
\pgfmathprintnumberto{\erefFloat}{\EREF}             
\pgfmathprintnumberto{\relimprFloat}{\RELIMPR}
\pgfmathprintnumberto{\relimprPctFloat}{\RELIMPRPCT}
\pgfmathprintnumberto{\ratioFloat}{\RATIO}
\pgfmathprintnumberto{\ordersFloat}{\ORDERS}

\pgfkeys{/pgf/fpu=false}
With these settings, we apply the quadratic penalty method to solve the fully coupled design optimization in \gls{qubo} form via \gls{qa} until the feasibility tolerance is met; this occurs at iteration $k=\num{26}$. 
\Cref{fig:historyQuadraticPenaltyMethod} illustrates the effectiveness of the adaptive strategy, showing the evolution of nodal coefficients, their ranges, and the resulting solution error in a combined view.
\par
The top and middle panels show the evolution of the nodal coefficients ($\basisCoeff_{\nodeIndex}$) and their corresponding range limits. This view makes the update mechanism apparent: the range limits progressively tighten from below or above depending on whether the solution decreases or increases, converging toward the final solution value over the iterations. The range‑extension rule is also evident: for $\basisCoeff_0$, the upper limit expands after the solution reaches the upper bound at iteration $\num{10}$ (see the magnified view).
\par
To quantify the impact of the adaptive strategy, we assess accuracy in the continuous field $\force(\x)$ using the relative $H^1$ error $\relErrorHOne[\force]$ with respect to the analytic solution $\forceAnalytic(\x)$. 
As a baseline, the fixed-range encoding yields an error of $\relErrorHOne^{\mathrm{ref}}=\num[exponent-mode = scientific,round-mode=places,round-precision = 2]{\EREFSTR}$. 
The error history for the iterative quadratic‑penalty method with adaptive ranges is shown in the bottom panel of \Cref{fig:historyQuadraticPenaltyMethod}.
For the first two iterations, the baseline error is reproduced because the ranges have not yet been updated. Thereafter, a pronounced reduction is observed, although temporary increases can occur when, due to the coarse discretization with $\numQubitsPerNode=3$, the representable numbers within the current range are poorly aligned with the underlying solution.
\par
A key insight emerges when comparing the panels: while the values of the nodal coefficients (top and middle panels) do not change substantially compared to their initial value, the solution accuracy (bottom panel) improves dramatically. 
This is an expected outcome of our approach. As a global optimizer acting on the fully-coupled problem, \gls{qa} finds the best possible solution within the discretization available at each step. 
This is confirmed in the bottom panel, where we additionally report the best attainable error under the current encoding for the best approximation; its coincidence with the \gls{qa} result indicates that the solver reaches the encoding‑optimal solution at each iteration. 
This shows that the initial iteration provides a strong first approximation of the solution, and the subsequent adaptive steps act as a refinement process to systematically increase its precision.
The effect of this refinement is substantial: compared with the baseline, the final error improves by more than three orders of magnitude and by \RELIMPRPCT\% in relative terms (see \Cref{tab:errorComparisonQuadraticPenaltyMethod}).
Therefore, a central benefit of our adaptive strategy is not a significant correction of coefficient values, but the striking increase in their encoding resolution and the resulting accuracy in the continuous variables.
This enables high-accuracy solutions for continuous variables in engineering applications, establishing \gls{qa} as a viable tool also for high-precision optimization tasks.
\begin{figure}
    \centering
    % This file was created with matplot2tikz v0.4.2.
\begin{tikzpicture}[spy using outlines={rectangle, red, magnification=3,
                       size=2cm, connect spies}]%[spy using overlays={size=12mm}]%[spy using outlines={circle, magnification=3, size=2cm, connect spies}]

\begin{groupplot}[group style={group size=1 by 3}]
\nextgroupplot[
    width=0.8\linewidth,
    height=0.45\linewidth,
legend cell align={left},
legend style={
  fill opacity=0.8,
  draw opacity=1,
  text opacity=1,
  at={(0.97,0.03)},
  anchor=south east,
  draw=oi-grey
},
tick align=outside,
tick pos=left,
x grid style={oi-grey},
xmin=-0.5, xmax=26.5,
xtick style={color=black},
xtick={0,2,4,6,8,10,12,14,16,18,20,22,24,26},
xticklabel=\empty,
y grid style={oi-grey},
ylabel={Value Node 0},
ymin=-0.05, ymax=1.05,
ytick style={color=black},
ytick={-0.2,0,0.2,0.4,0.6,0.8,1,1.2},
yticklabels={
  \(\displaystyle {\ensuremath{-}0.2}\),
  \(\displaystyle {0.0}\),
  \(\displaystyle {0.2}\),
  \(\displaystyle {0.4}\),
  \(\displaystyle {0.6}\),
  \(\displaystyle {0.8}\),
  \(\displaystyle {1.0}\),
  \(\displaystyle {1.2}\)
}
]
\addplot [very thick, oi-blue]
      table[x=iteration,y=node_0,col sep=comma] {data/structural_design_optimization/range_limits_over_iterations.csv};
\addlegendentry{$\basisCoeff_{0}$}
\addplot [very thick, black, dotted]
      table[x=iteration,y=node_0_min,col sep=comma] {data/structural_design_optimization/range_limits_over_iterations.csv};
\addlegendentry{$\basisCoeff_{0,\min}$}
\addplot [very thick, black, dashed]
      table[x=iteration,y=node_0_max,col sep=comma] {data/structural_design_optimization/range_limits_over_iterations.csv};
\addlegendentry{$\basisCoeff_{0,\max}$}
\spy [oi-grey, thick] on (4.7,4.7) in node at (4,2.75);
\nextgroupplot[
    width=0.8\linewidth,
    height=0.45\linewidth,
legend cell align={left},
legend style={fill opacity=0.8, draw opacity=1, text opacity=1, draw=oi-grey},
tick align=outside,
tick pos=left,
x grid style={color=oi-grey},
xmin=-0.5, xmax=26.5,
xtick style={color=black},
xtick={0,2,4,6,8,10,12,14,16,18,20,22,24,26},
xticklabel=\empty,
y grid style={color=oi-grey},
ylabel={Value Node 1},
% ymajorgrids,
ymin=-0.05, ymax=1.05,
ytick style={color=black},
ytick={-0.2,0,0.2,0.4,0.6,0.8,1,1.2},
]
\addplot [very thick, oi-blue]
      table[x=iteration,y=node_1,col sep=comma] {data/structural_design_optimization/range_limits_over_iterations.csv};
\addlegendentry{$\basisCoeff_{1}$}
\addplot [very thick, black, dotted]
      table[x=iteration,y=node_1_min,col sep=comma] {data/structural_design_optimization/range_limits_over_iterations.csv};
\addlegendentry{$\basisCoeff_{1,\min}$}
\addplot [very thick, black, dashed]
      table[x=iteration,y=node_1_max,col sep=comma] {data/structural_design_optimization/range_limits_over_iterations.csv};
\addlegendentry{$\basisCoeff_{1,\max}$}

\nextgroupplot[
    width=0.8\linewidth,
    height=0.45\linewidth,
xtick={0,2,4,6,8,10,12,14,16,18,20,22,24,26},
xmin=-0.5, xmax=26.5,
    xlabel={Iteration $k$},
    ylabel={Relative $H^1$ error $\relErrorHOne[\force]$},
    ymode=log,
    log basis y=10,
    grid=both,
    grid style={opacity=0.6},
    minor grid style={opacity=0.3},
    legend style={at={(0.02,0.02)}, anchor=south west, cells={anchor=west}},
    legend columns=1,
    tick align=outside,
    tick pos=left,
  ]
    % Best approximation error
    \addplot[only marks, mark=square, mark size=2.5pt, thick, color=oi-orange]
      table[x=iteration, y=error_ba, col sep=comma] {data/structural_design_optimization/error_over_iterations.csv};
    \addlegendentry{Best approximation}

    % Main error
    \addplot[mark=*, mark size=1.5pt, thick, color=oi-blue]
      table[x=iteration, y=error, col sep=comma] {data/structural_design_optimization/error_over_iterations.csv};
    \addlegendentry{Solution \acrshort{qa}}

    \addplot[mark=none, dashed, thick, color=black]
      table[
        x=iteration,
        % y expr=0.015873015873015817,
        y expr=\erefFloat,
        col sep=comma
      ] {data/structural_design_optimization/error_over_iterations.csv};
    % \addlegendentry{Reference~\cite{Key2024}}
    \addlegendentry{Reference}
\end{groupplot}

\end{tikzpicture}
    \caption{Structural design optimization for the composite rod: Top and Middle: evolution of nodal coefficients $\basisCoeff_{\nodeIndex}$ and their adaptively updated ranges during the quadratic penalty iterations. Bottom: Resulting history for the relative $H^1$ error of the \gls{qa} solution, together with the best‑approximation error and the reference value from literature. The figure highlights the direct relationship between the shrinking interval bounds and the reduction in solution error.}
    \label{fig:historyQuadraticPenaltyMethod}
\end{figure}
% \begin{figure}
%     \centering
%     \input{fig/results/design_optimization/2_elements/QA/history_error}
%     \caption{Structural design optimization for the composite rod: error history versus iteration for the quadratic penalty method, including the best‑approximation error, the relative $H^1$ error of the \gls{qa} solution, and the reference value from literature.}
%     \label{fig:errorHistoryQuadraticPenaltyMethod}
% \end{figure}
\begin{table}
\centering
\sisetup{round-mode=places, round-precision=6}
% \begin{tabular}{cccc}
% \toprule
%   {$\relErrorHOne$ (final)}
% & {$\relErrorHOne^{\mathrm{ref}}$ (reference)}
% & {Improvement (orders of magnitude)}
% & {Relative improvement [\%]}\\
% \midrule
%   {\EFINAL}
% & {\EREF}
% & {\ORDERS}
% & {\RELIMPRPCT\%}
% \\
% \bottomrule
% \end{tabular}

\begin{tabularx}{\textwidth}{CCCC}
\toprule
\multicolumn{2}{c}{Relative $H^1$ error} & \multicolumn{2}{c}{Improvement} \\
\cmidrule(lr){1-2}\cmidrule(lr){3-4}
{Final $\relErrorHOne$} & {Reference $\relErrorHOne^{\mathrm{ref}}$} & {Orders of magnitude} & {Relative [\%]} \\
\midrule
{\EFINAL} & {\EREF} & {\ORDERS} & {\RELIMPRPCT} \\
\bottomrule
\end{tabularx}

\caption{Structural design optimization for the composite rod: comparison of the final relative $H^1$ error $\relErrorHOne$ with the reference value $\relErrorHOne^{\mathrm{ref}}$ from Ref.~\cite{Key2024}, reporting the improvement in orders of magnitude $\log_{10}(\relErrorHOne^{\mathrm{ref}}/\relErrorHOne)$ and the relative improvement as a percentage $100\cdot(\relErrorHOne^{\mathrm{ref}}-\relErrorHOne)/\relErrorHOne^{\mathrm{ref}}$.}
\label{tab:errorComparisonQuadraticPenaltyMethod}
\end{table}
\bigskip\par
Having established performance under the baseline configuration where \gls{qa} consistently returned the best (encoding‑optimal) approximation at each iteration, we now assess the robustness and practical applicability of the proposed adaptive encoding strategy beyond this baseline scenario. First, we vary the relaxation factor $\relaxationFactor$ that governs range updates in the adaptive encoding to evaluate convergence behavior under more or less aggressive adaptations. Second, we probe sensitivity to the initial encoding ranges to test whether the method depends on careful prior scaling. Third, we reduce the number of reads $\numReads$ to increase the probability of suboptimal samples, quantifying how deviations from encoding‑optimal solutions influence the error convergence. 
Finally, we assess the scalability of the approach by applying it to a larger problem using a quantum-inspired solver.
\par
Because the adaptive ranges are path-dependent, the potential variability in the \gls{qa} samples causes different runs to follow distinct update trajectories and attain different approximations. We therefore report the median relative $H^1$ error with interquartile ranges across the runs, which reflects typical convergence behavior.
\added{This path dependence also qualifies the global-search interpretation of the method: at each iteration, the full coupled objective is optimized over the current encoded domain; however, this does not guarantee global optimality with respect to the original continuous-variable domain. In particular, if an intermediate \gls{qa} solution is suboptimal, an overly aggressive range update may contract the encoded domain around an incorrect region and potentially exclude the true optimum from subsequent searches. The relaxation factor therefore controls a trade-off between rapid refinement and the risk of premature contraction.}

\subsubsection{Effect of the Range Relaxation Factor in the Adaptive Encoding}
The relaxation factor $\relaxationFactor$ governs how aggressively the adaptive encoding updates variable ranges between penalty iterations.
\replaced{Small $\relaxationFactor$ values produce conservative updates, which can reduce the risk of premature contraction and preserve stability, but may slow convergence if the ranges lag behind the iterates and quantization remains too coarse. Large $\relaxationFactor$ values respond more strongly to recent samples and may accelerate refinement, but can increase the risk of contracting around a suboptimal region.}{Small $\relaxationFactor$ values make updates conservative, which can preserve stability but slow convergence if ranges lag the iterates and quantization remains too coarse. Large $\relaxationFactor$ values make updates reactive to recent samples, which can accelerate progress but risk settling on a tight, infeasible region too early. }
Given these competing effects, we assess how $\relaxationFactor$ influences convergence speed and final solution quality.
\par
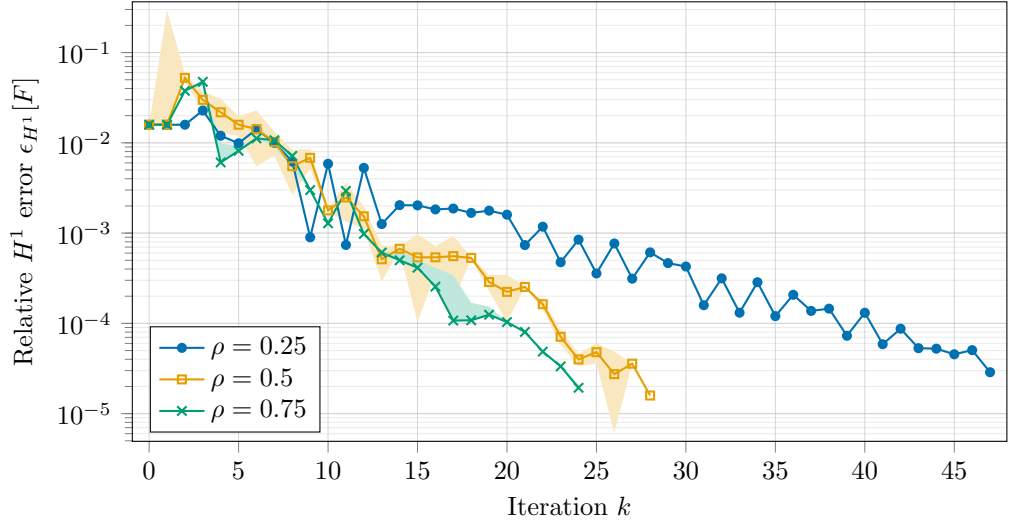
\begin{figure}
    \centering
    \begin{tikzpicture}
  \begin{axis}[
    width=0.8\linewidth,
    height=0.45\linewidth,
    xlabel={Iteration $k$},
    ylabel={Relative $H^1$ error $\relErrorHOne[\force]$},
    ymode=log,
    log basis y=10,
    grid=both,
    grid style={opacity=0.6},
    minor grid style={opacity=0.3},
    legend style={at={(0.02,0.02)}, anchor=south west, cells={anchor=west}},
    legend columns=1,
    tick align=outside,
    tick pos=left,
    enlargelimits=0.02,
  ]

    % 0.25 — shaded IQR band
    \addplot[name path=low025, draw=none, forget plot]
      table[x=iteration, y=error_q25, col sep=comma]
      {data/structural_design_optimization/relaxation_factor/error_over_iterations_aggregated_0_25.csv};
    \addplot[name path=high025, draw=none, forget plot]
      table[x=iteration, y=error_q75, col sep=comma]
      {data/structural_design_optimization/relaxation_factor/error_over_iterations_aggregated_0_25.csv};
    \addplot[oi-blue!25, forget plot] fill between[of=low025 and high025];
    % 0.25
    \addplot[mark=*, mark size=1.5pt, thick, color=oi-blue]
      table[x=iteration, y=error_median, col sep=comma] {data/structural_design_optimization/relaxation_factor/error_over_iterations_aggregated_0_25.csv};
    \addlegendentry{$\relaxationFactor=0.25$}

    % 0.5 — shaded IQR band
    \addplot[name path=low05, draw=none, forget plot]
      table[x=iteration, y=error_q25, col sep=comma]
      {data/structural_design_optimization/relaxation_factor/error_over_iterations_aggregated_0_5.csv};
    \addplot[name path=high05, draw=none, forget plot]
      table[x=iteration, y=error_q75, col sep=comma]
      {data/structural_design_optimization/relaxation_factor/error_over_iterations_aggregated_0_5.csv};
    \addplot[oi-orange!25, forget plot] fill between[of=low05 and high05];
    % 0.5
    \addplot+[mark=square, mark size=1.5pt, thick, color=oi-orange]
      table[x=iteration, y=error_median, col sep=comma] {data/structural_design_optimization/relaxation_factor/error_over_iterations_aggregated_0_5.csv};
    \addlegendentry{$\relaxationFactor=0.5$}

    % 0.75 — shaded IQR band
    \addplot[name path=low075, draw=none, forget plot]
      table[x=iteration, y=error_q25, col sep=comma]
      {data/structural_design_optimization/relaxation_factor/error_over_iterations_aggregated_0_75.csv};
    \addplot[name path=high075, draw=none, forget plot]
      table[x=iteration, y=error_q75, col sep=comma]
      {data/structural_design_optimization/relaxation_factor/error_over_iterations_aggregated_0_75.csv};
    \addplot[oi-green!25, forget plot] fill between[of=low075 and high075];
    % 0.75
    \addplot[mark=x, mark size=2.5pt, thick, color=oi-green]
      table[x=iteration, y=error_median, col sep=comma] {data/structural_design_optimization/relaxation_factor/error_over_iterations_aggregated_0_75.csv};
    \addlegendentry{$\relaxationFactor=0.75$}

  \end{axis}
\end{tikzpicture}
    \caption{Structural design optimization for the composite rod: median relative $H^1$ error $\relErrorHOne$ versus penalty iteration $\iter$ for relaxation factors $\relaxationFactor\in\{0.25,0.5,0.75\}$ across $\nRuns=10$ independent runs. Shaded bands denote the interquartile range.}
    \label{fig:errorHistoryRelaxationFactor}
\end{figure}

We sweep $\relaxationFactor$ over $\{0.25,0.5,0.75\}$ while holding other settings fixed. For each $\relaxationFactor$, we perform $\nRuns=10$ independent runs and report median and interquartile ranges.
As shown in \Cref{fig:errorHistoryRelaxationFactor}, smaller 
$\relaxationFactor$ values lead to slower convergence (more iterations to reach the feasibility criterion), while the final error remains comparable across the tested settings.
\subsubsection{Sensitivity to Initial Ranges}
\begin{figure}
    \centering
    \begin{tikzpicture}
  \begin{axis}[
    width=0.8\linewidth,
    height=0.45\linewidth,
    xlabel={Iteration $k$},
    ylabel={Relative $H^1$ error $\relErrorHOne[\force]$},
    ymode=log,
    log basis y=10,
    grid=both,
    grid style={opacity=0.6},
    minor grid style={opacity=0.3},
    legend style={at={(0.02,0.02)}, anchor=south west, cells={anchor=west}},
    legend columns=1,
    tick align=outside,
    tick pos=left,
    enlargelimits=0.02,
  ]

    % [0,1] — shaded IQR band
    \addplot[name path=low01, draw=none, forget plot]
      table[x=iteration, y=error_q25, col sep=comma]
      {data/structural_design_optimization/initial_range/error_over_iterations_aggregated_0_1.csv};
    \addplot[name path=high01, draw=none, forget plot]
      table[x=iteration, y=error_q75, col sep=comma]
      {data/structural_design_optimization/initial_range/error_over_iterations_aggregated_0_1.csv};
    \addplot[oi-blue!25, forget plot] fill between[of=low01 and high01];  
    % [0,1]
    \addplot[mark=*, mark size=1.5pt, thick, color=oi-blue]
      table[x=iteration, y=error_median, col sep=comma] {data/structural_design_optimization/initial_range/error_over_iterations_aggregated_0_1.csv};
    \addlegendentry{$[0,1]$}

    % [0,5] — shaded IQR band
    \addplot[name path=low05, draw=none, forget plot]
      table[x=iteration, y=error_q25, col sep=comma]
      {data/structural_design_optimization/initial_range/error_over_iterations_aggregated_0_5.csv};
    \addplot[name path=high05, draw=none, forget plot]
      table[x=iteration, y=error_q75, col sep=comma]
      {data/structural_design_optimization/initial_range/error_over_iterations_aggregated_0_5.csv};
    \addplot[oi-orange!25, forget plot] fill between[of=low05 and high05]; 
    % [0,5]
    \addplot+[mark=square, mark size=1.5pt, thick, color=oi-orange]
      table[x=iteration, y=error_median, col sep=comma] {data/structural_design_optimization/initial_range/error_over_iterations_aggregated_0_5.csv};
    \addlegendentry{$[0,5]$}

    % [0,10] — shaded IQR band
    \addplot[name path=low010, draw=none, forget plot]
      table[x=iteration, y=error_q25, col sep=comma]
      {data/structural_design_optimization/initial_range/error_over_iterations_aggregated_0_10.csv};
    \addplot[name path=high010, draw=none, forget plot]
      table[x=iteration, y=error_q75, col sep=comma]
      {data/structural_design_optimization/initial_range/error_over_iterations_aggregated_0_10.csv};
    \addplot[oi-green!25, forget plot] fill between[of=low010 and high010]; 
    % [0,10]
    \addplot[mark=x, mark size=2.5pt, thick, color=oi-green]
      table[x=iteration, y=error_median, col sep=comma] {data/structural_design_optimization/initial_range/error_over_iterations_aggregated_0_10.csv};
    \addlegendentry{$[0,10]$}

  \end{axis}
\end{tikzpicture}
    \caption{Structural design optimization for the composite rod: median relative $H^1$ error $\relErrorHOne$ versus penalty iteration $\iter$ for different initial ranges across $\nRuns=10$ independent runs. Shaded bands denote the interquartile range.}
    \label{fig:errorHistoryInitialRanges}
\end{figure}
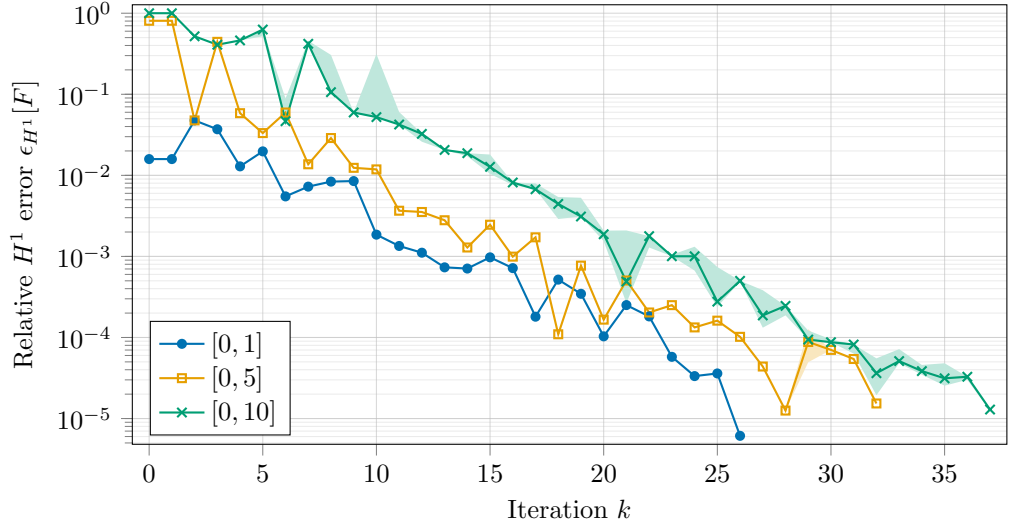

So far, the initial ranges for the nodal force coefficients $\basisCoeff_i$ were $[0,1]$, i.e., aligned with the expected order of magnitude observed in the results. We now examine how deviations from this scale affect convergence and solution quality.
To this end, we vary the initial per‑variable ranges over $[0,1]$, $[0,5]$, and $[0,10]$ while holding all other settings fixed. For each choice, we perform $\nRuns=10$ independent runs and report median and interquartile range for the relative $H^1$ error in \Cref{fig:errorHistoryInitialRanges}.
The plot shows that initializing with wider ranges ($[0,5]$ and $[0,10]$) yields a higher  initial relative $H^1$ error than $[0,1]$. This is expected since, with a fixed bit‑depth, larger ranges provide only a reduced local resolution in the region where the analytic solution lies (roughly $[0,1]$). As a result, more penalty iterations are required to reach the feasibility criterion for the wider ranges. Despite these differences in the transient phase, the final error after convergence remains comparable across all three initializations, indicating that the adaptive updates ultimately recover sufficient resolution around the relevant region.
\subsubsection{Impact of Suboptimal Solutions}
Finally, we evaluate robustness under reduced reads by performing $n=10$ independent runs of the full quadratic‑penalty procedure with $\numReads=400$ and $\numReads=800$. With fewer reads $\numReads$, stochastic variability increases, raising the likelihood that \gls{qa} returns suboptimal solutions. As shown in \Cref{fig:solutionHistoryQuadraticPenaltyMethodNumReads}, even with fewer reads, the median error converges at a rate comparable to the baseline and reaches similar final accuracy, indicating that the approach is resilient to suboptimal samples.
\begin{figure}
    \centering
    \begin{tikzpicture}
  \begin{axis}[
    width=0.8\linewidth,
    height=0.45\linewidth,
    xlabel={Iteration $k$},
    ylabel={Relative $H^1$ error $\relErrorHOne[\force]$},
    ymode=log,
    log basis y=10,
    grid=both,
    grid style={opacity=0.6},
    minor grid style={opacity=0.3},
    legend style={at={(0.02,0.02)}, anchor=south west, cells={anchor=west}},
    legend columns=1,
    tick align=outside,
    tick pos=left,
    enlargelimits=0.02,
  ]

    % num_reads 800 - shaded IQR band
    \addplot[name path=low025, draw=none, forget plot]
      table[x=iteration, y=error_q25, col sep=comma]
      {data/structural_design_optimization/num_reads/error_over_iterations_aggregated_800.csv};
    \addplot[name path=high025, draw=none, forget plot]
      table[x=iteration, y=error_q75, col sep=comma]
      {data/structural_design_optimization/num_reads/error_over_iterations_aggregated_800.csv};
    \addplot[oi-blue!25, forget plot] fill between[of=low025 and high025];
    % num_reads 800
    % \addplot[mark=*, mark size=1.5pt, thick, color=oi-blue]
    %   table[x=iteration, y=error_h1_rel, col sep=comma] {data/design_optimization/2_elements/QA/history.csv};
    % \addlegendentry{$\numReads=800$ (single run)}
    \addplot[mark=*, mark size=1.5pt, thick, color=oi-blue]
      table[col sep=comma, x=iteration, y=error_median]{data/structural_design_optimization/num_reads/error_over_iterations_aggregated_800.csv};
    \addlegendentry{$\numReads=800$}

    % num_reads 400 - shaded IQR band
    \addplot[name path=low025, draw=none, forget plot]
      table[x=iteration, y=error_q25, col sep=comma]
      {data/structural_design_optimization/num_reads/error_over_iterations_aggregated_400.csv};
    \addplot[name path=high025, draw=none, forget plot]
      table[x=iteration, y=error_q75, col sep=comma]
      {data/structural_design_optimization/num_reads/error_over_iterations_aggregated_400.csv};
    \addplot[oi-orange!25, forget plot] fill between[of=low025 and high025];
    % num_reads 400 - median error
    \addplot+[mark=square, mark size=1.5pt, thick, color=oi-orange]
      table[col sep=comma, x=iteration, y=error_median]{data/structural_design_optimization/num_reads/error_over_iterations_aggregated_400.csv};
    \addlegendentry{$\numReads=400$}

    % \addplot[mark=none, dashed, thick, color=black]
    %   table[
    %     x=iteration,
    %     % y expr=0.015873015873015817,
    %     y expr=\erefFloat,
    %     col sep=comma
    %   ] {data/structural_design_optimization/num_reads/error_over_iterations_aggregated_400.csv};
    % \addlegendentry{Reference~\cite{Key2024}}

  \end{axis}
\end{tikzpicture}
    % \caption{Median relative $H^1$ error $\relErrorHOne$ versus penalty iteration $\iter$ for $\numReads=800$ (single run) and $\numReads=400$ ($\nRuns=10$ independent runs). Shaded bands denote the interquartile range.}
    \caption{Structural design optimization for the composite rod: median relative $H^1$ error $\relErrorHOne$ versus penalty iteration $\iter$ for $\numReads=800$ and $\numReads=400$ across $\nRuns=10$ independent runs. Shaded bands denote the interquartile range.}
    \label{fig:solutionHistoryQuadraticPenaltyMethodNumReads}
\end{figure}
\bigskip\par
To summarize the parameter studies, the adaptive encoding strategy is robust across variations in the relaxation factor $\relaxationFactor$, initial range scales, and \gls{qa} read budgets: conservative $\relaxationFactor$ values delay feasibility attainment but do not degrade the final objective quality; wider initial ranges increase early quantization error yet converge to comparable final relative $H^1$ errors; and smaller read budgets raise run‑to‑run variability without materially affecting median outcomes. These trends delineate a broad envelope of practical hyperparameters for the penalty‑based \gls{qa} solver with adaptive encoding.
\subsubsection{Scalability Analysis with a Quantum-Inspired Solver}
Having established the method's effectiveness on hardware-sized problems, we now investigate its scalability to a larger, more complex instance. As established in \Cref{subsubsec:principleOfMinimumComplementaryEnergyForStructuralDesignOptimization}, the design optimization formulation naturally results in a cubic polynomial, leading to a third-order \gls{pubo} problem.
We previously noted that mapping this \gls{pubo} to a \gls{qubo} via degree reduction is a necessary step for current \gls{qa} devices. Since this process introduces a large number of auxiliary variables and a dense, highly-connected interaction graph that is notoriously difficult to embed, it is known to degrade optimization performance and solution quality significantly.
\par
To circumvent this critical bottleneck for our scalability study, we employ Toshiba's \texttt{SQBM+}
\footnote{\url{https://www.global.toshiba/ww/products-solutions/ai-iot/sbm.html}} 
solver, a quantum-inspired solver capable of optimizing higher-order problems directly. This allows us to work with the natural formulation of the problem without degree reduction. 
For this analysis, we used the same problem setup as above except for the rod being composed of $\nElem=15$ elements now. 
This increased number of elements makes the optimization landscape intrinsically more challenging. Specifically, because the total structural compliance is a global property calculated by summing contributions from all elements, a change in a single element's design has a proportionally smaller impact on the final objective value.
This requires the optimizer to become far more sensitive, which in turn informs the choice of settings in our adaptive strategy. First, a small relaxation factor, $\relaxationFactor=0.25$, is used to allow the search to proceed more deliberately over more iterations and avoid premature convergence to a design very similar to the optimal one. Second, we employ a larger number of binary variables ($\numQubitsPerNode=6$) from the outset. This ensures the initial encoding has sufficient resolution to represent the small variations in the continuous force variables that distinguish between designs with very similar compliance values.
Together with the binary variables for the design choices, the total number of binary variables is $\numQubits=105$.
So, the final coupled binary problem involves over $3\times10^4$ possible designs and more than $10^{31}$ solutions.
\bigskip\par
For the quadratic penalty method, we set the initial penalty weight to $\penaltyWeight^{(0)}=1$ with update factor $\penaltyWeightFactor=1.25$; the feasibility tolerance is $\tolFeasibility=\num{1e-8}$, and we allow up to $\maxIter=100$ iterations.
Within a penalty iteration, we run the \texttt{SQBM+} solver with a timeout of $\SI{5}{\second}$ and $400$ steps.
% \gls{sa} with $\numReads=800$
The results of this scalability analysis, presented in \Cref{fig:solutionHistoryScalingAnalysis}, validate that the benefit of the adaptive strategy is maintained when scaling to larger problems. As above, the figure plots the median relative $H^1$ error over $\nRuns=5$ independent runs, showing a clear convergence to the optimal design and the corresponding high-precision solution. The algorithm successfully navigates the challenging optimization landscape, reducing the median error from an initial value of around $0.5$ down to almost $\num{1e-4}$ in roughly $60$ iterations for all runs. The final low error and minimal variance across runs confirm that our approach is robust and can consistently find the optimal design and accurate solutions even for these larger, more difficult problems.
\begin{figure}
    \centering
    \begin{tikzpicture}
  \begin{axis}[
    width=0.8\linewidth,
    height=0.45\linewidth,
    xlabel={Iteration $k$},
    ylabel={Relative $H^1$ error $\relErrorHOne[\force]$},
    ymode=log,
    ymin=0.0001,
    ymax=1,
    log basis y=10,
    grid=both,
    grid style={opacity=0.6},
    minor grid style={opacity=0.3},
    legend style={at={(0.02,0.02)}, anchor=south west, cells={anchor=west}},
    legend columns=1,
    tick align=outside,
    tick pos=left,
    enlargelimits=0.02,
  ]

    % num_reads 800 - shaded IQR band
    \addplot[name path=low025, draw=none, forget plot]
      table[x=iteration, y=error_q25, col sep=comma]
      {data/structural_design_optimization/scaling_analysis/error_over_iterations_aggregated.csv};
    \addplot[name path=high025, draw=none, forget plot]
      table[x=iteration, y=error_q75, col sep=comma]
      {data/structural_design_optimization/scaling_analysis/error_over_iterations_aggregated.csv};
    \addplot[oi-blue!25, forget plot] fill between[of=low025 and high025];
    \addplot[mark=*, mark size=1.5pt, thick, color=oi-blue]
      table[col sep=comma, x=iteration, y=error_median]{data/structural_design_optimization/scaling_analysis/error_over_iterations_aggregated.csv};
    \addlegendentry{Solution (Toshiba \texttt{SQBM+})}

    % num_reads 400 - shaded IQR band
    % \addplot[name path=low025, draw=none, forget plot]
    %   table[x=iteration, y=error_q25, col sep=comma]
    %   {data/structural_design_optimization/num_reads/error_over_iterations_aggregated_400.csv};
    % \addplot[name path=high025, draw=none, forget plot]
    %   table[x=iteration, y=error_q75, col sep=comma]
    %   {data/structural_design_optimization/num_reads/error_over_iterations_aggregated_400.csv};
    % \addplot[oi-orange!25, forget plot] fill between[of=low025 and high025];
    % % num_reads 400 - median error
    % \addplot+[mark=square, mark size=1.5pt, thick, color=oi-orange]
    %   table[col sep=comma, x=iteration, y=error_median]{data/structural_design_optimization/num_reads/error_over_iterations_aggregated_400.csv};
    % \addlegendentry{$\numReads=400$}

  \end{axis}
\end{tikzpicture}
    \caption{Structural design optimization for the composite rod with $\nElem=15$ elements: median relative $H^1$ error $\relErrorHOne$ versus penalty iteration $\iter$ across $\nRuns=5$ independent runs. Shaded bands denote the interquartile range.}
    \label{fig:solutionHistoryScalingAnalysis}
\end{figure}

%% DISCUSSION %%
% \section{Discussion}
% \label{sec:discussion}
% \input{discussion}
%% CONCLUSION %%
\section{Conclusion}
In this work, we introduced an adaptive encoding strategy for continuous variables in \gls{qa} that decouples local resolution from total binary budget. By dynamically updating representable ranges while keeping the number of binary variables constant, the method concentrates precision where it matters without inflating binary counts or embedding complexity. 
Integrated into a quadratic penalty framework for the minimum complementary energy formulation in structural design optimization---serving as a representative of mixed‑variable engineering optimization---\replaced{the approach re-solves the full coupled problem at each iteration and performs a global search over the current encoded domain, rather than updating the design through decoupled incremental subproblems.}{the approach refines the solution to the full coupled problem at each iteration, preserving the global optimality of the target objective rather than progressing via incremental local updates.}
\bigskip\par
An empirical error analysis of a one-dimensional rod, modeled via the minimum total potential energy principle with a fixed encoding of the continuous displacement variables, clarified why adaptive encoding is crucial for performance on current hardware: increasing the bit depth in a fixed, uniform encoding improves the encoding‑optimal resolution but does not guarantee higher accuracy on hardware. Beyond a moderate bit depth, the solution quality does no longer improve due to size‑induced hardware effects such as \glspl{ice}, even when chain‑break fractions are low. This favors smaller logical problems with higher fidelity and, thus, motivates using adaptive encoding to achieve higher effective precision under a fixed binary budget.
\bigskip\par
On a published structural design test case, coupling the quadratic penalty method with adaptive encoding yielded the optimal design and improved the accuracy of the continuous field by more than three orders of magnitude under the same binary budget\added{, at the cost of multiple \gls{qa} solves across the penalty iterations}. Parameter studies further showed that the approach is robust: across moderate variations of the relaxation factor $\relaxationFactor$, initial encoding ranges, and number of \gls{qa} reads, the final relative $H^1$ error remains comparable to baseline choices. In practice, this enables selecting hyperparameters for stability and cost without sacrificing final accuracy.
Notably, using only $\numQubitsPerNode=3$ binary variables per continuous variable consistently yielded final relative $H^1$ errors in the range $10^{-5}\text{--}10^{-4}$, solely using \gls{qa}.
Furthermore, to confirm its scalability, we demonstrated that the method maintains this robust, high-precision convergence on a larger-scale problem.
\bigskip\par
Despite the strong empirical gains in accuracy and robustness delivered by the adaptive encoding strategy, our study’s scope naturally comes with limitations and provides avenues for future work. Our primary analysis focused on validating the method on a \textit{D‑Wave Advantage} system, ensuring comparability with existing literature. We then took a first step toward assessing scalability and portability by applying the method to a larger problem using a quantum-inspired solver.
\par
However, a systematic, head-to-head benchmark across multiple hardware generations (e.g., newer \textit{Advantage2} systems) and a broader range of alternative Ising-type platforms was beyond the scope of this initial study. Similarly, while we demonstrated scalability on a larger 1D problem, applying the methodology to more complex structural topologies, such as 2D or 3D trusses, remains an important next step. Consequently, future work will focus on these areas: further scaling the approach, conducting extensive cross-platform performance comparisons, and extending the methodology to new classes of structural design problems.
\par
Nonetheless, within these constraints, the results point to broader impact. The proposed framework applies to a wide class of mixed discrete–continuous engineering problems where continuous fields couple to discrete design variables. 
By targeting the full coupled objective at each iteration, \replaced{the method allows \gls{qa} to search globally over the current encoded domain rather than solving separate design and field subproblems, while adaptive encoding improves accuracy under a fixed binary budget.}{it retains the global search advantages of \gls{qa}, while the adaptive encoding improves accuracy under a fixed binary budget.}
These properties suggest practical benefits for design tasks, even beyond structural optimization, and motivate further studies to evaluate applicability and performance gains in different applications.

%%%%%%%%%%%%%%%%
%% REFERENCES %%
%%%%%%%%%%%%%%%%
\printbibliography

\end{document}